\documentclass[fleqn,usenatbib]{mnras}

\usepackage{newtxtext,newtxmath}
\usepackage{ulem}

\usepackage[T1]{fontenc}
\usepackage{ae,aecompl}
\usepackage{color}

\usepackage{graphicx}	
\usepackage{amsmath}	
\usepackage{amssymb}	

\title[Spatially Resolved Stellar Initial Mass Function]{SDSS-IV MaNGA: The Spatially Resolved Stellar Initial Mass Function in $\sim$400 Early-Type Galaxies}

\author[T. Parikh et al.]
{Taniya Parikh$^{1}$\thanks{E-mail: taniya.parikh@port.ac.uk},
Daniel Thomas$^{1}$,
Claudia Maraston$^{1}$,
Kyle B. Westfall$^{2}$,
\newauthor{Daniel Goddard$^{1}$,
Jianhui Lian$^{1}$,
Sofia Meneses-Goytia$^{1}$,
Amy Jones$^{3}$,
Sam Vaughan$^{4}$}	
\newauthor{Brett H. Andrews$^{5}$,
Matthew Bershady$^{6}$,
Dmitry Bizyaev$^{7}$,
Jonathan Brinkmann$^{7}$,}
\newauthor{Joel R. Brownstein$^{8}$,
Kevin Bundy$^{2}$,
Niv Drory$^{9}$,
Eric Emsellem$^{10}$,
David R. Law$^{11}$,}
\newauthor{Jeffrey A. Newman$^{12}$,
Alexandre Roman-Lopes$^{13}$,
David Wake$^{14}$,
Renbin Yan$^{15}$,}
\newauthor{Zheng Zheng$^{16}$}
\\
$^{1}$Institute of Cosmology and Gravitation, University of Portsmouth, 1-8 Burnaby Road, Portsmouth PO1 3FX, UK\\
$^{2}$UCO/Lick Observatory, University of California, Santa Cruz, 1156 High St. Santa Cruz, CA 95064, USA\\
$^{3}$Max-Planck-Institut f\"ur Astrophysik, Karl-Schwarzschild-Str. 1, D-85748 Garching, Germany\\
$^{4}$Sub-department of Astrophysics, Department of Physics, University of Oxford, Denys Wilkinson Building, Keble Road, Oxford OX1 3RH\\
$^{5}$PITT PACC, Department of Physics and Astronomy, University of Pittsburgh, Pittsburgh, PA 15260, USA\\
$^{6}$Department of Astronomy, University of Wisconsin-Madison, 475N. Charter St., Madison WI 53703, USA\\
$^{7}$Apache Point Observatory, P.O. Box 59, Sunspot, NM 88349\\
$^{8}$Department of Physics and Astronomy, University of Utah, 115 S. 1400 E., Salt Lake City, UT 84112, USA\\
$^{9}$McDonald Observatory, The University of Texas at Austin, 1 University Station, Austin, TX 78712, USA\\
$^{10}$European Southern Observatory, Karl-Schwarzschild-Str. 2, 85748 Garching, Germany\\
$^{11}$Space Telescope Science Institute, 3700 San Martin Drive, Baltimore, MD 21218, USA\\
$^{12}$PITT PACC, Department of Physics and Astronomy, University of Pittsburgh, Pittsburgh, PA 15260, USA\\
$^{13}$Departamento de F\'isica, Facultad de Ciencias, Universidad de La Serena, Cisternas 1200, La Serena, Chile\\
$^{14}$Department of Physics, University of North Carolina Asheville, One University Heights, Asheville, NC 28804, USA\\
$^{15}$Department of Physics and Astronomy, University of Kentucky, 505 Rose Street, Lexington, KY 40506, USA\\
$^{16}$National Astronomical Observatories of China, Chinese Academy of Sciences, 20A Datun Road, Beijing 100012, China\\
}

\date{Accepted XXX. Received YYY; in original form ZZZ}

\pubyear{2017}

\begin{document}
\label{firstpage}
\pagerange{\pageref{firstpage}--\pageref{lastpage}}
\maketitle

\begin{abstract}
MaNGA provides the opportunity to make precise spatially resolved measurements of the IMF slope in galaxies owing to its unique combination of spatial resolution, wavelength coverage and sample size. We derive radial gradients in age, element abundances and IMF slope analysing optical and near-infrared absorption features from stacked spectra out to the half-light radius of 366 early-type galaxies with masses $9.9 - 10.8\;\log M/M_{\odot}$. We find flat gradients in age and [$\alpha$/Fe] ratio, as well as negative gradients in metallicity, consistent with the literature. We further derive significant negative gradients in the [Na/Fe] ratio with galaxy centres being well enhanced in Na abundance by up to 0.5 dex. Finally, we find a gradient in IMF slope with a bottom-heavy IMF in the centre (typical mass excess factor of 1.5) and a Milky Way-type IMF at the half-light radius. This pattern is mass-dependent with the lowest mass galaxies in our sample featuring only a shallow gradient around a Milky Way IMF. Our results imply the local IMF-$\sigma$ relation within galaxies to be even steeper than the global relation and hint towards the local metallicity being the dominating factor behind the IMF variations. We also employ different stellar population models in our analysis and show that a radial IMF gradient is found independently of the stellar population model used. A similar analysis of the Wing-Ford band provides inconsistent results and further evidence of the difficulty in measuring and modelling this particular feature.
\end{abstract}




\section{Introduction}
The stellar initial mass function (IMF) is key in characterising stellar populations with implications for a wide range of problems in astro-physics from star formation and stellar evolution to galaxy formation and evolution. Several decades ago, \citet{Salpeter1955} first proposed a single power law for the IMF with a slope of $2.35$. A Milky Way-like IMF, instead, has a turnover to flatter slopes at low masses ($0.5\; M_{\odot}$) as in \citet{Kroupa2001} and \citet{Chabrier2003}. The universality of the IMF has been controversially discussed in the literature ever since.

Studies of resolved stellar populations in different environments within our own Galaxy find little or no deviation from this universal IMF \citep{Scalo1986, Kroupa2001, Bastian2010, Kroupa2013}. An alternative to this approach is to constrain the IMF in unresolved stellar populations through either a combination of stellar population modelling with dynamical mass measurements or stellar populations modelling alone using gravity-sensitive absorption features in the spectrum. The obvious advantage is the extension to other galaxies and stellar systems allowing a statistical assessment of possible variations of the IMF with galaxy properties. The downside is that the measurement is challenging and hugely dependent on analysis methods.

Rapid advances have been made in this regard in recent years. Some consensus appears to be emerging in the recent literature from various different methods that the IMF is not universal with higher fractions of low-mass dwarf stars being detected in more massive galaxies \citep{Treu2010, ThomasJ2011, Conroy2012b, Cappellari2012, Spiniello2012, Ferreras2013,LaBarbera2013,Lyubenova2016}. However, this conclusion remains controversial. \citet{Smith2013} and \citet{Smith2015a} present a handful of nearby strongly lensed massive galaxies with particularly accurate mass determinations that contain Milky Way-type IMFs and hence do not fit into this trend. Moreover, there appears to be no correlation between the various IMF slope variations inferred from different approaches on a galaxy-by-galaxy basis \citep{Smith2014}. It should be noted though that this analysis made use of different data taken over different apertures, and others have found consistency between stellar population and dynamics \citep{Lyubenova2016} and stellar population and lensing \citep[e.g.][]{Spiniello2012}. Still, this issue highlighted by \citep{Smith2014} casts doubts on the robustness of the evidence for a non-universality of the IMF.

Progress in all aspects of the analysis from data quality to modelling techniques is required to resolve this controversy. One way of shedding more light on the issue is to expand the experiment and to study possible variations of the IMF within galaxies. Radial gradients will allow us to investigate whether any IMF variations are driven by local or global properties of galaxies. The possible presence of a gradient in the IMF would also have implications for galaxy formation models, setting further constraints on galaxy growth and outside-in vs inside-out formation mechanisms. In addition, dynamical mass estimates are rather sensitive to IMF gradients, so quantifying gradients is a crucial step in quantifying the stellar mass density \citep{Bernardi2018}.

Spatially resolved measurements of the IMF in galaxies based on stellar population indicators have only just started appearing in the literature. It is time-consuming and expensive to reach the necessary data quality at significant radii and so far, no consensus has been reached. Some studies show the presence of a radial gradient with strongly bottom-heavy IMFs in the centres \citep{Martin-Navarro2015,LaBarbera2016,LaBarbera2017,vanDokkum2016,Conroy2017a}, while other studies find no compelling evidence for radial gradients in the IMF \citep{Vaughan2017, Zieleniewski2015, Zieleniewski2017,Alton2017}. 

The aim of the present paper is to contribute to this effort by harnessing the opportunity provided by data from the survey Mapping Nearby Galaxies at Apache Point Observatory \citep[MaNGA;][]{Bundy2015} to carry out precise spatially resolved measurements of IMF slope in galaxies as a function of galaxy mass and other key parameters. The MaNGA survey is particularly suited for this purpose because of its unique combination of spatial resolution, wavelength coverage, and sample size. We study the IMF and other stellar population parameters for stacked spectra of 366 galaxies, representing the largest sample on which a radial study of the IMF has been carried out so far. We study previously known key absorption features in the spectra and compare these with stellar population models to untangle age, element abundance and IMF effects.

The paper is structured as follows. In Section~2 we give details of the data and stacking procedure along with a brief description of the models used in our analysis. The results of the analysis are presented in Section~3. We discuss our results in the context of the literature in Section~4, followed by our conclusion and scope for further work in Section~5.

\section{Data and analysis tools}
In this section we describe the observational data and the analysis tools used in the present study. In summary, spatially resolved spectroscopy from the MaNGA survey is used. Spectra are stacked in radial bins and across galaxies within mass bins to achieve high S/N. Key absorption features are then measured and analysed with stellar population models.

\subsection{MaNGA}
MaNGA, part of the Sloan Digital Sky Survey IV \citep{Blanton2017}, aims to obtain spatially resolved spectroscopy for 10,000 nearby galaxies at a spectral resolution of $R\sim 2000$ in the wavelength range $3,600-10,300\;$\AA, upon completion in 2020. It makes use of integral-field units (IFUs) to collect this information, with 17 simultaneous observations of galaxies from the output of independent fibre-bundles \citep{Drory2015}. These fibres are fed into the BOSS spectrographs \citep{Smee2013} on the Sloan $2.5\;$m telescope \citep{Gunn2006}. Such high precision spectroscopy for different regions of a galaxy and for such a large sample has never been provided before. MaNGA targets are chosen from the NASA Sloan Atlas catalogue \citep[NSA,][]{Blanton2005} such that there is a uniform distribution in mass \citep{Wake2017}.

At the redshift of a galaxy, optical fibre bundles of different sizes are chosen to ensure the galaxy is covered out to at least $1.5 R_\mathrm{e}$ for the 'Primary' and 'Color-enhanced' samples, and to $2.5 R_\mathrm{e}$ for the 'Secondary' sample \citep{Wake2017}. The Color-enhanced sample supplements colour space that is otherwise under-represented relative to the overall galaxy population. The spatial resolution is $1 - 2\;$kpc at the median redshift of the survey ($z\sim 0.03$), and the $r$-band S/N is $4-8\;$\AA$^{-1}$, for each 2\arcsec\ fibre, at the outskirts of MaNGA galaxies. For more detail on the survey we refer the reader to \citet{Law2015} for MaNGA's observing strategy, to \citet{Yan2016a} for the spectrophotometry calibration, to \citet{Wake2017} for the survey design, and to \citet{Yan2016b} for the initial performance.

\begin{figure}
 	\includegraphics[width=\linewidth]{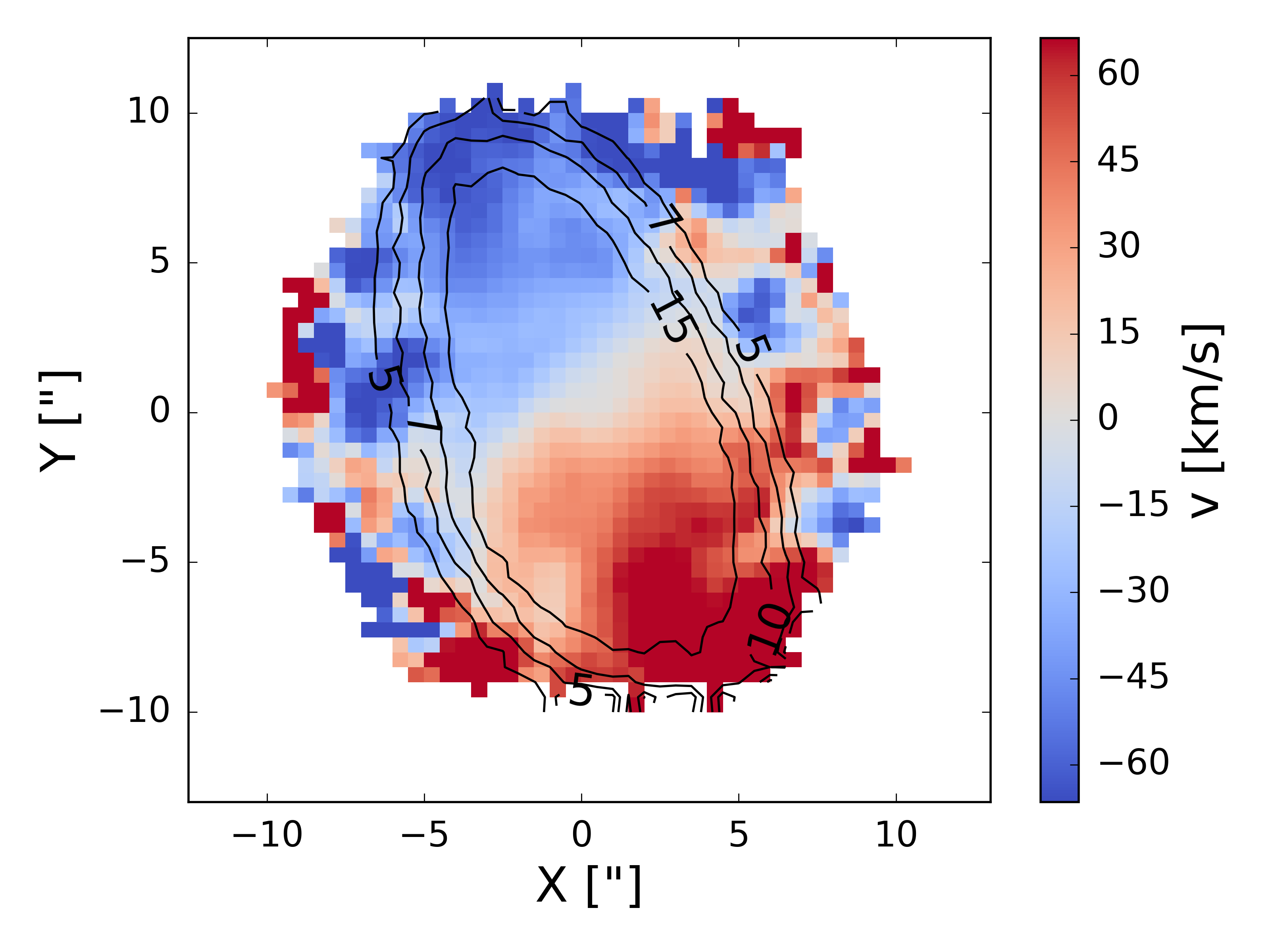}
    \caption{S/N contours (at  $5, 7, 10, 15$\; pixel$^{-1}$) plotted over the stellar velocity map for an example galaxy. The random velocities seen at the edges are due to the spectra being too noisy for a good measurement. We choose spectra with S/N \textgreater \ 7\; pixel$^{-1}$ to achieve the desired balance between ensuring accuracy in v and not loosing too much data.}
    \label{fig:vel_snmaps}
\end{figure}
\subsubsection{Galaxy sample}
We make use of data taken during the first two years of survey operations, equivalent to SDSS's fourteenth data release \citep[DR14]{Abolfathi2017} containing $2812$ datacubes. The galaxy masses range from $\sim 10^{9}-10^{11}\; M_{\odot}$. The Primary and Color-enhanced samples of MaNGA, together known as the Primary$+$ sample, contains 1700 galaxies, out of which 970 are early-type. These were selected using Galaxy Zoo \citep{Lintott2011, Willett2013} morphologies and visually inspected when necessary \citep{Goddard2017}.

\subsubsection{Data reduction}
The Data Reduction Pipeline \citep[DRP,][]{Law2016} processes all exposures taken for each galaxy into data-cubes. This involves spectral extraction, flux calibration, subtracting sky lines and continuum from the raw observed spectra and using astrometric information to resample each wavelength channel into a full datacube. The resulting datacube records information as a function of wavelength at each on-sky sample, hence it is composed of spatial pixels (spaxels). From the DRP we use the LOGCUBE files, in which the wavelength vector has been logarithmically binned \citep[][see Appendix B for data model]{Law2016}. The files provide flux, noise, bitmasks and spectral resolution as a function of $x$, $y$, and $\lambda$.

MaNGA's Data Analysis Pipeline (DAP, Westfall et al., in prep) fits the stellar continuum and nebular emission lines of each spectrum, providing the kinematics of both components, as well as emission-line fluxes and equivalent widths. This analysis is carried out for different binning schemes within galaxies as well as for each individual spaxel. We use the stellar velocities and velocity dispersions from the DAP for each spectrum in our analysis. In order to convert from RA and DEC to semi-major axis polar coordinates on the IFU we use the $r$-band isophotal ellipticity and position angle from the NSA catalogue. The latter consists of galaxy images and astrometric and photometric parameters. We also adopt the galaxy mass \citep[based on a Chabrier IMF,][]{Chabrier2003} and half-light radius from this catalogue, always using Elliptical Petrosian quantities.

\subsection{Stacking Spectra}
The typical $r$-band S/N of $4-8\;$\AA$^{-1}$ in individual 2\arcsec\ fibre toward the outer radii is too small for reliable kinematic or stellar population analyses. The necessary S/N is generally obtained through combining the spaxels of individual galaxies in e.g.\ Voronoi cells or radially in annuli.

However, the resulting S/N is not sufficiently high for IMF studies requiring S/N ratios of at least $100\;$\AA$^{-1}$ \citep{Ferreras2013}. This is because dwarf stars only contribute little to the integrated light of stellar populations owing to their high M/L ratios \citep{Maraston1998,Maraston2005}. Hence the IMF-sensitive features, which generally are strong in the spectra of dwarf stars, drop to the level of a few percent in the integrated light of a stellar population \citep{Conroy2012a}. We therefore need very high S/N to make measurements which are precise enough to detect small changes in these features. This can be achieved by stacking spectra of several galaxies.

On top of this, the majority of IMF sensitive features are found in the near-infrared (NIR) region of the spectrum, which suffers from contamination by sky lines. As we will show, the technique of stacking galaxies mitigates this problem successfully.

\begin{figure*}
  \centering
  \includegraphics[width=0.36\textwidth]{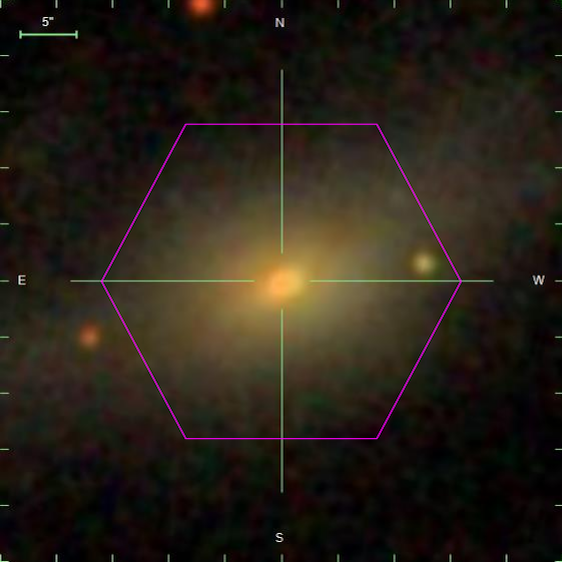}
  \includegraphics[width=0.5\textwidth]{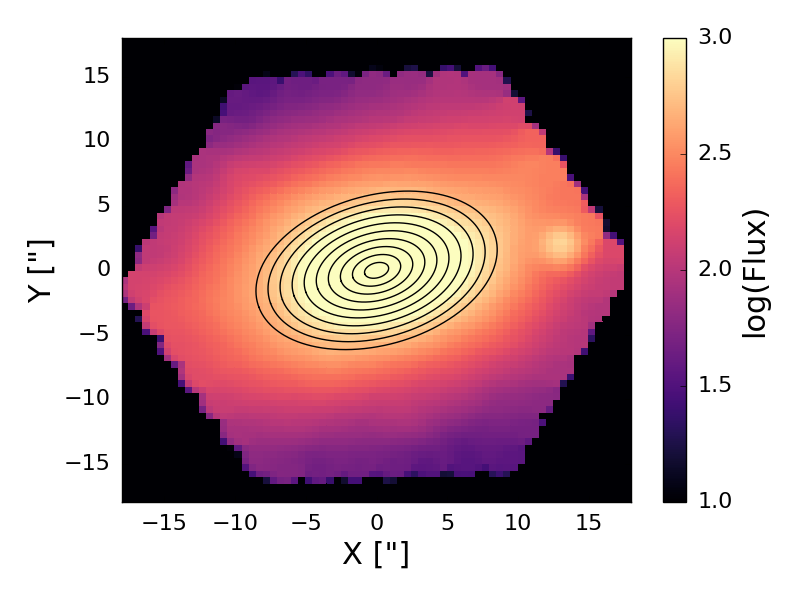}
\caption{{\it Left-hand panel}: Galaxy (MaNGA-ID 1-24476) image with the magenta hexagon representing MaNGA's IFU bundle. {\it Right-hand panel}: Galaxy flux map with ellipses plotted from 0.1 - 1 $R_\mathrm{e}$ at 0.1 $R_\mathrm{e}$ intervals, representing our radial bins.}
\label{fig:galimg}
\end{figure*}
\subsubsection{S/N cut}
Before stacking, wavelengths are converted to rest-frame for each spectrum using the equation,
\begin{equation}
\lambda_{\rm rest} = \frac{\lambda_{\rm obs}}{(1 + z)\left(1 + \frac{v}{c}\right)},
\end{equation}
where $z$ is the galaxy's NSA redshift and $v$ is the internal stellar velocity for each spectrum as measured by the DAP. The chosen threshold for the accuracy in the measurement of $v$ is critical for the success of this stacking procedure. We carefully analysed the data and decided that a conservative value of an S/N of 7\; pixel$^{-1}$ provides the optimal balance between accuracy in $v$ and data volume.

We illustrate this with Fig.~\ref{fig:vel_snmaps} where we plot a galaxy's stellar velocity with an overlay of S/N contours of $5, 7, 10, 15$\; pixel$^{-1}$. Here, the measurements come from the DAP where each spaxel has been analysed independently. The figure shows that the velocity field becomes noisy beyond the ${\rm S}/{\rm N}=7$\; pixel$^{-1}$ contour, suggesting that the measured velocities are not reliable below this threshold in S/N. We include all galaxies in our stack that contain at least one spectrum in each radial bin after applying the S/N cut.

\subsubsection{Binning procedure}
We stack galaxies in radial bins using each galaxy's elliptical radius along the semi-major axis. An example early type galaxy's (MaNGA-ID 1-24476) image and flux map are shown in Fig.~\ref{fig:galimg}. The elliptical annuli used in our binning method are plotted over the flux map in steps of 0.1 $R_\mathrm{e}$, going out to 1 $R_\mathrm{e}$.

The binning procedure is carried out in two steps: i) spaxels of each galaxy are binned in elliptical annuli, ii) annuli of different galaxies are combined together. These steps are made consistent so that the binning within one galaxy and across multiple galaxies is carried out in exactly the same way. We bin full spaxels such that each spaxel is assigned a unique radial bin, and is located in the annulus within which its centre falls.

Flux, error, and spectral resolution are interpolated to a common wavelength grid so that the average value at each wavelength can be calculated. Using a median stack of the interpolated spectra, we minimise contamination by emission lines, bad pixels and sky residuals. Indeed, we chose a median stack instead of a sigma-clipped mean precisely because of its better performance in removing these artefacts from the stacked spectra, which is particularly important toward the red limit of the spectral range (near the FeH band). The error in the median is $1.25 \sigma/N^{1/2}$ (i.e., 25\% larger than the error on the mean), where $\sigma$ is the standard deviation in the stack of $N$ spectra \citep{Lupton1993}. The spectral resolution of each stacked spectrum is the quadratic mean of $\sigma_{\rm inst}(\lambda)$, the dispersion of the line-spread function as a function of wavelength.  This can be derived by considering the second moment of the sum of a set of Gaussian functions. Similarly, we calculate the expected velocity dispersion for the stacked spectrum as the quadratic mean of the velocity dispersions measured from each spectrum in the stack.

The high quality masks produced by the DRP and DAP for the flux and kinematic quantities are propagated while stacking. The mask corresponding to the 3D data-cube indicates problems such as low/no fibre coverage, foreground star contamination etc. Hence, on top of applying the S/N cut ($7$\; pixel$^{-1}$ for each unbinned spaxel, as described in the previous section) to entire spectra, masked individual pixels in the spectra and masked $v$ and $\sigma$ values are not included in the stack.

To accommodate the variation in fluxes from galaxy to galaxy, spectra are normalised at a wavelength window in the continuum, $6780-6867\;$\AA\ before stacking and then re-multiplied by the mean of the normalisation factor. From the 970 Primary$+$ early-type galaxies in DR14/MPL-5, 611 remain after the S/N cut, losing objects at both the lowest and highest mass ends. This is because achieving the necessary high S/N ratio requires a large galaxy sample for stacking, which leads to tails in the distribution at the low mass and high mass ends due to the S/N cut. High mass galaxies have steeper surface brightness profiles, hence their outer regions have lower surface brightnesses and therefore lower S/N. Therefore in this work, we focus on galaxies with $\log M/M_{\odot}$ between 9.85 and 10.80.

\begin{figure}
 	\includegraphics[width=\linewidth]{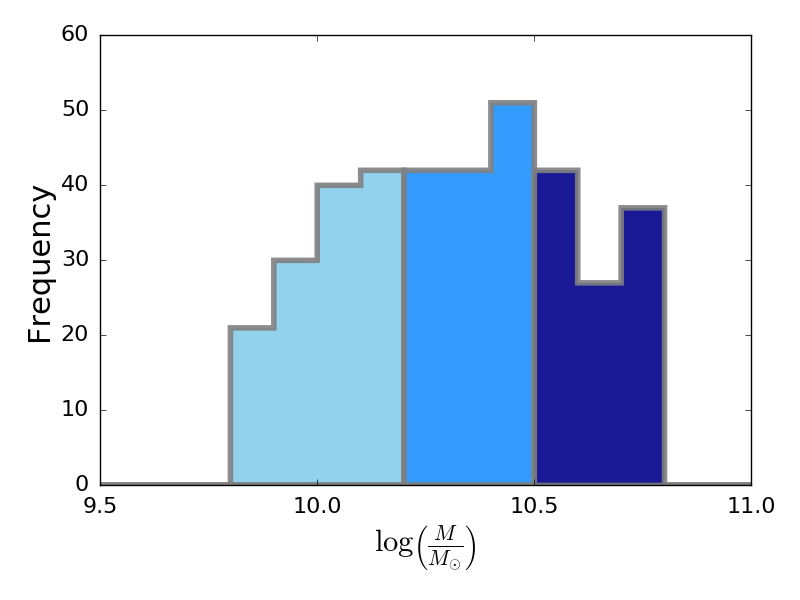}
    \caption{Masses of our sample of 366 early type galaxies, split into 3 bins containing equal numbers of galaxies.}
    \label{fig:masshist}
\end{figure}

The mass distribution and the three resulting mass bins are shown in Fig.~\ref{fig:masshist} and listed in Table~\ref{tab:bins}. The mass bins, containing 122 galaxies each, are centred on $\log M/M_{\odot}$ of 10, 10.4, and 10.6, and a central velocity dispersion $\sigma$ of 130, 170, and $200\;$km/s, respectively. Note that this mass range covers the lower mass end compared to samples in other IMF studies in the recent literature.

\subsubsection{Final S/N}
\begin{figure*}
  \includegraphics[width=.49\linewidth]{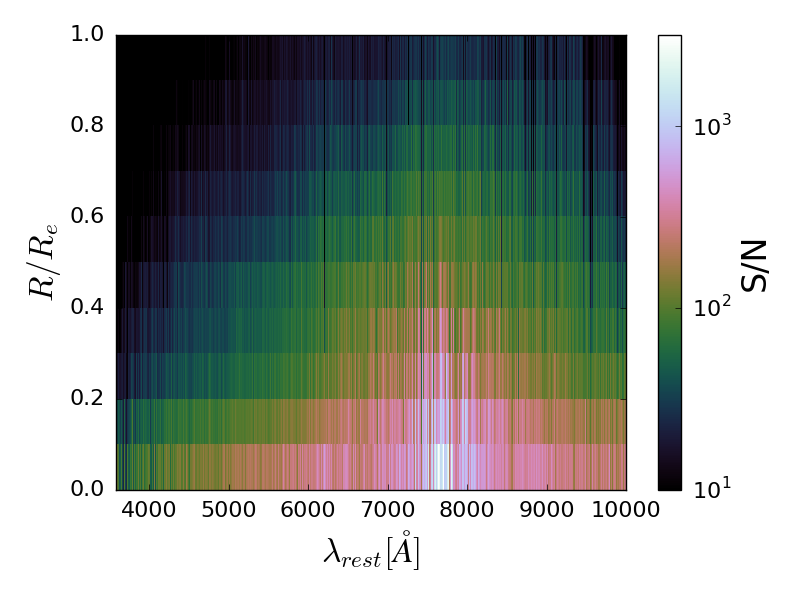}
  \includegraphics[width=.49\linewidth]{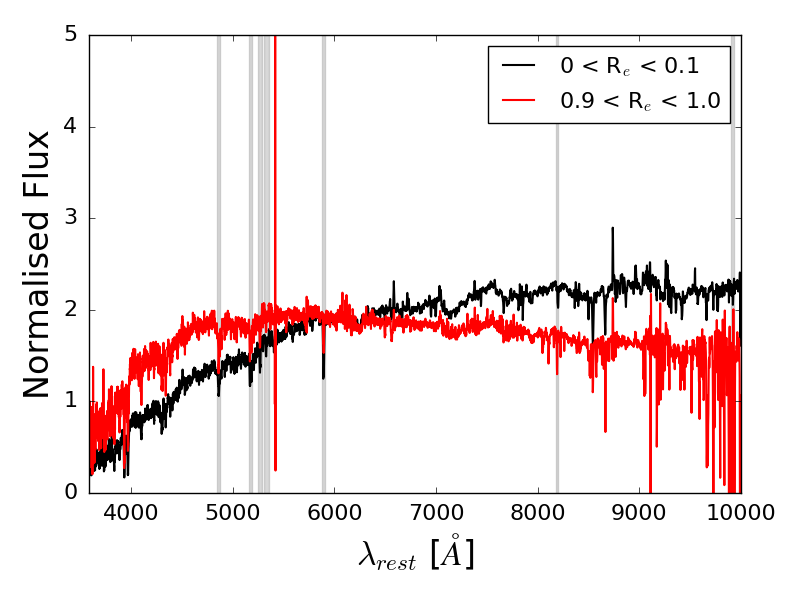}
  \includegraphics[width=.49\linewidth]{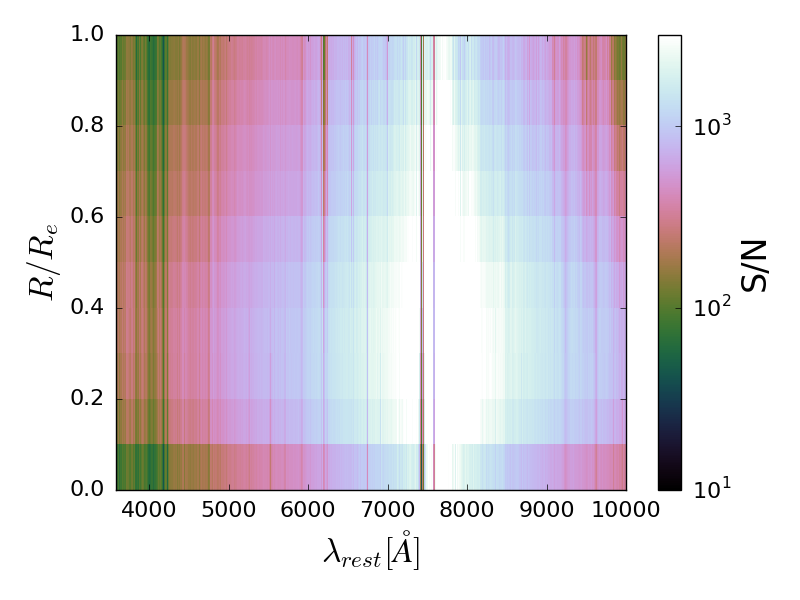}
  \includegraphics[width=.49\linewidth]{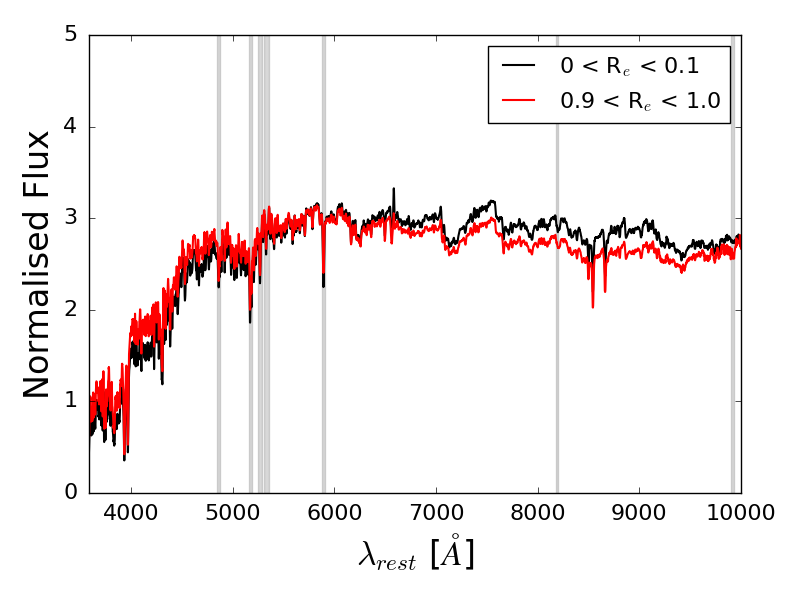}
\caption{{\it Top row}: S/N as a function of wavelength and radial bin for the galaxy from Fig.~\ref{fig:galimg} (left). Radially binned spectra from the centre of the galaxy in black and at 1 $R_\mathrm{e}$ in red (right). {\it Bottom row}: Same for the mass bin $\log M/M_{\odot}=10.2 - 10.5$ containing 122 galaxies. Grey shaded regions highlight the features used in this work.}
\label{fig:stack}
\end{figure*}

In Fig.~\ref{fig:stack} we show an image of the S/N ratio as a function of wavelength and galaxy radius (left-hand panels) and the corresponding spectra for the innermost bin (red) and around the half-light radius (black, right-hand panels). The top and bottom rows shows an individual galaxy and the stack of 122 galaxies, respectively.

The figure illustrates how S/N depends on both wavelength and radius, and how S/N improves in the stack. The highest S/N is found at the centre of the galaxy. In individual MaNGA galaxies the S/N is of the order $\sim 200-300$\; pixel$^{-1}$, and drops to $\sim 30$\; pixel$^{-1}$ at $1 R_\mathrm{e}$ (top left-hand panel). A wavelength dependence for S/N can also be seen: S/N is highest between $6,500-8,500\;$\AA\ and decreases at the blue and red boundaries of MaNGA's coverage. This is reflected in the spectra which are noisy especially in the NIR (top right-hand panel). In particular, contamination from sky residuals is significant, and noise is dominating the spectrum in the outer radial bin (red line). The S/N is clearly too low for a reliable stellar populations analysis of the near-IR part of the spectrum.

The bottom row of Fig.~\ref{fig:stack} shows the same map and spectra for the mass bin $\log M/M_{\odot}=10.2 - 10.5$ containing 122 early type galaxies (including the galaxy from Fig.~\ref{fig:galimg}).  The S/N is significantly higher, and an S/N of at least $400$\; pixel$^{-1}$ around $1 R_\mathrm{e}$ is achieved at all wavelengths, except at the very edges. The interplay between two competing effects when moving outward from the centre of a galaxy, i) increase in the number of observed spectra ii) decrease in surface brightness hence S/N, lead to the shape of the S/N map \citep[see also][]{Goddard2017}. Hence S/N in the stacked spectra reaches its maximum not at the centre but around $0.2-0.4\; R_{\rm e}$. 

The corresponding spectra in the bottom, right-hand panel further illustrate this substantial improvement. Most importantly, also the spectrum of the outer bin around the half-light radius is now of high quality, and the noise and sky residuals in the near-IR have been eliminated successfully. Key absorption features that serve as IMF indicators are now detected well. Also, the spectra from the centre and at $1 R_\mathrm{e}$ are now of comparable quality.

It is worth noting that stacking leads to a further crucial benefit, namely the elimination of sky line residuals, which is particularly important in the near-IR wavelength range. They get averaged out through stacking when individual galaxy spectra are corrected to the rest-frame wavelength.

The final set of stacked spectra, after emission line fitting and removal (see Section~\ref{sec:TAP}), for the various mass and radial bins used in our analysis are shown in Appendix~\ref{sec:app_spectra}.

\begin{table}
	\centering
	\caption{Median velocity dispersion, effective radius and redshift of the mass terciles.}
	\label{tab:bins}
	\begin{tabular}{cccc} 
		\hline
		Mass range ($\log M/M_{\odot}$) & $\sigma$ (km/s) & $R_\mathrm{e}$ (kpc) & z\\
		\hline
		$9.9-10.2$ & 130 & 2.43 & 0.027\\
		$10.2-10.5$ & 170 & 3.11 & 0.030\\
		$10.5-10.8$ & 200 & 4.38 & 0.034\\
		\hline
	\end{tabular}
\end{table}

\subsection{Spectral analysis}
Stellar population parameters are derived through fitting stellar population models to spectra. This can be done either through fitting the full spectrum \citep[e.g.,][]{Fernandes2005, Ocvirk2006, Koleva2009, Conroy2014, Cappellari2017, Goddard2017, Conroy2017b, Wilkinson2017} or selected absorption line features \citep[e.g.,][]{Trager2000, Proctor2002, Thomas2005, Shiavon2007, Thomas2010, Johansson2012}. In the current work we use the latter approach and focus on a few key absorption features.

\label{sec:TAP}
\subsubsection{Stellar kinematics and emission line removal}
Before measuring absorption features on the spectrum for the analysis, the broadening of absorption line features caused by the stellar velocity dispersion $\sigma$ needs to be measured and, ideally, the emission line component of the spectrum removed. To this end we follow the approach used in \citet{Thomas2010} and \citet{Johansson2012} and perform initial full spectrum fitting on the stacked spectra in order to measure $\sigma$ and to extract the emission line spectra.

This is done using the DAP's pPXF wrapper (\citealt{Cappellari2004}; Westfall et al., in prep). The fit is carried out using the stellar library, MILES-HC, which is a hierarchical clustering of MILES \citep{Sanchez-Blazquez2006}. The average DAP velocity dispersion measurement obtained for the individual spectra, weighted and combined for each spectrum in a bin, is used as a guess value for $\sigma$.

\begin{figure*}
  \includegraphics[width=\linewidth]{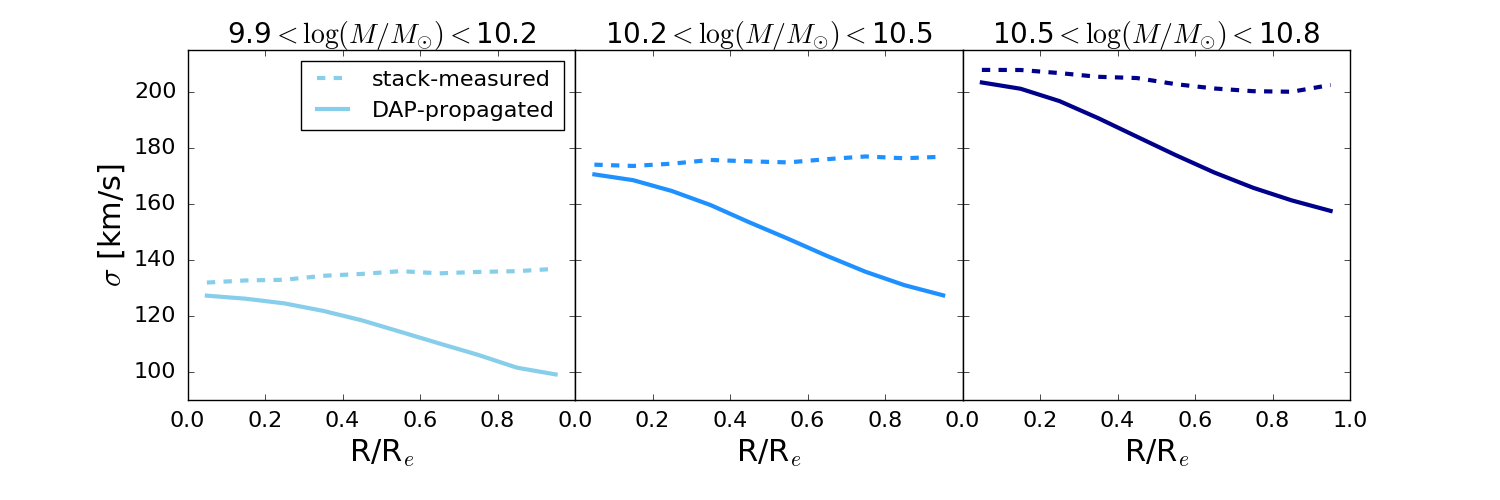}
\caption{Velocity dispersion profiles i) expected values as propagated from unstacked spectra (solid) and ii) values measured directly on stacked spectra (dashed), left to right represents increasing mass bin. The stack-measured values suffer at large radii due to large errors leading to an effective dispersion becoming convolved with the spectrum.}
\label{fig:dispersions}
\end{figure*}

This first fitting run provides us with the nominal stellar velocity dispersions of the stacked spectra, which we can compare to the DAP-propagated mean $\sigma$ from the measurements on the individual spectra. Both of these $\sigma$ values are shown as a function of galaxy radius for each mass bin in Fig.~\ref{fig:dispersions}. Comparing the DAP-propagated $\sigma_{\rm mean}$ (solid lines) and the stack-measured $\sigma_{\rm stack}$ (dashed lines) provides an assessment of the success in velocity registration of the stacked spectra. As can be seen from the figure, there is a marked difference with $\sigma_{\rm stack}$ being systematically larger than $\sigma_{\rm mean}$, except at the smallest radii. This ought to be expected, as the measurement error in radial velocity leads to an error in velocity registration, which adds an additional component in line broadening. In fact the difference is minimal at the centre and increases with increasing radius because of the generally lower S/N and hence lower accuracy in velocity measurements at those radii \citep{Westfall2011, Cappellari2017}. The consequence is that the $\sigma_{\rm stack}$ profiles are artificially flat, while $\sigma_{\rm mean}$ displays the well-known negative gradient \citep[e.g.][]{Jorgensen1999,Mehlert2000,Mehlert2003,Emsellem2004}.

The quantity $\sigma_{\rm stack}$ derived here directly from the stacked spectrum will be used in our future analysis to characterise the effective velocity dispersion affecting the absorption line broadening. Note, however, that $\sigma_{\rm mean}$, instead, should be used for any kinematic analysis.

Finally, after subtracting the continuum and absorption line component from the spectra, we fit for the emission lines using a list of 21 lines\footnote{\href{url}{http://physics.nist.gov/PhysRefData/ASD/Html/help.html}} and a preliminary version of the DAP emission-line fitting routine. This emission-line fit is then subtracted from the stacked spectra to produce emission line-free spectra for the analysis.

\subsubsection{Absorption indices}
\label{sec:abs_ind}
\begin{figure*}
  \includegraphics[height=0.9\textheight]{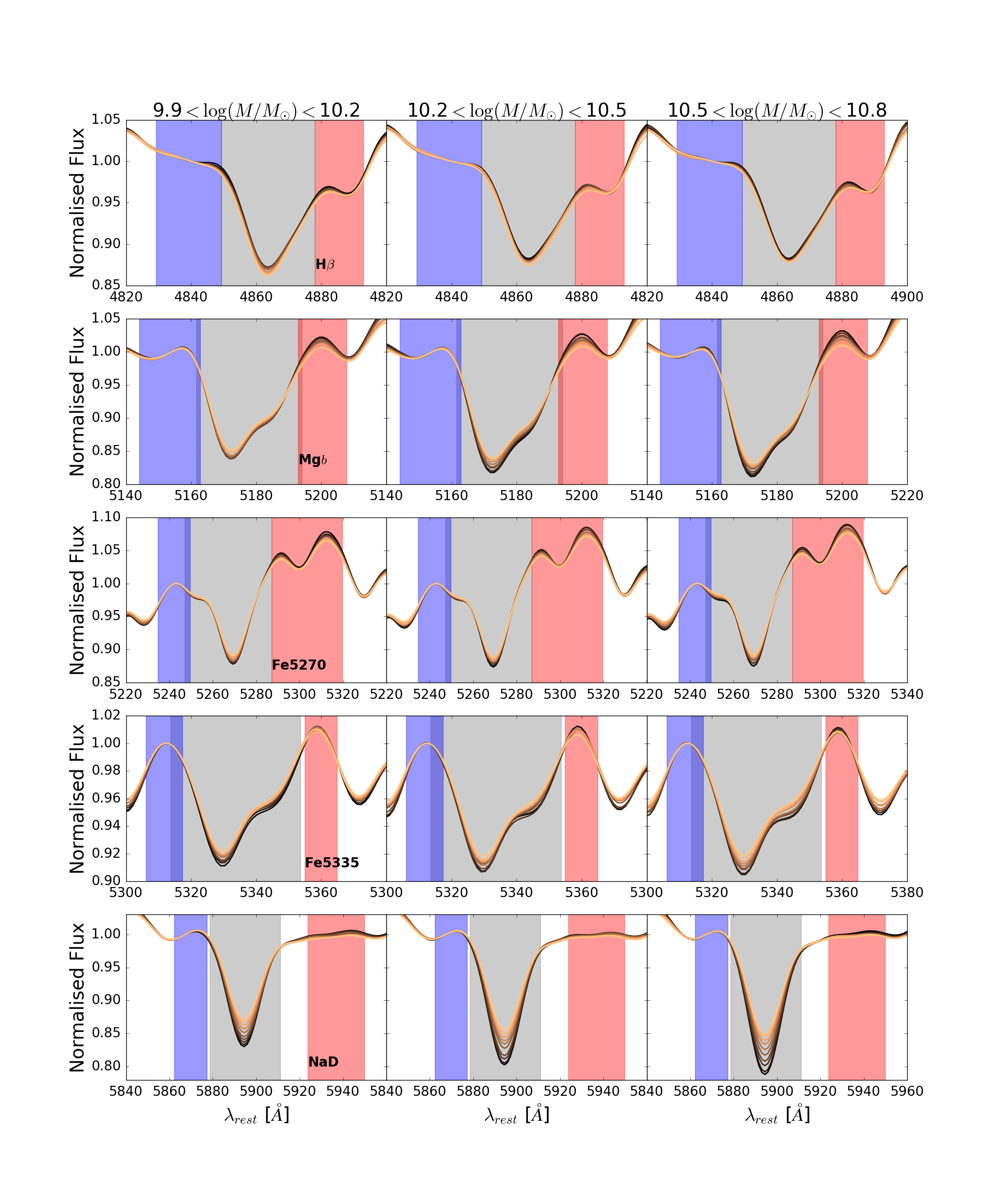}
  \caption{Spectral cut-outs of the optical absorption features used in this study. Left to right is increasing mass bin, the colours represent radial bins going from the centre (black) to $1 R_\mathrm{e}$ (orange). All spectra have been smoothed to 300km/s and normalised in the blue bandpass for visualisation purposes. The index bandwidth is indicated by the grey box, the blue and red pseudo-continua by the blue and red boxes, respectively}
\label{fig:features_optical}
\end{figure*}
\begin{figure*}
  \includegraphics[width=\linewidth]{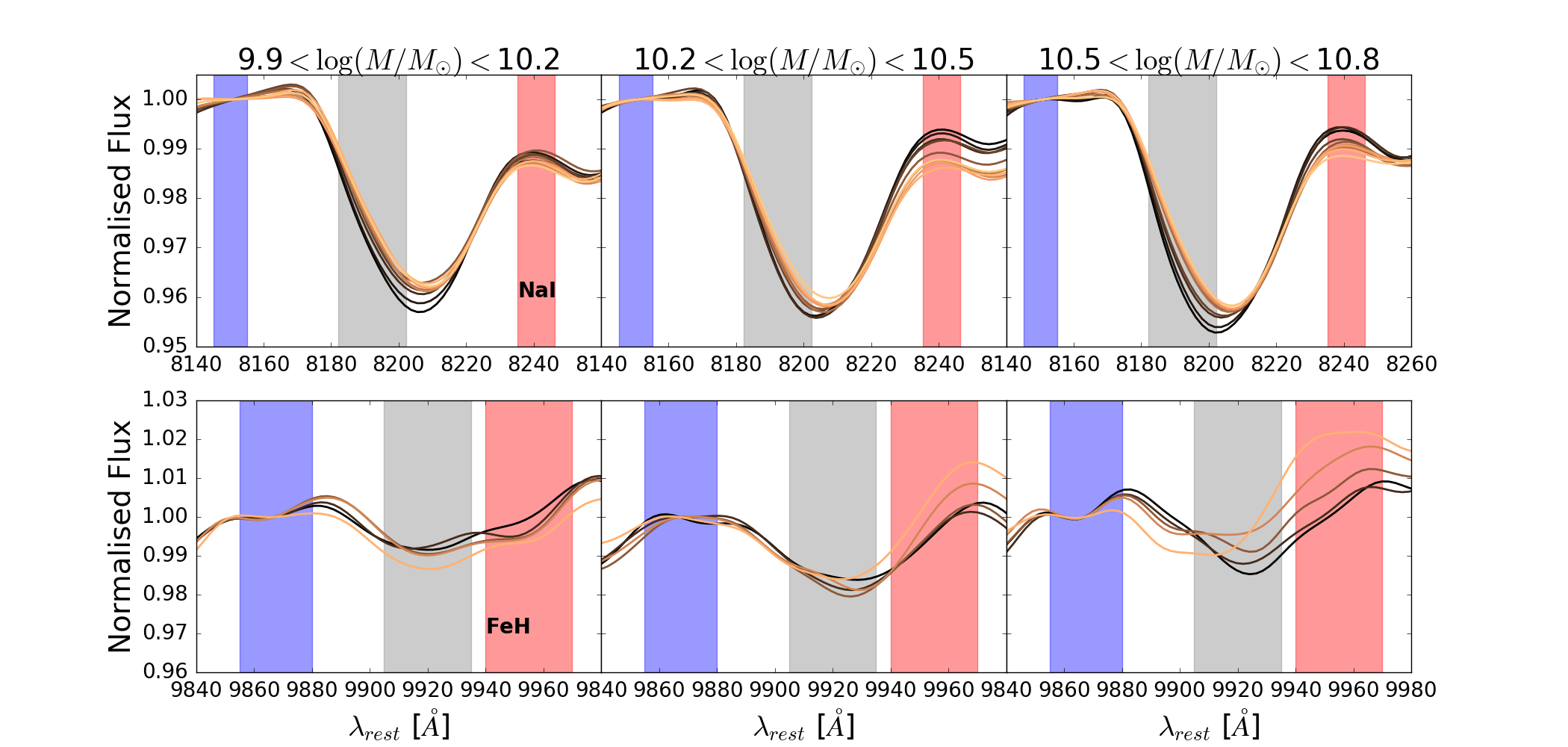}
\caption{Same as Fig.~\ref{fig:features_optical} for the near-IR, IMF-sensitive absorption features.}
\label{fig:features_NIR}
\end{figure*}
An absorption index is determined by defining two pseudo-continuum bandpasses on either side of a spectral feature, which is contained within the index bandpass. Equations 2 and 3 in \citet{Worthey1994} show how an equivalent width is measured in angstroms for atomic features and in magnitudes for molecular bands. Most IMF studies in the recent literature use gravity-sensitive spectroscopic features directly \citep[e.g.][]{vanDokkum2012, Martin-Navarro2015, LaBarbera2017, Alton2017}. 

Our choice of indices is as follows: the combination of H$\beta$, Mg$b$, Fe5270, and Fe5335 are suitable to constrain age, metallicity and [$\alpha$/Fe] abundance ratio. The NaD index is crucial for constraining [Na/Fe]. For the IMF slope, we include the gravity-sensitive indices NaI and FeH, in the near-IR, separately. Table~\ref{tab:indices} lists the indices and their definitions as used in our analysis.
\begin{table}
	\centering
	\caption{Index definitions for the absorption features used in this work. All wavelengths are in vacuum. The optical indices are defined in \citet{Trager1998}; NaI in \citep{LaBarbera2013}; and FeH in \citet{Conroy2012a}.}
	\label{tab:indices}
	\begin{tabular}{lccc}
		\hline
		Index & Blue Continuum & Feature & Red continuum\\
		\hline
		H$\beta$ & 4829.2 - 4849.2 &  4849.2 - 4878.0 & 4878.0 - 4893.0\\
		Mg$b$ & 5144.1 - 5162.8 & 5161.6 - 5194.1 & 5192.8 - 5207.8\\
		Fe5270 & 5234.6 - 5249.6 & 5247.1 - 5287.1 & 5287.1 - 5319.6\\
		Fe5335 & 5306.1 - 5317.4 & 5313.6 - 5353.6 & 5354.9 - 5364.9\\
		NaD & 5862.2 - 5877.3 & 5878.5 - 5911.0 & 5923.8 - 5949.8\\
		NaI & 8145.2 - 8155.2 & 8182.2 - 8202.3 & 8235.3 - 8246.3\\
		FeH & 9855.0 - 9880.0 & 9905.0 - 9935.0 & 9940.0 - 9970.0\\
		\hline
	\end{tabular}
\end{table}

Once measured on the observed galaxy spectrum, the index values are corrected to a common velocity dispersion. Here we correct to the resolution of the stellar population models in the following way \citep[see also][]{Johansson2012}: For each stacked spectrum we take the model templates from the initial full spectral fit and measure the index strength two times. First at the spectral resolution of the data from the best fit model directly convolved with the derived $\sigma_{\rm stack}$. Then we generate the same model template from the model library at the resolution that we want to correct to. The ratio between these measurements in the correction factor, which is multiplied by the index measurements from the data to give the corrected index \citep{Kuntschner2004}

We then calculate Monte-Carlo based errors for our indices. To do so, we perturb the flux at each wavelength with a number randomly drawn from a Gaussian which has a standard deviation equal to the error in the flux at each pixel. The indices are measured on perturbed spectra from 100 such realisations and the error is measured as the standard deviation of these measurements, divided by $\sqrt{2}$ in order to account for the fact that we are perturbing spectra that already have an error. 

Covariance between spatially adjacent spaxels can lead to noise being underestimated when they are binned together. In our radial binning scheme, the number of adjacent spaxels being combined is very small. Also, the MaNGA reconstructed point-spread function, and hence the correlation matrix, is only a weak function of wavelength \citep{Law2016}. Therefore we use individual, pixel-based errors to perturb the spectra because we do not expect strong correlations of uncertainties in the wavelength or spatial directions.

In Figs.~\ref{fig:features_optical} and ~\ref{fig:features_NIR} we show the spectral cut-outs of the stacked spectra around the seven features measured for the different radial and mass bins. All spectra have been smoothed to $300\;$km/s and normalised for visualisation purposes. Mass increases from left to right, and the colours represent radial bins going from the centre (black) to $1 R_\mathrm{e}$ (orange). The index bandwidth is indicated by the grey box, the blue and red pseudo-continua by the blue and red boxes, respectively. As mentioned in the previous section, the edges of the stacked spectra suffer from lower S/N, and at large radii this is particularly the case for the longest wavelengths. As a consequence, the S/N in the region of reddest feature we study, FeH, drops below the values in the central radial bin of at least $200$\; pixel$^{-1}$. Therefore, we only measure and analyse this out to $0.5 R_\mathrm{e}$.

\begin{figure*}
  \includegraphics[height=0.9\textheight]{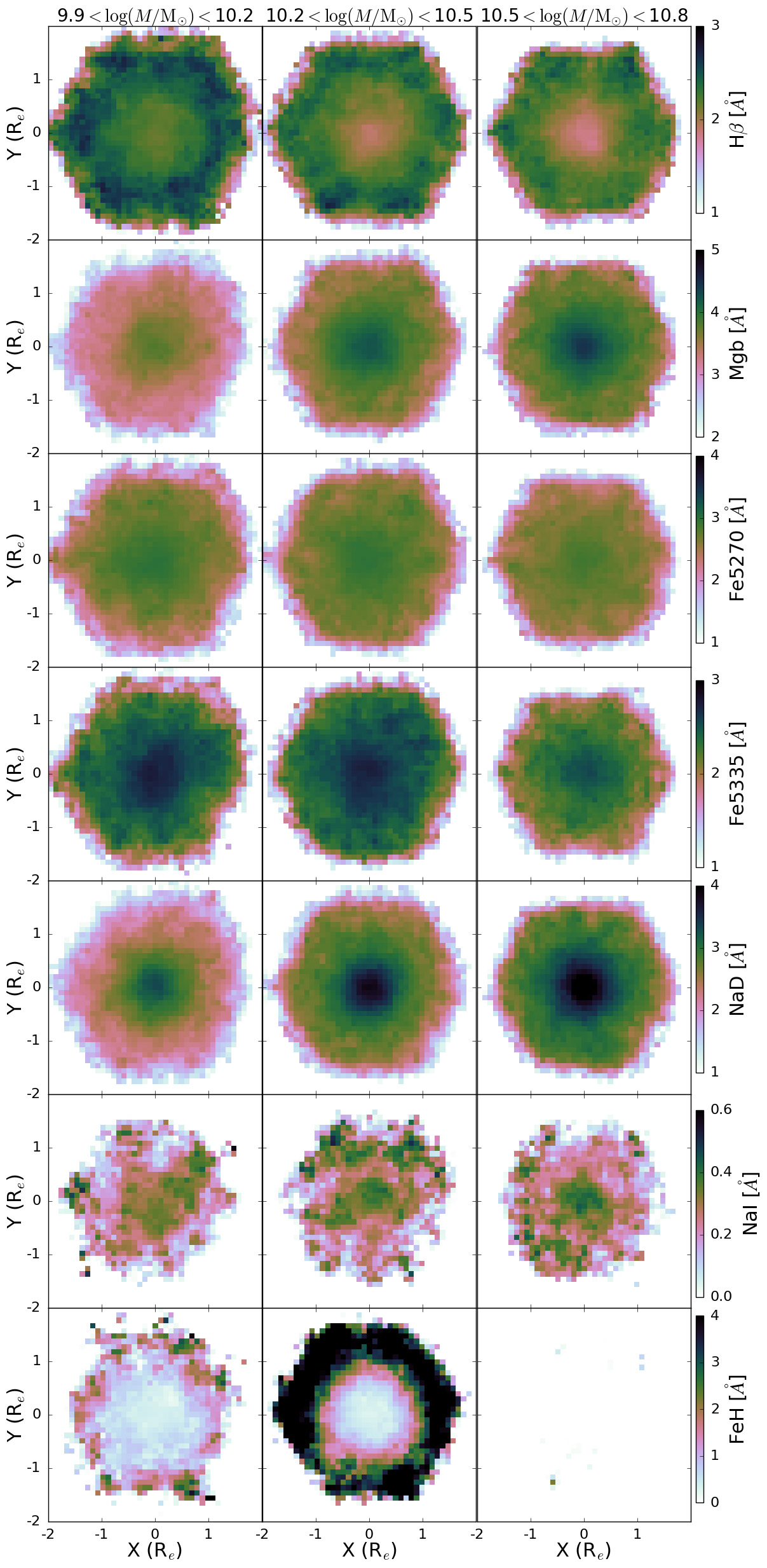}
\caption{Stacked maps, in units of $R_\mathrm{e}$, for the indices we use in our analysis. Left to right is increasing mass bin.}
\label{fig:stackedmaps}
\end{figure*}

It can be seen clearly from Figs.~\ref{fig:features_optical} and ~\ref{fig:features_NIR} how the various absorption features will depend on galaxy mass and radius. The absorption depths of the optical metallic features Mg$b$, Fe5270, Fe5335, and NaD increase both with decreasing galaxy radius (orange to black lines) and increasing galaxy mass (panels left to right). The IMF-sensitive absorption feature NaI behaves similarly (top panels in Fig.~\ref{fig:features_NIR}), and displays the well-known pattern of increasing depth accompanied by a shift of the minimum to bluer wavelengths as galaxy mass increases \citep{Ferreras2013}. We show clearly here that this same pattern (increase in depth and blue-shift) is seen with decreasing galaxy radius (orange to black lines). In contrast to NaI, the Wing-Ford band FeH is more complex, and no clear pattern is visible from the spectra (bottom panels in \ref{fig:features_NIR}). If at all, the radial trend appears to be reversed with a stronger FeH absorption at larger radii. Any possible dependence on galaxy radius and/or mass appears to be dominated by the observed flux in the red pseudo-continuum band. We will discuss this point in detail in Section~\ref{sec:results} when comparing measurements of FeH with stellar population model predictions.

Finally, for a spatial visualisation of the data, in Fig.~\ref{fig:stackedmaps} we present 2D maps of the seven indices for the three mass bins. Mass increases from left to right, and each row corresponds an index. These maps are created using stacks of the index measurements for {\it individual} spaxels in individual galaxies after spatial scaling. Hence they are not based on the stacked spectra used in the present analysis and although systematic errors remain in these maps, random noise is reduced so they are able to serve as an illustration of the data properties. Gradients are clearly visible for all the features. The maps further support our first interpretation of the data based on Figs.~\ref{fig:features_optical} and ~\ref{fig:features_NIR}. The strengths of Mg$b$, Fe5270, Fe5335, NaD, and NaI increase, while the strengths of H$\beta$ and FeH decrease toward the galaxy centre. These profiles generally get stronger with increasing galaxy mass, a trend that is most pronounced in NaD.

\subsection{Models}
The analysis of galaxy spectra naturally is sensitive to the stellar population model used. \citet{Kuntschner2010} and \citet{Conroy2017b} compare models by \citet{Thomas2003,ThomasD2011b}, \citet{Schiavon2007}, and \citet{Conroy2017b} and find encouraging agreement for optical absorption features. Modelling of near-IR features including the IMF-sensitive indices, instead, is more challenging \citep{Lyubenova2010,Lyubenova2012,Spiniello2014,Baldwin2017}. We therefore use a combination of different model sets in the present analysis as outlined below.

\label{sec:models}
\subsubsection{TMJ and M11-MARCS}
We use the \citet[][hereafter TMJ]{ThomasD2011b} stellar population models of the optical Lick absorption indices at the $2.5\;$\AA MILES resolution \citep{Beifiori2011,Falcon-Barroso2011,Prugniel2011} in our analysis. In brief, the TMJ models are an update and extension of the earlier models by \citet{Thomas2003,Thomas2004} and based on the \citet{Maraston1998,Maraston2005} evolutionary synthesis code. The update uses new empirical calibrations by \citet{Johansson2010} based on the MILES stellar library \citep{Sanchez-Blazquez2006}. Element response functions from \citet{Korn2005} are adopted. The models are carefully calibrated with galactic globular clusters and reproduce the observations well for the spectral featured used in this study \citep{Thomas2011a}. The models are available for different ages, metallicities, variable element abundance ratios, in particular [$\alpha$/Fe] and [Na/Fe], and a Salpeter IMF.  Element variation are calculated at constant total metallicity, hence the TMJ models enhance the $\alpha$-elements and suppress the Fe-peak elements according to Equations 1-3 in \citet{Thomas2003}. The models are provided for Lick and MILES spectral resolution. We use the latter in the present work because it is well matched to the MaNGA resolution over the relevant wavelength range.

\begin{figure}
  \includegraphics[width=\linewidth]{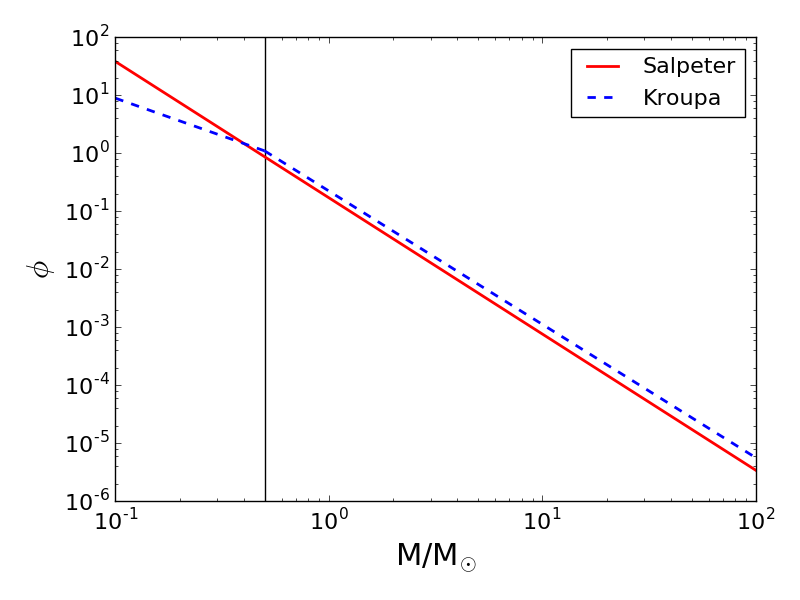}
\caption{With a lower mass cut-off of $0.1\; M_{\odot}$ for M11 models and $0.08\; M_{\odot}$ for VCJ, a Salpeter IMF has a slope of 2.35 whereas a Kroupa IMF has a similar slope of 2.3 above $0.5\; M_{\odot}$ and a shallower slope of 1.3 for lower stellar masses. The IMFs have been normalised to have the same mass.}
\label{fig:sketch}
\end{figure}

In combination with these, we use the \citet[M11]{Maraston2011} models based on the theoretical MARCS \citep{Gustafsson2008} library. The theoretical library MARCS allows us to extend our analysis to NaI and to FeH without loss of stellar parameter coverage. These models are available for Kroupa and Salpeter IMFs. Using the power law form $\phi\propto M^{-x}$, a Salpeter IMF refers to a single power-law with $x = 2.35$ and a Kroupa IMF refers to a double-power law with $x = 1.3$ between $0.1-0.5\; M_{\odot}$ (shallower than Salpeter) and $x = 2.3$ between $0.5-100\; M_{\odot}$. These are shown in Fig.~\ref{fig:sketch}, normalised to the same mass. We note that the turnover to a shallower slope at low stellar masses is also present in other versions of the IMF like \citet{Chabrier2003} or the recently widely discussed integrated galactic IMF \cite{Kroupa2003,Weidner2005,Kroupa2013,deMasi2017}. We use the M11-MARCS models to calculate the TMJ models for a Kroupa IMF, and the VCJ response functions (described below) to calculate element abundance variation for M11-MARCS models.

\subsubsection{VCJ}
We also make use of the stellar population models based on the extended IRTF stellar library by \citet[][hereafter VCJ]{Villaume2017}. The library, complemented with M-dwarf stars from \citet{Mann2015}, consists of stars present in the MILES library \citep{Sanchez-Blazquez2006}, giving continuous coverage for each star across the MaNGA wavelength range. Due to the different instrumental resolutions of the MILES and IRTF libraries, the models are smoothed to a common dispersion of 100km/s. Note that the dwarf stars from \citet{Mann2015} are at a lower spectral resolution than both MILES and IRTF libraries, and these spectra were deconvolved in VCJ. The empirical SSPs are available for a range of triple-power law IMFs. The low mass cut-off is $0.08M\; _{\odot}$ and the slope above $1\; M_{\odot}$ is fixed at the Salpeter slope. The slopes for $0.08-0.5\; M_{\odot}$ and $0.5-1\; M_{\odot}$ can be varied from 0.5 to 3.5. Hence we can obtain models for the Kroupa and Salpeter IMFs mentioned above.

We multiply the theoretical response functions for element abundance variation \citep{Conroy2017b} to whichever base models (M11-MARCS or VCJ) we use in the analysis. These functions are available for individual elements spanning a range in age and metallicity. We individually vary the elements Na and Fe as well as the $\alpha$ and other light elements (N, O, Ne, S, Mg, Ca, Na, Si, Ti). We calculate an $\alpha$-enhanced model following the same procedure as the TMJ models, for a consistent comparison.

\section{Results}
\label{sec:results}
\begin{figure*}
  \includegraphics[width=.48\linewidth]{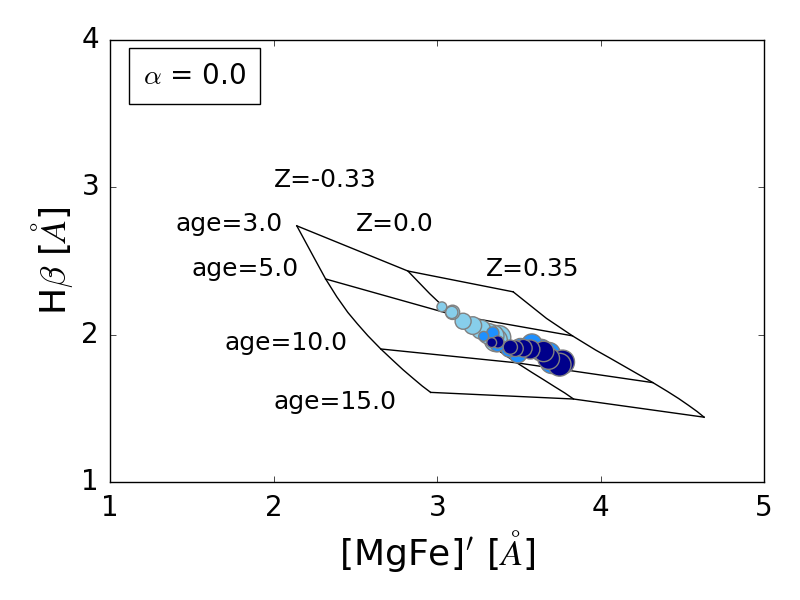}
  \includegraphics[width=.48\linewidth]{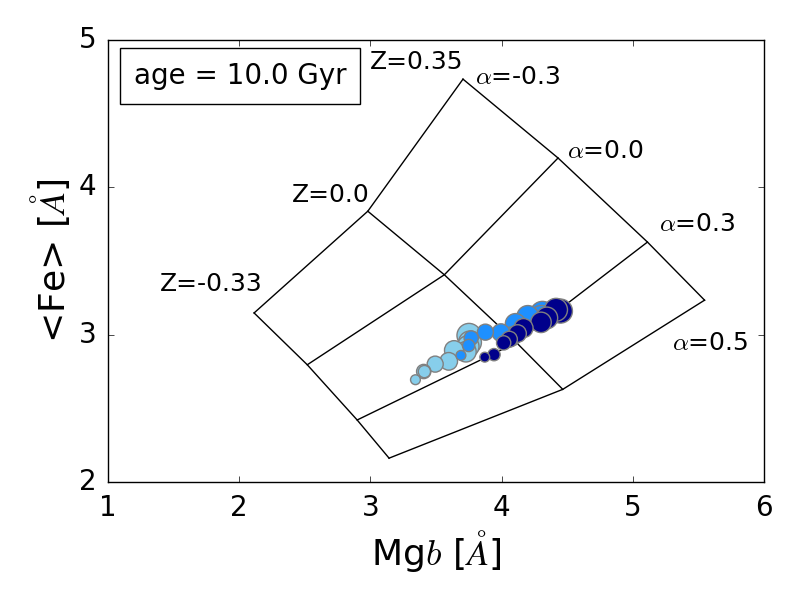}
\caption{Optical indices for the three mass bins against the high-resolution TMJ model index grids. The age and metallicity dependence plotted for models with [$\alpha$/Fe] = 0 (left-hand panel) and metallicity against [$\alpha$/Fe] for models at a fixed age of 10~Gyr (right-hand panel). Dark to light shades of blue represent the mass bins in decreasing order; decreasing size represents increasing radius.}
\label{fig:grids}
\end{figure*}

\begin{figure*}
  \includegraphics[width=.32\linewidth]{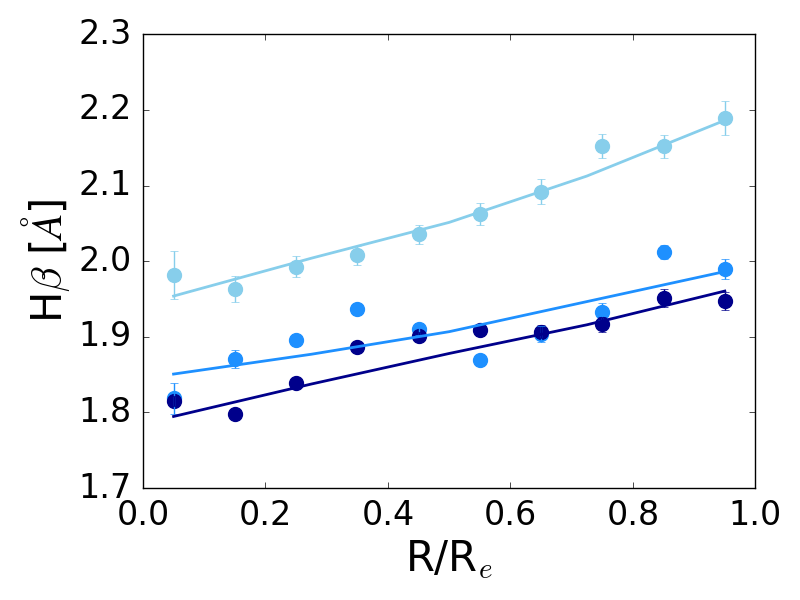}
  \includegraphics[width=.32\linewidth]{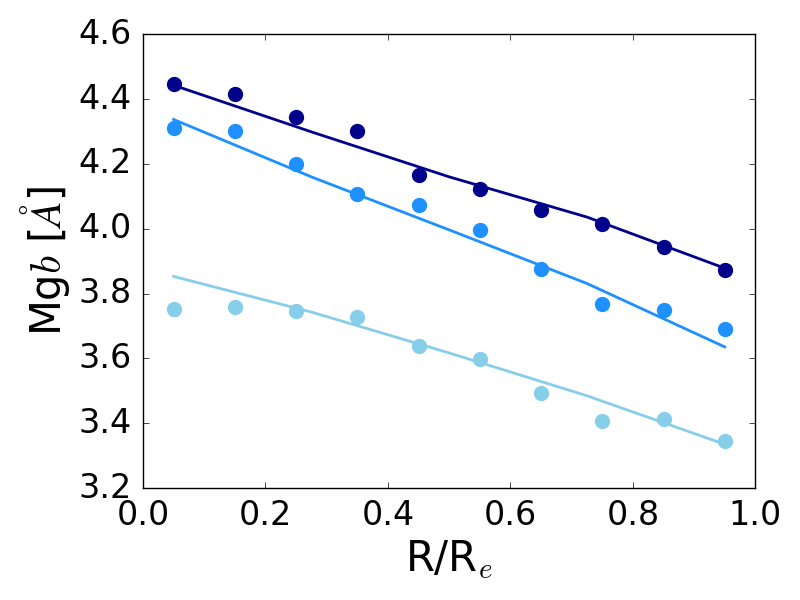}
  \includegraphics[width=.32\linewidth]{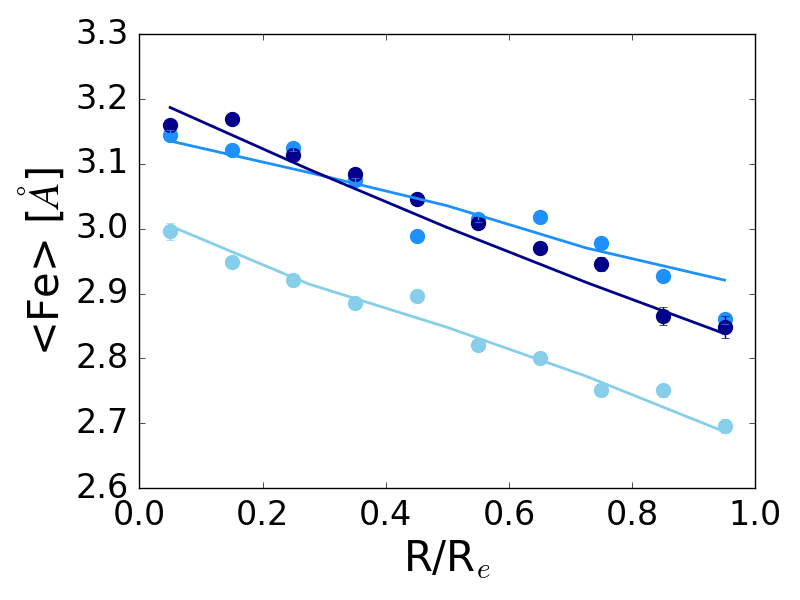}
  \includegraphics[width=.32\linewidth]{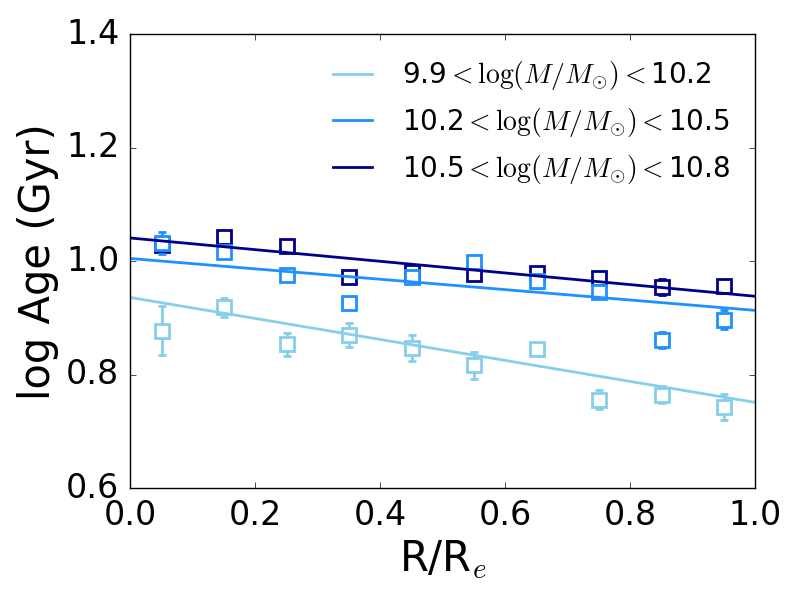}
  \includegraphics[width=.32\linewidth]{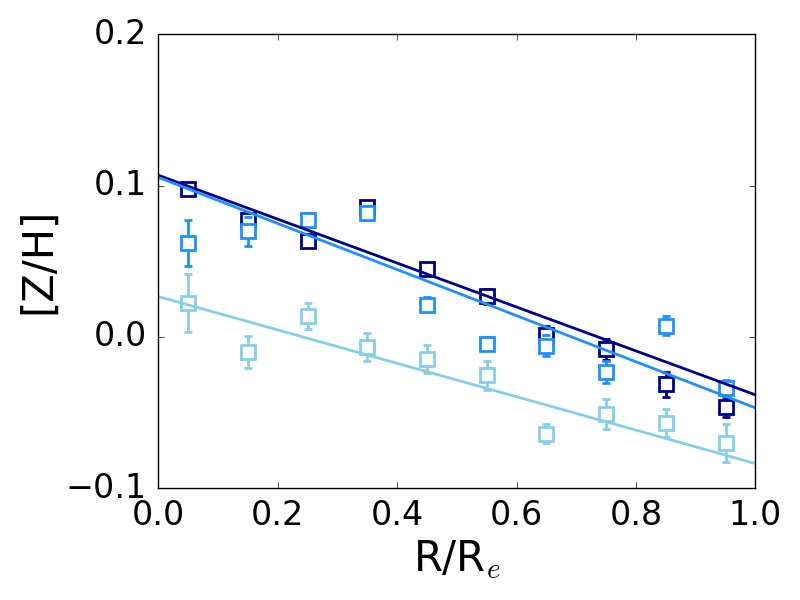}
  \includegraphics[width=.32\linewidth]{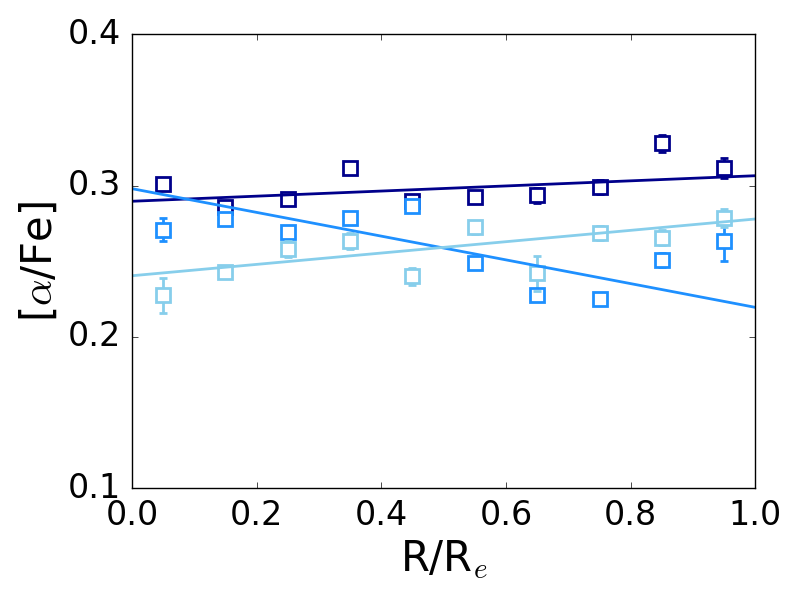}
\caption{{\it Top row}: H$\beta$, Mg$b$ and <Fe> for each mass bin as a function of radius.
Line-strengths are measured on the stacked spectra with 1-$\sigma$ errors calculated using a Monte Carlo-based analysis. The lines are TMJ model fittings derived through chi-squared minimisation (see text).{\it Bottom row}: TMJ-derived age, metallicity and [$\alpha$/Fe] as a function of radius for different mass bins. The lines are calculated by fitting a straight line to the derived parameter at each radius in order to match the data}
\label{fig:indices}
\end{figure*}

We now turn to the analysis of the index-strength measurements. Stellar population parameters are derived from fittings using the models described in Section~\ref{sec:models}. We use a chi-squared minimisation code to simultaneously fit the optical indices H$\beta$, Mg$b$, Fe5270, Fe5335, and NaD and either NaI or FeH to derive age, metallicity, [$\alpha$/Fe] ratio, Na abundances, and IMF slopes. The TMJ and M11-MARCS, and VCJ model grids are linearly interpolated in this 5-dimensional parameter space. For clarity, we present the results in three steps: First we show the fits to H$\beta$, Mg$b$, Fe5270, Fe5335 and the parameters age, metallicity, [$\alpha$/Fe] ratio in (Section~\ref{sec:age_elements}). We then show the fits to the NaD index with and without Na abundance and the derived [Na/Fe] gradients (Section~\ref{sec:na}). Finally we present the IMF-sensitive NaI and FeH indices, with and without IMF variation and the derived IMF slopes in (Section~\ref{sec:imf}).

We provide all our index measurements and errors for the various mass and radial bins used in our analysis in Appendix~\ref{sec:app_tables} as well as all the derived parameters.

\subsection{Age, metallicity and [$\alpha$/Fe]}
\label{sec:age_elements}
To illustrate the derivation of stellar population parameters form these index line-strengths we plot them for all mass bins against the high-resolution TMJ index model grids in Fig.~\ref{fig:grids}. The age and metallicity dependence is plotted for [$\alpha$/Fe] = 0 in the left panel showing [MgFe]$^\prime$ vs H$\beta$. The right panel illustrates metallicity against [$\alpha$/Fe] at a fixed age of 10~Gyr showing Mg$b$ vs <Fe>. <Fe> is the mean of the two Fe indices, Fe5270 and Fe5335, \citep{Gonzalez1993} and [MgFe]$^\prime$ is a mean between Mg$b$ and both iron indices as defined in \citet{Thomas2003}. Dark to light shades of blue represent the mass bins in decreasing order; decreasing size represents increasing radius. The model grid shows stellar populations for fixed age and metallicity and [$\alpha$/Fe] ratio as labelled.

It can be seen from these plots that our results are consistent with findings in the literature. Our data suggest that all three parameters, age, metallicity and [$\alpha$/Fe] ratio, correlate with galaxy mass \citep[e.g.][]{Kuntschner2000,Thomas2005,Thomas2010}. Moreover, there is a significant metallicity gradient, especially at higher masses, while gradients in age, except for the low mass bin, and [$\alpha$/Fe] are generally flat \citep{Mehlert2003, Goddard2017}.

\begin{figure*}
  \includegraphics[width=.32\linewidth]{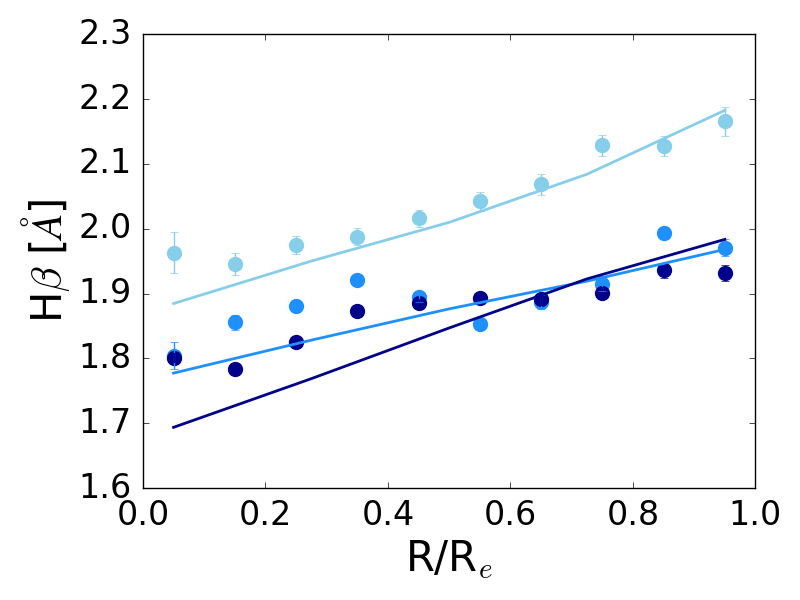}
  \includegraphics[width=.32\linewidth]{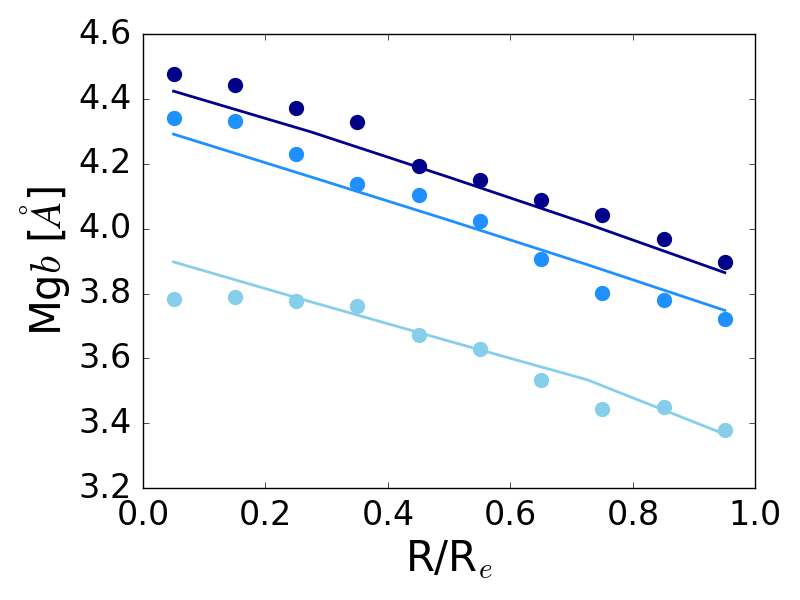}
  \includegraphics[width=.32\linewidth]{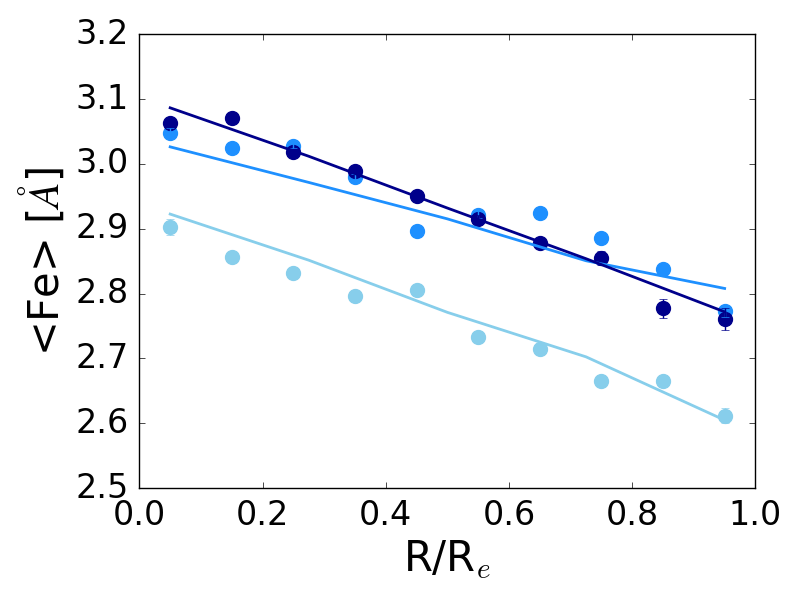}
  \includegraphics[width=.32\linewidth]{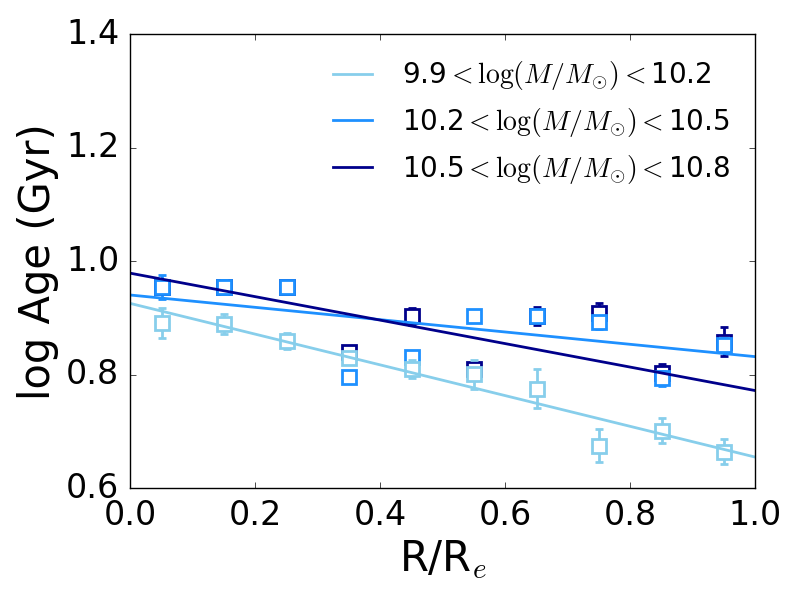}
  \includegraphics[width=.32\linewidth]{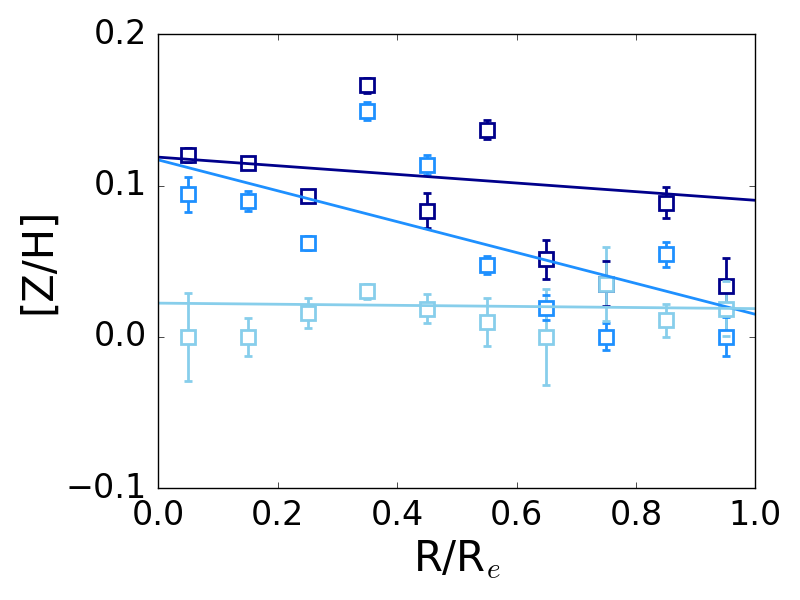}
  \includegraphics[width=.32\linewidth]{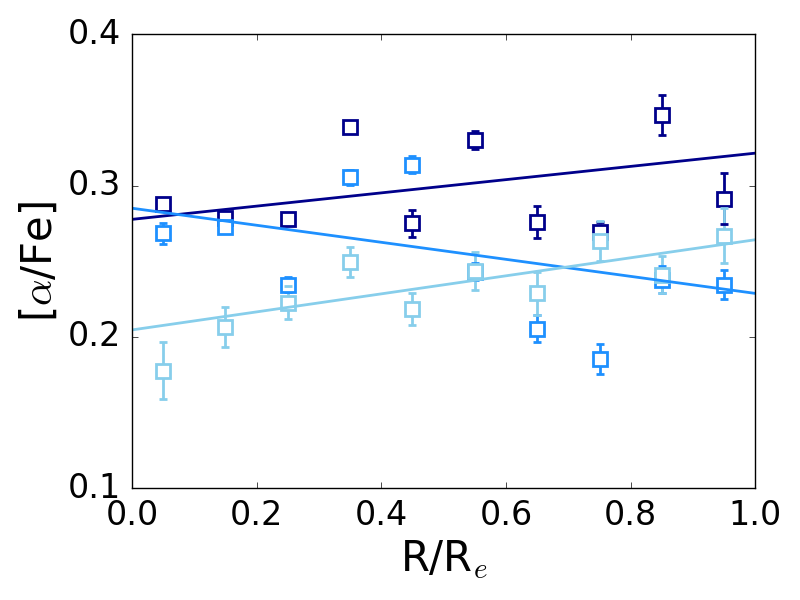}
\caption{Same as Fig.~\ref{fig:indices} for VCJ models.}
\label{fig:indices_Conroy}
\end{figure*}

In the top panel of Fig.~\ref{fig:indices}, we plot the measurements of the absorption-line indices H$\beta$, Mg$b$ and <Fe> for each mass bin as a function of radius. These indices are sensitive to the age, total metallicity, and [$\alpha$/Fe] ratio of a stellar population. As well known from the literature, more massive galaxies have weaker H$\beta$ and stronger metallic indices \citep[e.g.][]{Bender1992, Trager1998} which means they are older and more metal rich \citep[e.g.][]{Thomas2005}. Also, the observed profiles are in line with previous literature exhibiting positive radial gradients in H$\beta$ and negative radial gradients in both metallic indices \citep{Davies1993,Carollo1993,Mehlert2000}.

The best fit parameter and error at each radial bin is derived through the process described above. We fit a straight line to the parameters obtained at each point, taking the error into account, and derive the radial gradients summarised in Table \ref{tab:gradients}. The uncertainties in the gradients account for index errors and the scatter of the points around a straight line. The resulting model fittings to the index profiles are shown in the top panel of Fig.~\ref{fig:indices} as lines. The observed index values are reproduced well through our stellar population model fittings. Note that these models already include the inferred [Na/Fe] and IMF slopes shown in the next sections.

The bottom panels of Fig.~\ref{fig:indices} show the final resulting radial profiles of these three parameters. We detect negative metallicity gradients in all mass bins, with a steeper negative slope in the higher mass bins in agreement with \citet{Goddard2017, Zheng2017}. The age gradients are comparable in being shallow, except for the lowest mass-bin where we find a steeper negative gradient ($-0.18\pm 0.04$\ dex) than \citet{Goddard2017} in the equivalent mass range ($-0.03\pm 0.03$). We further find the [$\alpha$/Fe] gradients to be negligible at all masses in agreement with previous findings in the literature \citep{Mehlert2003,Kuntschner2010}.

We also derive these stellar population parameters using the VCJ models; the fits and parameters are shown in the top and bottom panels of Fig.~\ref{fig:indices_Conroy} respectively. In order to be consistent with the TMJ analysis, we create an index grid for VCJ models of different parameters and use the same chi-squared minimisation code to simultaneously fit H$\beta$, Mg$b$, Fe5270, Fe5335, NaD and NaI or FeH. A larger degree of scatter is found, hence the resulting gradients, though similar to TMJ models (apart from the highest mass bin), have larger uncertainties. See \citet{Conroy2017b} for a comparison of their optical indices with TMJ. 

\begin{table}
	\centering
	\caption{Gradients and errors for age, metallicity and abundances in dex/$R_e$.}
	\label{tab:gradients}
	\begin{tabular}{cccccc}
		\hline
		Mass bin & TMJ Age & Metallicity & [$\alpha$/Fe]\\
		\hline
		$9.9 - 10.2$ & $-0.18 \pm 0.04$ & $-0.11 \pm 0.02$ & $0.04 \pm 0.01$\\
		$10.2 - 10.5$ & $-0.09 \pm 0.06$ & $-0.15 \pm 0.03$ & $-0.08 \pm 0.03$\\
		$10.5 - 10.8$ & $-0.10 \pm 0.02$ & $-0.15 \pm 0.02$ & $0.02 \pm 0.02$\\
		\hline
		Mass bin & VCJ Age & Metallicity & [$\alpha$/Fe]\\
		\hline
		$9.9 - 10.2$ & $-0.27 \pm 0.02$ & $-0.00 \pm 0.02$ & $0.06 \pm 0.02$\\
		$10.2 - 10.5$ & $-0.11 \pm 0.07$ & $-0.10 \pm 0.06$ & $-0.06 \pm 0.05$\\
		$10.5 - 10.8$ & $-0.21 \pm 0.07$ & $-0.03 \pm 0.05$ & $0.04 \pm 0.04$\\
		\hline
	\end{tabular}
\end{table}

\subsection{[Na/Fe]}
\label{sec:na}
The sodium doublet, NaI, is strongly sensitive to changes in the IMF but is also affected by sodium abundance variations. In this work we lift this degeneracy by including the NaD doublet as main Na abundance indicator. NaD is highly sensitive to Na abundance \citep{Thomas2011a}, and its dependence on the IMF slope is negligible \citep{Conroy2012a, Spiniello2012, LaBarbera2013}.

\begin{figure*}
  \includegraphics[width=.32\linewidth]{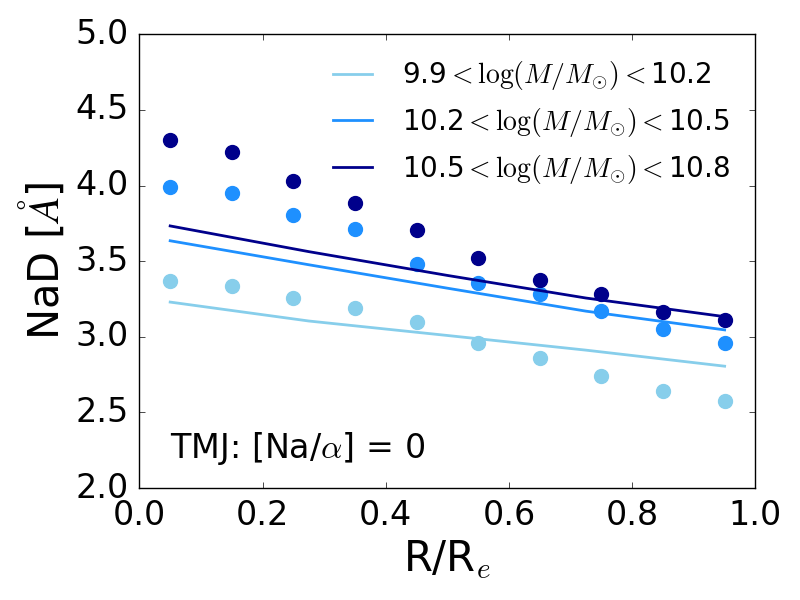}
  \includegraphics[width=.32\linewidth]{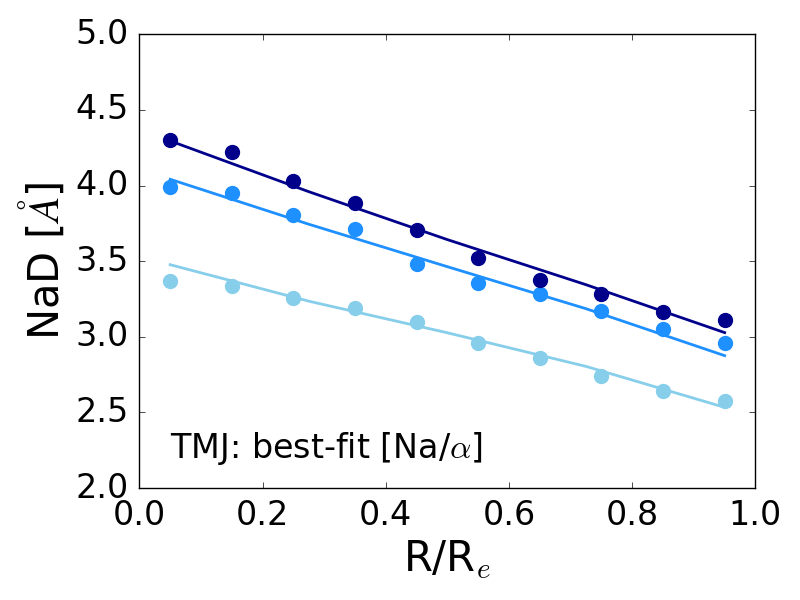}
  \includegraphics[width=.32\linewidth]{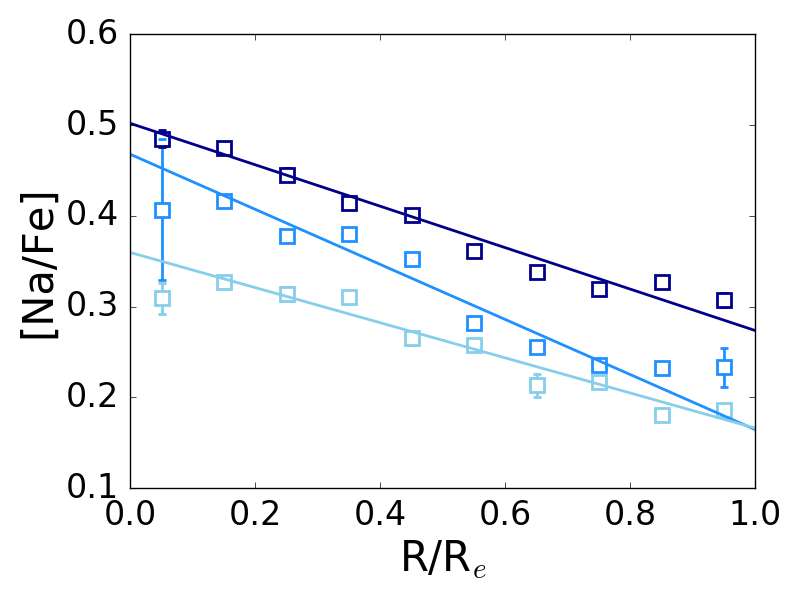}
  \includegraphics[width=.32\linewidth]{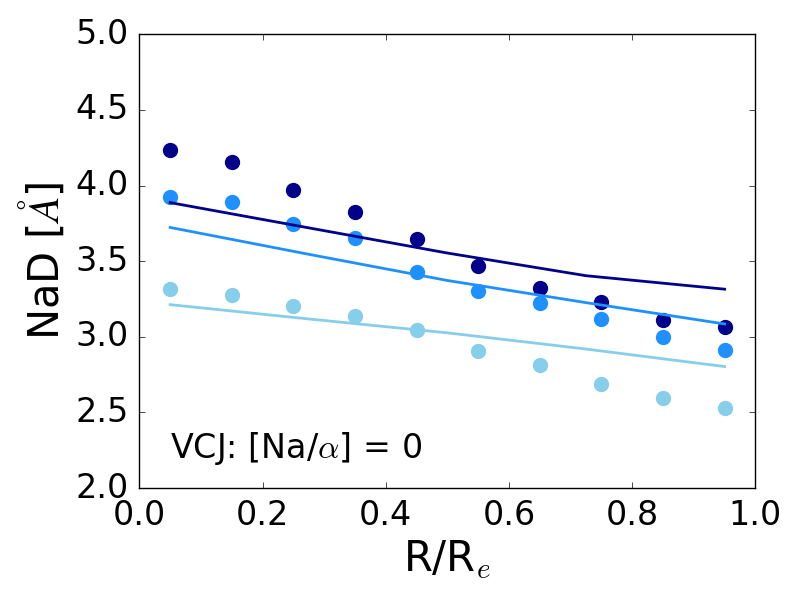}
  \includegraphics[width=.32\linewidth]{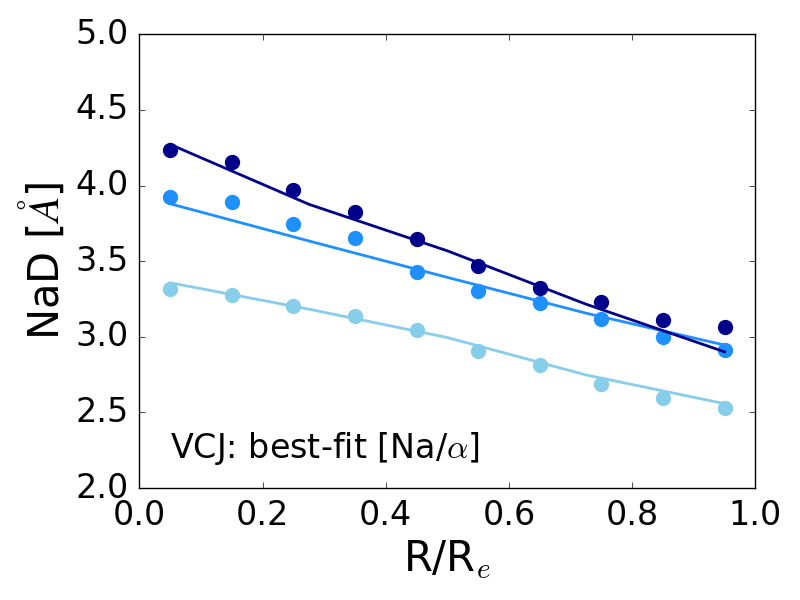}
  \includegraphics[width=.32\linewidth]{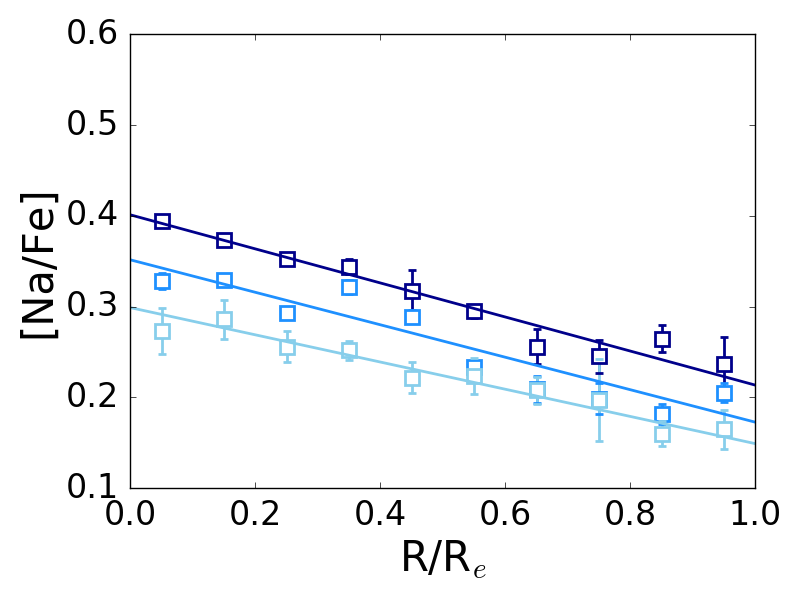}
\caption{NaD equivalent width as a function of radius for the mass bins, with 1-$\sigma$ errors calculated using a Monte Carlo-based analysis, and stellar population fittings using TMJ (top panels) and VCJ (bottom panels). Left: Lines are models with the previously derived parameters of age, metallicity and [$\alpha$/Fe]. Centre: [Na/Fe] abundance in models altered in order to match the data. Right: [Na/Fe] abundance (including [$\alpha$/Fe] given before) for the best-fit models.}
\label{fig:indices_nad}
\end{figure*}

The NaD profiles measured on the stacked spectra are shown as filled circles in the left-hand and middle panels of Fig.~\ref{fig:indices_nad}. NaD displays a strong negative gradient, slightly steepening with increasing galaxy mass. The left-hand panel also shows as lines the TMJ model predictions (top row) and VCJ model predictions (bottom row) based on our derived parameters using the corresponding models but without any Na abundance variation. It can be seen that the NaD gradients are not reproduced, and NaD strength is underestimated at small radii. This effect is strongest for the highest mass bin.

We then show both models (lines) in the middle panels of Fig.~\ref{fig:indices_nad} with the derived Na abundance. The data are now reproduced well. Note that these models, with and without Na abundance variation, already include the IMF variation shown in the next section.

The resulting [Na/Fe] gradients based on TMJ and VCJ are shown in the right-hand top and bottom panels of Fig.~\ref{fig:indices_nad}, respectively. The derived [Na/Fe] gradients, listed in Table \ref{tab:nagradients}, are around -0.24 dex/$R_\mathrm{e}$ for the TMJ models and -0.17 dex/$R_\mathrm{e}$ for the VCJ models, with no clear dependence on galaxy mass for both models. The overall value of Na-enhancement shows a very significant correlation with mass though, with the highest mass bin exhibiting the highest Na-enhancement with central values around $[{\rm Na}/{\rm Fe}]\sim 0.4 - 0.5\;$dex. The VCJ models predict stronger NaD indices and hence require less Na-enhancement, therefore leading to slightly lower central [Na/Fe] values and shallower negative gradients. These models also predict a more metal rich and alpha enhanced population for the high mass bin at large radii, which affects the fits for NaD and NaI in this region.

The detection of significant negative radial gradients in [Na/Fe] is in good agreement with other findings in the recent literature \citep{McConnell2016, vanDokkum2016, Alton2017, Vaughan2017}. The values derived here are slightly lower than in some of the work quoted above, where [Na/Fe] values as high as 0.9~dex in the most massive galaxies are reported.

Finally it should be noted that the NaD index could in principle be affected by contamination from interstellar medium absorption. While we expect this effect to be small in early-type galaxies, we explore this possibility and determine the reddening of our stacked spectra through the full spectrum fitting code FIREFLY \citep{Wilkinson2015,Wilkinson2017}. The results are presented in Appendix~\ref{sec:app_dust}. In summary, we find $E(B-V)$ values lower than 0.1, and a shallow gradient in $E(B-V)$ with a slightly negative slope around $-0.05$ per $R_\mathrm{e}$, consistent with the analysis in \citet{Goddard2017}. We conclude that the dust content in these object is small and the effect on NaD is negligible.

\begin{table}
	\centering
	\caption{Gradients and errors for [Na/Fe] from TMJ and VCJ models.}
	\label{tab:nagradients}
	\begin{tabular}{cccccc}
		\hline
		Mass bin & TMJ [Na/Fe] & VCJ [Na/Fe]\\
		\hline
		$9.9 - 10.2$ & $-0.19 \pm 0.01$ & $-0.15 \pm 0.01$\\
		$10.2 - 10.5$ & $-0.30 \pm 0.03$ & $-0.18 \pm 0.03$\\
		$10.5 - 10.8$ & $-0.23 \pm 0.02$ & $-0.19 \pm 0.01$\\
		\hline
	\end{tabular}
\end{table}

\subsection{IMF slope}
\label{sec:imf}
Having shown the stellar population parameters age, metallicity, [$\alpha$/Fe], and [Na/Fe], we now present the IMF slopes and NIR indices. We focus on NaI and FeH, which are both primarily gravity-sensitive indices and therefore good indicators of the IMF-dependent dwarf-to-giant star ratio \citep{vanDokkum2012}. As such, they have become standard features used for IMF studies in the literature.

For clarity and because of the different results provided by each index, we study the FeH and NaI IMF implications separately. We discuss the results obtained from both the M11-MARCS and VCJ models (see Section~\ref{sec:models}).

\subsubsection{NaI}
We use the NaI index definition by \citet{LaBarbera2013}. It is a narrow passband which excludes TiO and keeps the pseudo-continua blueward and redward of the feature in such a way that contamination from TiO is avoided (see Fig.~\ref{fig:features_NIR}).

\begin{figure*}
 \includegraphics[width=0.32\linewidth]{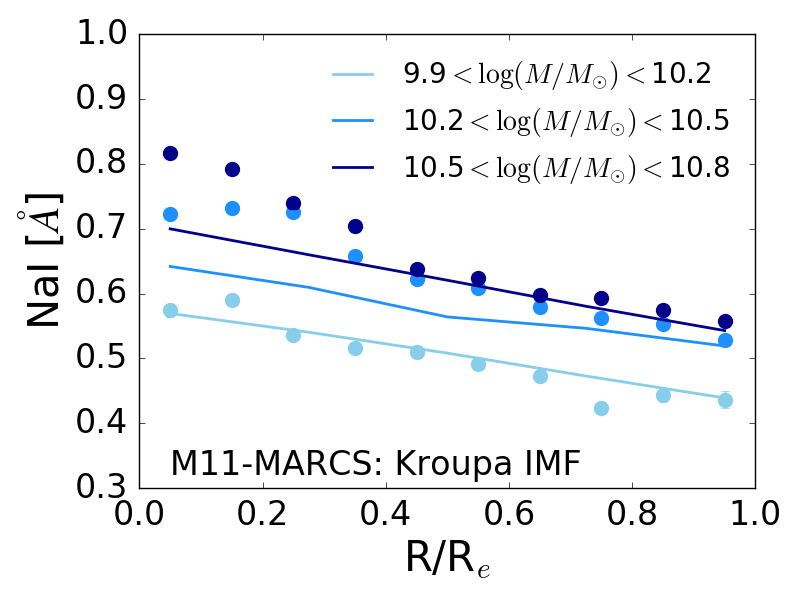}
 \includegraphics[width=0.32\linewidth]{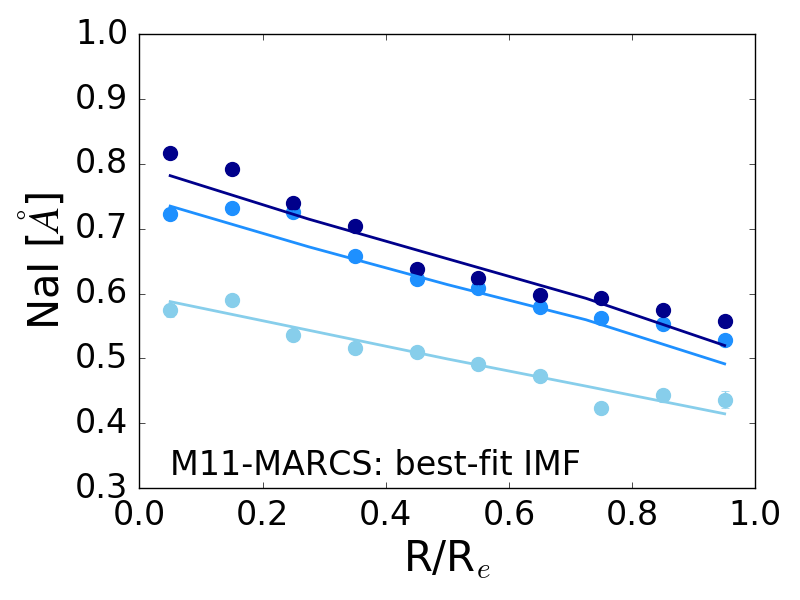}
 \includegraphics[width=0.32\linewidth]{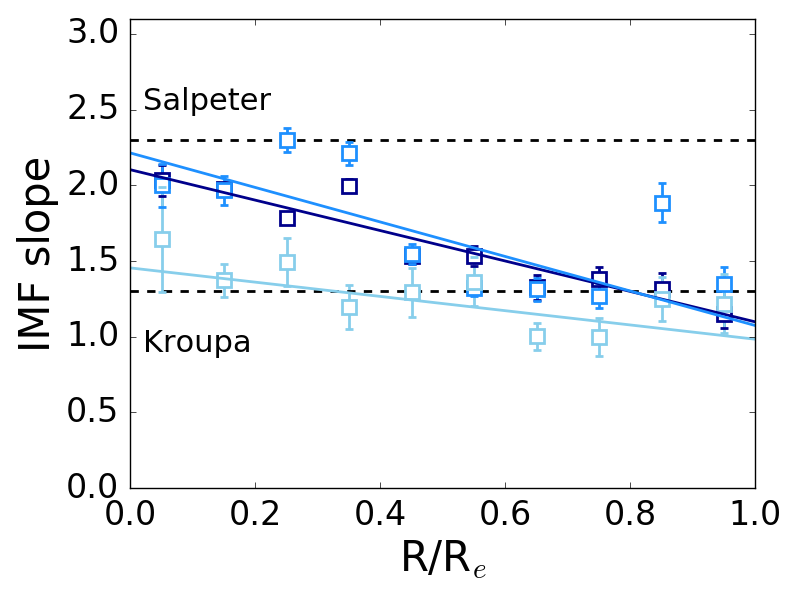}
  \includegraphics[width=.32\linewidth]{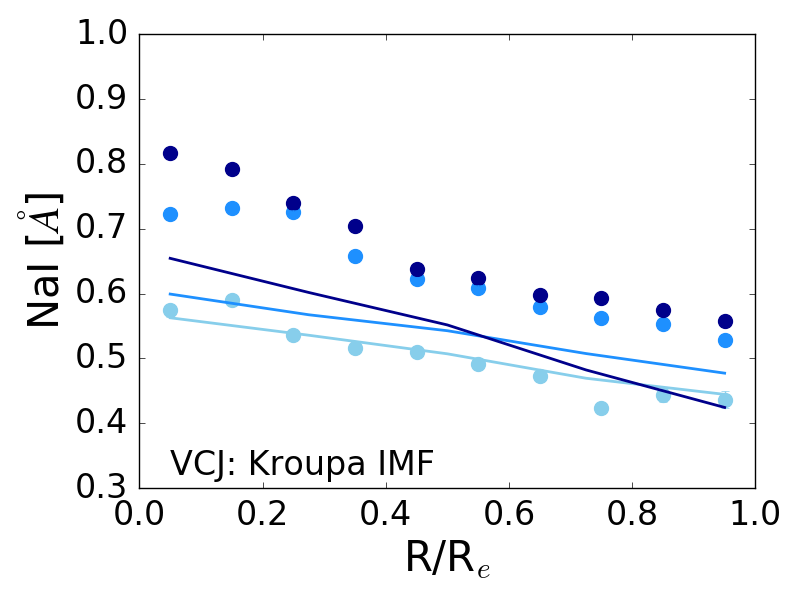}
  \includegraphics[width=.32\linewidth]{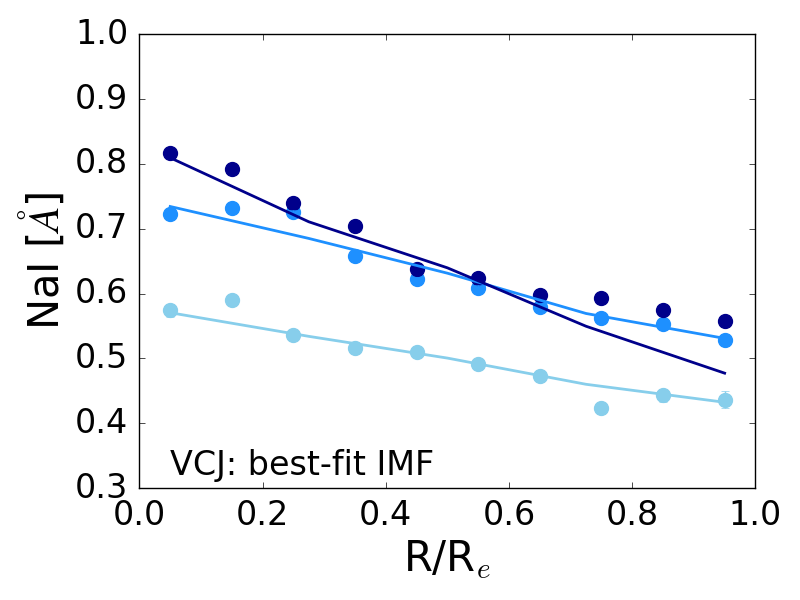}
  \includegraphics[width=.32\linewidth]{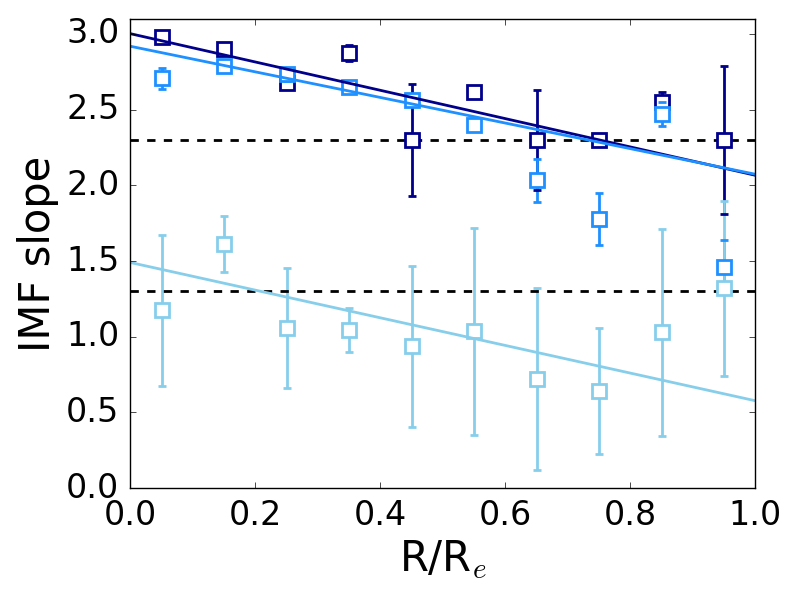}
\caption{NaI line-strengths for the mass bins as a function of radius, with 1-$\sigma$ errors calculated using a Monte Carlo-based analysis, and stellar population model fittings using M11-MARCS (top panels) and VCJ (bottom panels). {\it Left-hand panels}: Models shown as lines for a Kroupa IMF with the age, metallicity, [$\alpha$/Fe] and [Na/Fe] as derived in Sections~\ref{sec:age_elements} and \ref{sec:na}. {\it Middle panels}: Models (lines) with varying IMF. {\it Right-hand panels}: Open squares show the IMF slopes of the best-fit models shown in the middle panels.}
\label{fig:marcs}
\end{figure*}

The radial profiles of the NaI measurements are shown as filled circles in the left-hand and middle panels of Fig.~\ref{fig:marcs}. The negative gradient is evident with stronger NaI absorption toward the centre. The index strength is weakest at lowest masses, but very similar for the two high mass bins giving a generally weak dependence on galaxy mass within $\log M/M_{\odot}=10.2 - 10.8$. Also, the slope is only mildly mass-dependent with a slightly stronger rise in the centre for the more massive bins.

\paragraph*{M11-MARCS models:}
Using the derived age, metallicity, [$\alpha$/Fe] and [Na/Fe] from the previous sections, we plot the predictions from the M11-MARCS models with Kroupa IMF as lines in the top left-hand panel of Fig.~\ref{fig:marcs}. The models are well consistent with the data for the lowest mass bin at all radii, as well as at large radii in the two high mass bins. In the latter the model implies a shallower profile in NaI, though, and NaI strength is under-estimated in the centre, if no additional IMF variation is considered.

We now show the models with the best-fit IMF slope. To this end we extrapolate the M11-MARCS models in log space, which allows us to determine the best fit IMF slope in the mass range $0.1 - 0.5\; M_{\odot}$ (see Fig.~\ref{fig:sketch}). We fix the lower and upper bounds at $0.5$ and $3.5$, so IMFs shallower or steeper than these values are not allowed. The resulting stellar population models are shown in the top middle panel of Fig.~\ref{fig:marcs}. The observed NaI strengths are now well reproduced at all radii for all mass bins.

The corresponding IMF slopes are shown in the top right-hand panel of Fig.~\ref{fig:marcs}. The horizontal dashed lines illustrate the slopes of Salpeter and Kroupa IMFs (see Fig.~\ref{fig:sketch}). The fit results show that the NaI absorption we measure is roughly consistent with a Kroupa IMF, hence Milky Way-type IMF, and no significant enhancement in dwarf stars is required for the lowest mass bin ($\log M\sim 10$, $\sigma\sim 130\;$km/s). Moreover, only a shallow gradient is detected. The other two mass bins, instead, corresponding to $\log M\sim 10.4$ ($\sigma\sim 170\;$km/s) and $\log M\sim 10.6$ ($\sigma\sim 200\;$km/s), respectively, exhibit an enhancement of NaI in the centres requiring a slightly steeper IMF slope of around 2. The IMF slope required at around the half-light radius is Kroupa again, though, for both mass bins resulting in an IMF slope gradient. The derived radial gradients in IMF slope are listed in Table~\ref{tab:imfgradients_nai}. The linear fit is shown as lines in the top right-hand panel of Fig.~\ref{fig:marcs}.

The variation toward a more bottom-heavy IMF half-way between Kroupa and Salpeter (corresponding to a mass excess factor, $\alpha$ of about $1.5$) in the centres of galaxies in the range $170<\sigma/{\rm km/s}<200$ is in good agreement with previous results in the recent literature \citep{Cappellari2012,Conroy2012b,Ferreras2013,Spiniello2014,Smith2015b}. We find here that this variation is restricted to the centre, and a normal Milky Way-type IMF is again found in the outer parts. A more comprehensive discussion of these results including a full comparison with the literature is given in Section 4.

\begin{figure*}
  \includegraphics[width=.32\linewidth]{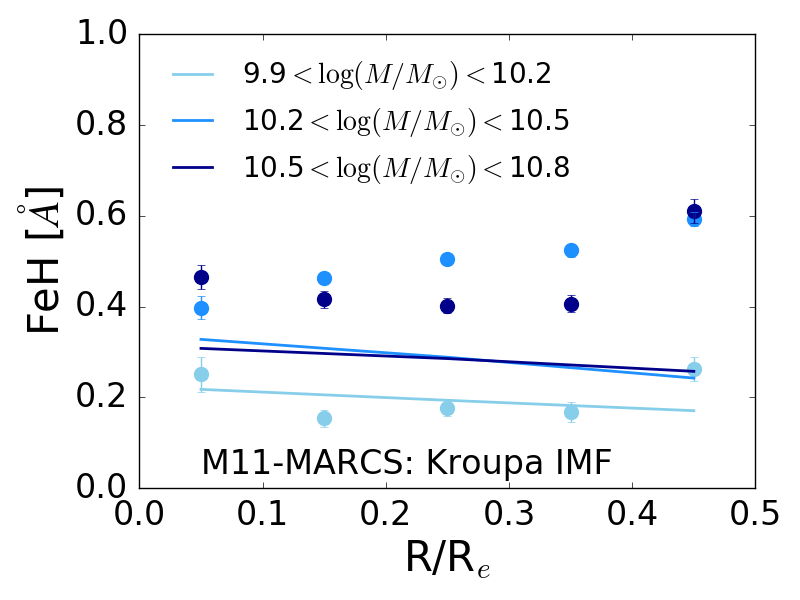}
  \includegraphics[width=.32\linewidth]{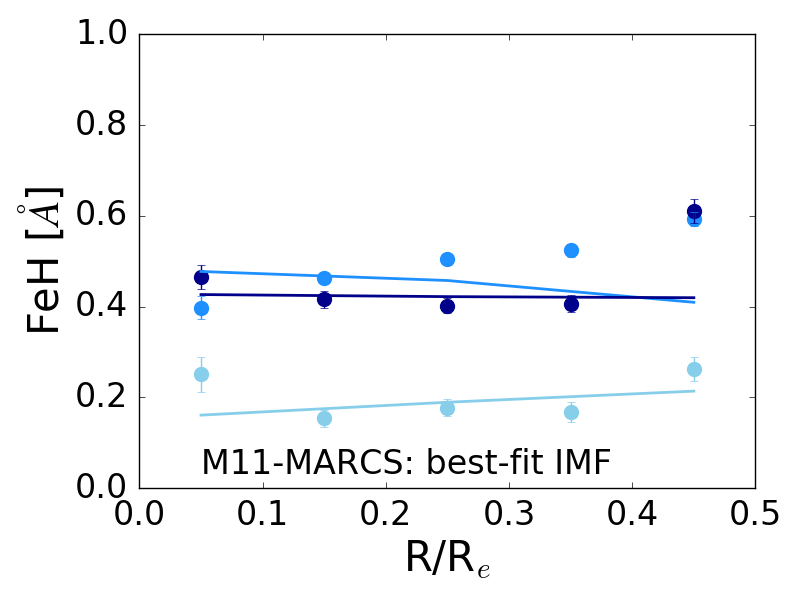}
  \includegraphics[width=.32\linewidth]{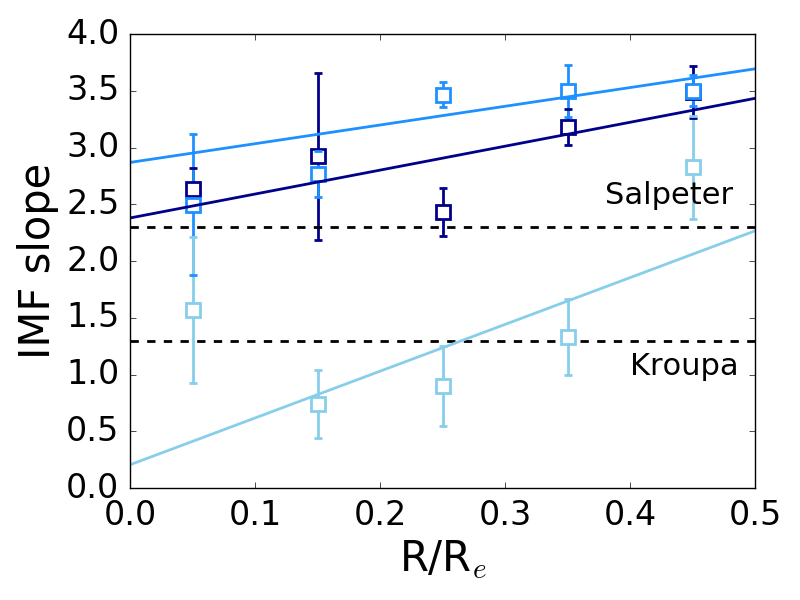}
    \includegraphics[width=.32\linewidth]{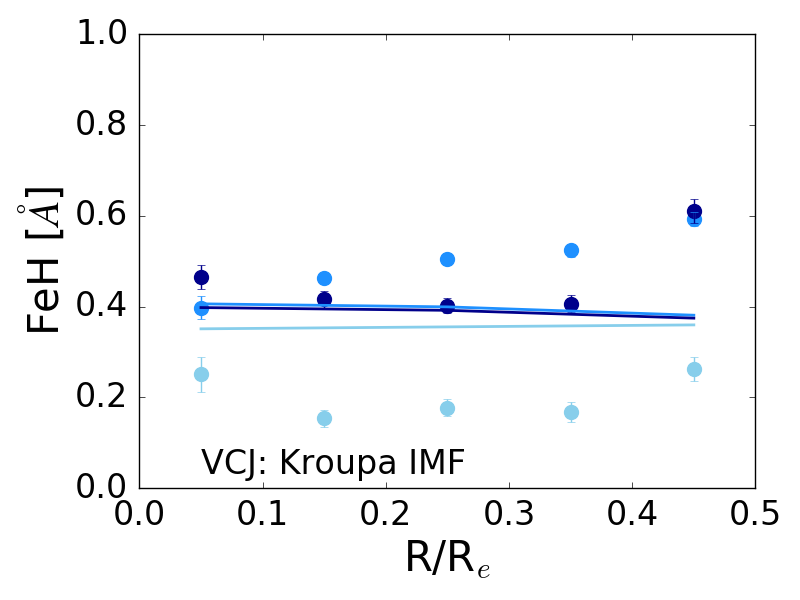}
  \includegraphics[width=.32\linewidth]{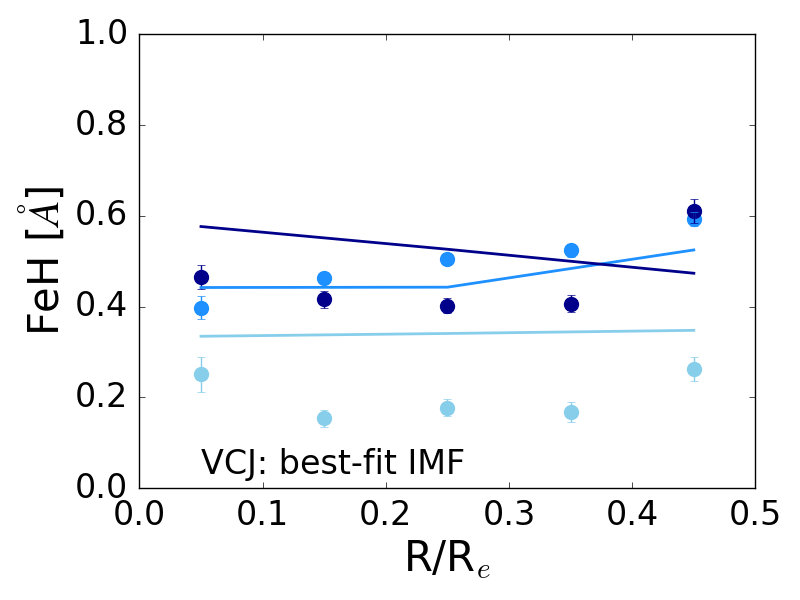}
  \includegraphics[width=.32\linewidth]{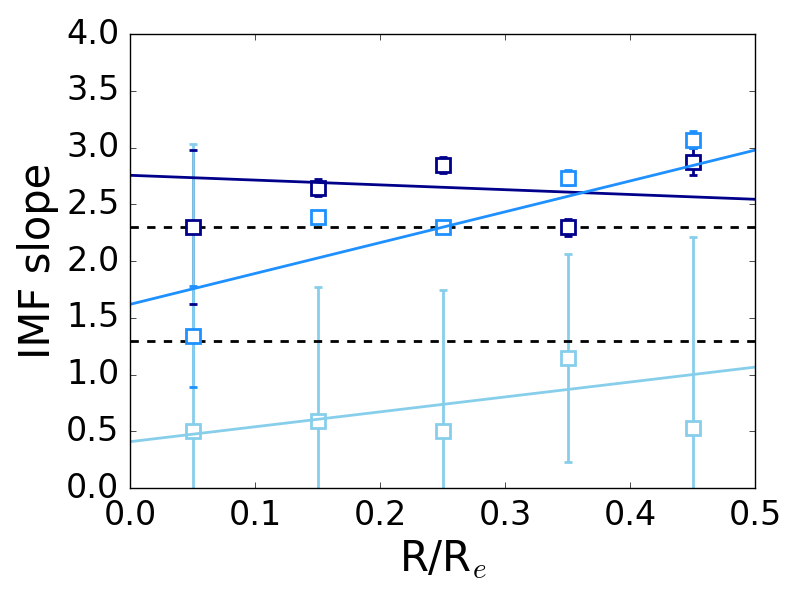}
\caption{FeH line-strengths for the mass bins as a function of radius, with 1-$\sigma$ errors calculated using a Monte Carlo-based analysis, and stellar population model fittings using M11-MARCS (top panels) and VCJ (bottom panels). {\it Left-hand panels}: Models shown as lines for a Kroupa IMF with the age, metallicity, [$\alpha$/Fe] and [Na/Fe] as derived in Sections~\ref{sec:age_elements} and \ref{sec:na}. {\it Middle panels}: Models (lines) with varying IMF. {\it Right-hand panels}: Open squares show the IMF slopes of the best-fit models shown in the middle panels.}
\label{fig:marcs_feh}
\end{figure*}

\begin{table}
	\centering
	\caption{Gradients and errors for IMF slope between 0.1-0.5M$_{\odot}$ per $R_e$, as derived from NaI.}
	\label{tab:imfgradients_nai}
	\begin{tabular}{cccc}
		\hline
		Mass bin & M11-MARCS & VCJ\\
		\hline
		$9.9 - 10.2$ & $-0.47 \pm 0.19$ & $-0.91 \pm 0.38$\\
		$10.2 - 10.5$ & $-1.14 \pm 0.45$ & $-0.85 \pm 0.20$\\
		$10.5 - 10.8$ & $-1.00 \pm 0.18$ & $-0.94 \pm 0.07$\\
		\hline
	\end{tabular}
\end{table}

\paragraph*{VCJ models:}
Similar to the procedure with M11-MARCS, we also model the NaI feature with the VCJ models. The lower and upper bounds on the IMF slope are the same as before. The bottom panels in Fig.~\ref{fig:marcs} show the results. We plot the predictions from the VCJ models with Kroupa IMF as lines in the bottom left-hand panel of Fig.~\ref{fig:marcs}. 

Just like with the M11-MARCS models, the observations are generally well reproduced for the low mass bin, and the biggest difference is at the centres in the higher mass bins. This difference becomes smaller as radius increases, however it is non-negligible at the half-light radius. Again, the fits can be further improved by including the IMF slope as an additional variable, and the resulting models and the corresponding IMF slopes are shown in the bottom middle and right-hand panels of Fig.~\ref{fig:marcs}, respectively.

The adjustments to the IMF slope yield a better representation of the data. This is particularly true at the centres where the enhanced NaI absorption requires a steepening of the IMF slope. Consistent with the analysis using the M11-MARCS models, a radial gradient in the IMF slope is found. The gradients derived with VCJ models are in agreement, to 1-$\sigma$, with the gradients derived through M11-MARCS. However the absolute values of the IMF slope are systematically higher for the VCJ models, requiring super-Salpeter slopes in the centre and a slope of around $2$ (between Kroupa and Salpeter) at the half-light radius. This discrepancy is caused by the lower [Na/Fe] ratios obtained with the VCJ models compared to TMJ (see Section~\ref{sec:na}).

The discrepancy in gradient appears most extreme in the lowest mass bin, where a sub-Kroupa IMF slope at the half-light radius and a significant negative gradient in IMF slope, albeit with large errors, are found with the VCJ model, in contrast to the shallow gradient around a Kroupa IMF derived with M11-MARCS. The origin of this discrepancy is unclear. We point out, though, that the VCJ Kroupa IMF model still appears to be a good fit to the data as can be seen from the bottom left panel of Fig.~\ref{fig:marcs}, which highlights the low sensitivity of the NaI index as the IMF gets more bottom-light i.e. the IMF needs to become significantly more bottom-light to cause a small change in the index.

\begin{table}
	\centering
	\caption{Gradients and errors for IMF slope between 0.1-0.5M$_{\odot}$ per $R_e$, as derived from the Wing-Ford band.}
	\label{tab:imfgradients_feh}
	\begin{tabular}{ccc}
		\hline
		Mass bin & M11-MARCS & VCJ\\
		\hline
		$9.9 - 10.2$ & $4.12 \pm 2.33$ & $1.31 \pm 1.37$\\
		$10.2 - 10.5$ & $1.65 \pm 1.04$ & $2.72 \pm 1.24$\\
		$10.5 - 10.8$ & $2.11 \pm 0.95$ & $-0.42 \pm 1.33$\\
		\hline
	\end{tabular}
\end{table}
 
\subsubsection{FeH}
The Wing-Ford band, FeH, is generally considered a strong IMF slope indicator because of the virtual absence of this feature in giant stars \citep{Conroy2012a}. The radial profiles of the FeH measurements are shown as filled circles in the left-hand and middle panels of Fig.~\ref{fig:marcs_feh}, which is the FeH equivalent of Fig.~\ref{fig:marcs}. Note that when discussing the Wing-Ford band we only consider data within 0.5~$R_\mathrm{e}$ as opposed to 1~$R_\mathrm{e}$, because the S/N ratio falls below the central value of $200$\; pixel$^{-1}$, well below the values compared the rest of the spectra beyond 0.5~$R_\mathrm{e}$ (see Section~\ref{sec:abs_ind} and Fig.~\ref{fig:stack}).

It is evident that the profiles of FeH are very different from NaI. They tend to be flat for the lowest mass bin, positive for intermediate masses, and a significant upturn around 0.5~$R_\mathrm{e}$ for the highest mass bin. Note that in Fig.~\ref{fig:features_NIR} we show that much of this radial change in the FeH index is driven by the red pseudo-continuum.

Results obtained with the M11-MARCS and VCJ models are again shown in the top and bottom panels, respectively. The parameters for these models have been obtained by simultaneously fitting FeH (instead of NaI) and the optical indices. Hence we obtain slightly different ages, metallicities and element abundances. These parameters, as well as the fits to the optical indices, are shown in Appendix~\ref{sec:app_feh}.

\paragraph*{M11-MARCS models:}
The M11-MARCS model sets (lines) do not reproduce the observed FeH strengths without adjustment of the IMF slope, as can be seen from the top left-hand panel of Fig.~\ref{fig:marcs_feh}. The models display a negative slope in FeH owing to the metallicity gradient measured, in contrast to the positive slope of the data. While the match between models and data in the lowest mass bin is still acceptable, FeH is clearly underestimated, in particular around 0.5~$R_\mathrm{e}$. After inclusion of the IMF slope as a free parameter, the fit improves but is still unable to reproduce the strengths of the high mass bins at large radii (top middle panel of Fig.~\ref{fig:marcs_feh}).

The resulting IMF slopes are shown in the top right-hand panel of Fig.~\ref{fig:marcs_feh}. The general pattern is similar to and consistent with the result obtained from NaI for the central radial point: the lowest mass bin is consistent with Kroupa within less than 1-$\sigma$, while a steeper IMF is needed for both higher mass bins. It is certainly encouraging that measurements through NaI and FeH roughly agree at the centre.

However, there is a marked difference with respect to the radial dependence. The FeH strength increases with radius for the two higher mass bins, and the resulting IMF consequently remains steep as radius increases. As a result, a more bottom-heavy super-Salpeter IMF, restricted by the model bounds, is obtained at 0.5~$R_\mathrm{e}$ in contrast to the result derived from NaI (see Fig.~\ref{fig:marcs}). The radial gradients in IMF slope are shown in Table~\ref{tab:imfgradients_feh}. The scatter in the points around the straight lines result in large errors on the gradients.

\paragraph*{VCJ models:}
The picture is similar when using the VCJ models as shown by the bottom panels in Fig.~\ref{fig:marcs_feh}. FeH in the VCJ models appears to be less sensitive to total metallicity, and the model lines lie much closer together before adjustment through IMF variation (bottom left-hand panel in Fig.~\ref{fig:marcs_feh}). Once a variable IMF slope is included, though, VCJ models reproduce the observed FeH strengths for the intermediate mass bin but fail elsewhere. For the lowest mass bin, the predicted FeH strengths are too high even with the most bottom-light IMF available (bottom middle panel in Fig.~\ref{fig:marcs_feh}).

The final constraint on the IMF slope is again similar to the result obtained form the M11-MARCS model. A positive gradient in IMF slope is being derived from the FeH measurements in disagreement with the results obtained from NaI (bottom right-hand panel in Fig.~\ref{fig:marcs_feh}). There are large errors on the IMF slope derived for the low mass bin at all radii due to the grid boundary and the inability to reproduce such low strengths.

We comment in detail on the reliability of our FeH analysis, both in terms of the measurements and the modelling, in the Discussion below.

\section{Discussion}
In this paper we derive radial gradients of IMF slope alongside stellar ages, metallicities, and [$\alpha$/Fe] and [Na/Fe] element abundance ratios. To this end we measure the key absorption line-strengths, H$\beta$, Mg$b$, <Fe>, NaD, NaI and FeH, on stacked spectra from the MaNGA survey providing us with spatially resolved information within the half-light radius. We then analyse these features with state-of-the-art stellar population models and derive gradients in IMF slope. The MaNGA data allows us to explore three mass bins centred on $\log M/M_{\odot}=10, 10.4, 10.6$ corresponding to central velocity dispersions of $\sigma = 130, 170, 200\;$km/s.

In the following we discuss our results in light of findings in the recent literature. We also provide a discussion on model dependence.

\subsection{Comparison with the literature}
Overall our results are consistent with the current literature in that the gravity-sensitive absorption feature at red wavelengths favour some variation in the IMF slope. We find a Milky-Way IMF (Kroupa) for our lowest mass-bin, and slightly more bottom-heavy IMFs (close to Salpeter) for the next higher mass bins in agreement with previous work \citep{Cappellari2012,Conroy2012b,Ferreras2013,Spiniello2014,Smith2015b,Li2017}. 

The principle aim of this work is to study the radial gradients, though, and here the literature is painting a more controversial picture. Some work reports the presence of a radial gradient with strongly bottom-heavy IMFs in the centres \citep{Martin-Navarro2015,LaBarbera2016,LaBarbera2017,vanDokkum2016,Conroy2017a}, while other studies find no compelling evidence for radial gradients in the IMF \citep{Zieleniewski2015, Zieleniewski2017,Alton2017}. If IMFs are indeed much more bottom-heavy in the centres, then dynamical mass estimates in the literature \citep[e.g.][]{Cappellari2013, Li2017} have been overestimated \citep{Bernardi2018}.

\subsubsection{Evidence for an IMF gradient}
\begin{figure*}
 \includegraphics[width=.7\linewidth]{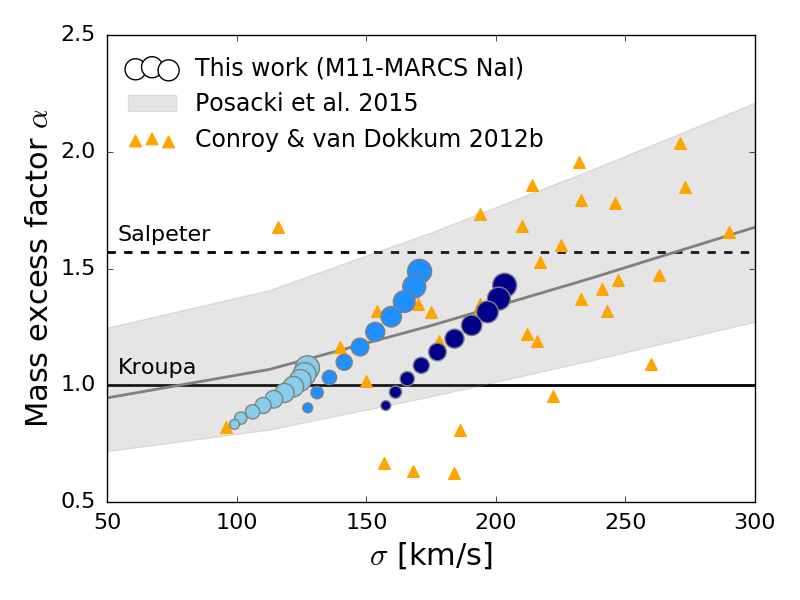}
\caption{Mass excess factor, $\alpha$ = ($M/L$)/($M/L_{\rm Kroupa}$) , versus velocity dispersion $\sigma$. Shaded circles are our results based on the NaI index using the M11-MARCS models for galaxies with $\log M/M_{\odot}=9.9 - 10.2$, $\log M/M_{\odot}=10.2 - 10.5$ and $\log M/M_{\odot}=10.5 - 10.8$. Dark to light shades of blue represent decreasing bins of mass and decreasing size represents increasing radius going from the centre to the half-light radius and the typical error on $\alpha$ is 0.1. The grey line is the \citet{Posacki2015} relation for the combined ATLAS$^{\rm 3D}$ and SLACS sample, with the shaded region representing 1$\sigma$ errors. Finally, the orange triangles are the galaxies from \citet{Conroy2012b}, with typical fractional error $\sim$ 7\%.}
\label{fig:lit_comparison}
\end{figure*}
\citet{Martin-Navarro2015} present radial constraints on the IMF for three early-type galaxies - one comparable to our low mass bin and the other two more massive than our galaxies. Although their observations of NaD and NaI are contaminated by sky emission and telluric absorption, they show that the gradients are consistent with a radial variation of the IMF for the massive galaxies. \citet{LaBarbera2016} use the Wing-Ford band to rule out a single power law IMF. This and TiO indices lead to a radial variation of the IMF in a massive $\sigma$ $\sim$ 300km/s galaxy. \citet{vanDokkum2016} also find strong radial variation in the IMF for six massive early type galaxies where after including abundance gradients they require an IMF variation to simultaneously fit NaD, NaI and FeH. \citet{LaBarbera2017} use a modified version of the stellar population models by \citet{Vazdekis2015} to match NaD, NaI and two further red NaI indices and still obtain radial IMF variations consistent with their previous result in \citet{LaBarbera2016}. Finally, \citet{Conroy2017a} use the observations for one of the galaxies from \citet{vanDokkum2016} to derive an extremely bottom-heavy IMF at the centre of the galaxy.

\subsubsection{Evidence against an IMF gradient}
\citet{Alton2017}, on the other hand, find no significant radial variation of the IMF in their galaxy sample. They study stacked and individual spectra of eight early-type galaxies with observations in the infrared. These galaxies mostly have $\sigma\sim 250\;$km/s. They find that the radial trends in these and other infrared indices can be accounted for by abundance variations (mostly in [Na/Fe]) rather than IMF changes. Likewise, \citet{Vaughan2017} also find that trends can be accounted for by abundance variations, although they cannot conclusively rule out the presence of an IMF gradient, and \citet{Zieleniewski2015,Zieleniewski2017} derive constant IMF slopes as a function of radius.

\subsubsection{This work}
The present study suggests that part of the discrepancies in the literature may well stem from a dependence on galaxy mass. We find a clear dependence of the radial IMF gradient with galaxy mass. However, we also show that the absolute value of the IMF slope is modelling dependent. We emphasise again that the galaxies studied here are on the low mass end of the galaxies typically studied in literature.

We derive a significant IMF gradient for the two larger mass bins ($\log M/M_{\odot}=10.4 - 10.6$, $\sigma = 170-200\;$km/s). This result is independent of the stellar population models used. For the M11-MARCS models, the derived IMF is Kroupa around the half-light radius and then slightly bottom-heavy (close to Salpeter) in the centre. The lowest mass bin ($\log M/M_{\odot}=10$, $\sigma = 130\;$km/s), instead, shows only a mild radial gradient in the IMF around a Kroupa slope. This result would suggest that the correlation of IMF slope with galaxy mass reported in the literature is generated by an increasingly stronger enhancement of dwarf stars in galaxy centres as the galaxy mass increases.

If we use the VCJ model, instead, we find departures from Kroupa IMF at all radii, more so in the centres, resulting in overall steeper IMF slope (super-Salpeter with a slope of around $3$ in the centre and $2$ at the half-light radius) for galaxies with masses above $\log M/M_{\odot}=10.4$ ($\sigma = 170\;$km/s). We derive a negative gradient for the lowest mass-bin ($\log M/M_{\odot}=10.0$ ($\sigma = 130\;$km/s), though, with a significantly sub-Kroupa bottom-light IMF at large radii, which seems somewhat contrived and incompatible with Milky Way measurements.

\subsubsection{Local vs. Global IMF-$\sigma$ relation}
Fig.~\ref{fig:lit_comparison} shows a comparison of our radial IMF results from the M11-MARCS models using the NaI index (the top panel of Fig.~\ref{fig:marcs}) with previous work from dynamical and lensing analysis \citep[][Equation 6]{Posacki2015} and stellar population analysis \citep[][Table 2]{Conroy2012b}. Note that the aperture size for the former can be up to 1~$R_\mathrm{e}$ and the latter is within $R$ < $R_\mathrm{e}$/8. We do not correct for aperture sizes because we wish to compare our radial gradient with the global trend rather than the absolute values.

From our IMF slopes, we can derive the corresponding mass excess factor, $\alpha$ = (M/L)/(M/L$_{Kroupa}$), obtaining the M/L values from \citet{Maraston2005}, based on the age and metallicity in Section~3.1. The velocity dispersion used here is $\sigma_{\rm mean}$ (solid line in Fig.~\ref{fig:dispersions}), allowing a comparison of the IMF-$\sigma$ relation within galaxies with the global relation. The three mass bins appear to follow the \citet{Conroy2012b} points. Very interestingly, in comparison to the combined dynamical and lensing relation (grey line), the gradient is steeper, implying that the relation within galaxies is steeper than the global IMF-$\sigma$ relation.

\begin{figure*}
 \includegraphics[width=\linewidth]{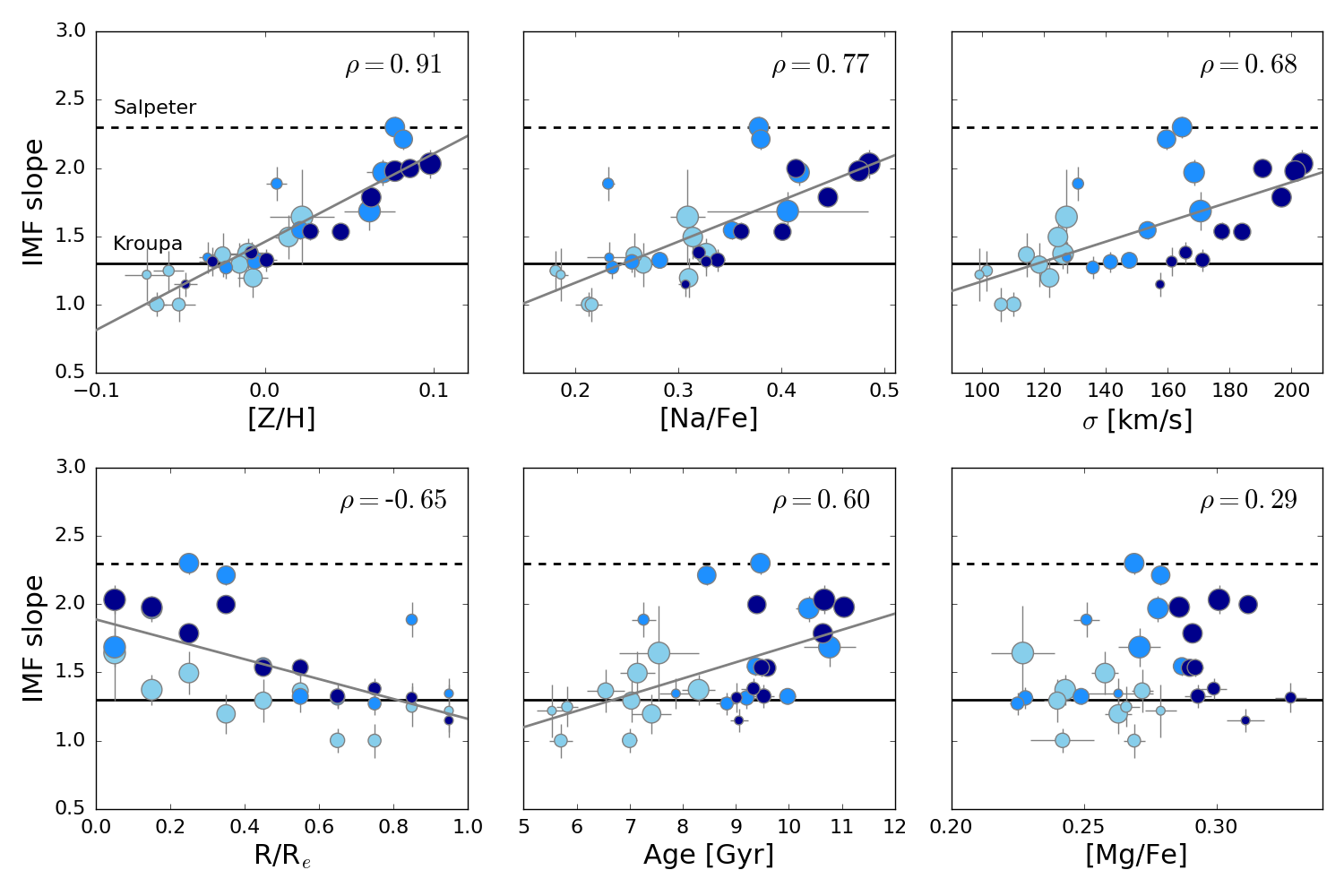}
\caption{IMF slope versus various parameters: metallicity [Z/H], [Na/Fe], velocity dispersion $\sigma$, radius, age, and [Mg/Fe]. These parameters and 1-$\sigma$ errors, derived for each radial bin, are based on the analysis using the TMJ and M11-MARCS models for galaxies with $\log M/M_{\odot}=9.9 - 10.2$, $\log M/M_{\odot}=10.2 - 10.5$ and $\log M/M_{\odot}=10.5 - 10.8$ (Section~\ref{sec:results}). Dark to light shades of blue represent decreasing bins of mass and decreasing size represents increasing radius going from the centre to the half-light radius. The Spearman correlation coefficient for each parameter is shown and a best-fit line is plotted when correlations are found.}
\label{fig:imf_params}
\end{figure*}

\subsubsection{What drives the IMF variation?}
The main contenders proposed for driving the IMF variation of galaxies are velocity dispersion \citep[e.g.,][]{Cappellari2012}, [Mg/Fe] \citep{Conroy2012b}, and metallicity \citep{MartinNavarro2015}. In order to test this with our results, we plot our IMF slopes versus all derived parameters: metallicity [Z/H], [Na/Fe], velocity dispersion $\sigma$, radius, age, and [Mg/Fe] (equivalent to our [$\alpha$/Fe]) in Fig.~\ref{fig:imf_params}. These parameters, along with the errors, were derived using the TMJ and M11-MARCS models (Section~\ref{sec:results}).

Each panel shows the Spearman correlation coefficient, $\rho$, and the panels are ordered from strongest to weakest correlation. The tightest relation is found between IMF slope and metallicity, and no correlation is found with [Mg/Fe], as also seen in \citet{MartinNavarro2015, vanDokkum2016}. We confirm this radial trend for a large sample of 366 galaxies. There are still some significant correlations found with [Na/Fe] ($\rho$ = 0.77), velocity dispersion ($\rho$ = 0.68) and stellar population age ($\rho$ = 0.60), alongside the negative correlation with radius ($\rho$ = -0.65). While the IMF gradient may well be a consequence of these correlations, the ultimate driver of IMF variations may still be a combination of metallicity, [Na/Fe], velocity dispersion, and age. Our results suggest however, in agreement with previous claims, that metallicity appears to be the dominating factor with metal-rich environments favouring the formation of low-mass stars. This is plausible astrophysically but solid theoretical support for this picture through star formation simulations remains yet to be found. While simulations may suggest IMF variations with bottom-heavier IMFs in environments of higher turbulence and star-formation rate \citep{Kroupa2013,Chabrier2014}, there is no clear evidence for a metallicity dependence along the lines found here \citep{Bate2014}.


\subsection{IMF diagnostics}
The derivation of the IMF slope in unresolved populations is naturally challenging. The approach discussed in this paper is matching stellar population models to observed absorption line-strengths, harnessing the fact that some features, particularly in the red optical wavelength regions are particularly sensitive to the ratio of dwarf to giant stars. Here we address some of the pitfalls and systematic uncertainties of this approach.

\subsubsection{Choice of IMF indicator}
A variety of absorption features are gravity sensitive and are used as IMF indicators in the current literature (see references in the introduction), including several Na features, the Ca triplet, the TiO bands and the Wing-Ford band. Ideally all these features should be recovered with the final best-fit model. However, systematics in measurement and modelling are different for different features, and not all of them are necessarily available in a particular set of observations. Therefore, many studies use a subset. In this work we focus on NaI and FeH as IMF indicators, modelling them separately, in order to constrain the complexity of the problem. Other features readily available in the MaNGA data, like CaT and TiO, are subject of future work.

The NaI index is widely used in the literature as an IMF indicator and our modelling performs well in matching this particular feature. Here we adopt the definition from \citet{LaBarbera2013} to minimise the effects of TiO contamination. The major caveat remains its dependence on Na abundance, even if relatively minor. We address this problem by deriving Na abundance independently from the NaD feature. This works well, but it means that any result we derive on the IMF slope directly depends on our ability to correctly infer Na abundance through NaD. In case of contamination from interstellar absorption due to dust in galaxy centres (see Appendix~\ref{sec:app_dust}), we would derive lower Na abundances. This in turn would lead to even steeper IMFs in galaxy centres and consequently steeper IMF gradients.

The Wing-Ford band, FeH, is considered an alternative IMF indicator, as it is strongly present in the spectra of dwarf stars. While this feature is less affected by Na abundance (however see \citealt{Alton2017}), it is highly sensitive to Fe-abundance, and hence the inferred [$\alpha$/Fe] ratio. Again we address this by deriving [$\alpha$/Fe] from optical indices independently. Nevertheless, the results based on FeH are contrived. Quite surprisingly, FeH displays a {\it positive} radial gradient, a measurement that is also seen by \citet{Alton2017}. The consequence is that we derive steeper IMFs, hence higher dwarf-to-giant ratios at large radii, which is inconsistent with our results obtained from NaI.

The FeH region of the spectrum is known to contain significant telluric absorption and sky emission, hence any residuals from imperfect subtraction of telluric lines and/or sky background will affect the FeH measurement. MaNGA observes multiple standard stars through fiber bundles simultaneously with the science targets. Slightly different wavelength sampling provided by different stars is combined to sample the high frequency variation providing an accurate telluric correction (Yan et al. 2016a). Also, there are no strong telluric features red-ward $9800\;$\AA\ rest-frame for the MaNGA sample with a median redshift of 0.03. Hence overall we do not expect telluric absorption to be a major problem in measuring the FeH feature on MaNGA spectra. There is reduced detector sensitivity at the wavelength edge, and the RMS variability in the telluric absorption measurements increases from $\sim 1$ per cent to $\sim 5$ per cent \citep[see also Fig. 5 of][]{Yan2016a}, however, this noise is random, which we expect to mitigate through stacking.

Therefore, the most likely contributor to uncertainties in FeH measurements from MaNGA spectra may be sky subtraction residuals. The DRP aims to reach Poisson-limited performance between $4000-10,000\;$\AA\ such that stacked spectra are generally not limited by systematic sky subtraction residuals (Law et al. 2016). We would therefore expect in principle that faint features like FeH can be measured accurately in a stacked, high-S/N MaNGA spectrum. However, the sky lines in the FeH region are very bright and pervasive, hence small uncertainties in sky subtraction will not cancel out in the stacked spectrum as effectively as at other wavelengths.

We note that similar caveats may also apply to NaI, however, the sky lines in this region are less bright and more distinct, hence we can expect that they are eliminated effectively during the stacking procedure. Note that, again in the rest frame for MaNGA galaxies, there are no significant telluric features in the NaI wavelength range.

As well as the problems in measuring FeH, it has been noticed before it is indeed difficult to model \citep{Vaughan2017,Alton2017}, so might not actually serve as a reliable IMF indicator. We remind the reader that most of the variation in FeH is driven by the red pseudo-continuum (see Fig.~\ref{fig:features_NIR}), which may not be well understood in current stellar population models. In fact FeH does not actually correlate well with galaxy velocity dispersion as can be seen in the original work by \citet{vanDokkum2012}. In this paper we are therefore inclined to discard the results obtained from FeH, concluding that our understanding of this feature is still limited.

\subsubsection{Stellar population models}
Part of the systematic uncertainties in IMF diagnostics come from differences in stellar populations modelling. We try to shed some light on this issue by using different model sets (TMJ/M11-MARCS and VCJ). While we find good agreement regarding the need for an IMF that varies with galaxy mass and with radius for more massive galaxies, discrepancies remain significant for more detailed aspects of the analysis. Progress in the field certainly relies on further improvements on the stellar population modelling side and in particular a better understanding of the systematic uncertainties involved.

\subsubsection{Constraints from chemical evolution}
It is well known that the IMF slope significantly affects chemical enrichment in a galaxy \citep[e.g.,][]{Thomas1999}. In fact, the finding of a dwarf-dominated population in massive galaxies is difficult to reconcile with its $\alpha$-enhanced populations \citep{Ferreras2013,Martin-Navarro2016,deMasi2017} because of the resulting lack of $\alpha$-enriching Type~Ia supernovae in the chemical enrichment process. This challenge ultimately needs to be overcome should the finding of the presence of bottom-heavy IMFs in massive galaxies prevail.

A further concern are some of the specific element abundance ratios derived. For instance, the Ca abundance required to match the CaT index (with a bottom-heavy IMF) is inconsistent with the Ca abundance inferred from the optical index Ca4227 \citep{LaBarbera2013}. 

Likewise, and more relevant to the present study, the Na abundance generally derived in conjunction with using NaI as an IMF indicator is puzzling. While the [Na/Fe] ratio derived here ($<0.5\;$dex, three times the solar value) is less extreme than some other much higher values found in the literature \citep{Vaughan2017,Alton2017}, it remains unclear whether these results can be reproduced with models of chemical evolution. Type~II supernovae do not produce [Na/Fe] ratios high enough \citep{Woosley1995}. From observations of halo and disc stars within the solar neighbourhood no stars have been found with $[{\rm Na}/{\rm Fe}] > +0.2\;$dex to date, and metal-poor halo stars even seem to have sub-solar [Na/Fe] \citep{Bensby2014}. Na is not enriched in lock-step with Mg and other $\alpha$-elements. However, [Na/Fe] does rise to about $\sim 0.3\;$dex for stars with super-solar metallicity both in the disc and in the bulge \citep{Bensby2017}, and [Na/Fe] clearly correlates with metallicity. The strong Na-enhancement detected in galaxy centres may therefore be the consequence of a metallicity-dependent Na yield \citep{Alton2017}, even though it remains difficult to accommodate the high values derived here and other studies in the recent literature with [Na/Fe] abundance ratios of individual stars in our Galaxy.

\subsection{Future work}
We will extend the present analysis in several aspects in future work. Firstly, we will include further diagnostics available from MaNGA spectroscopy in the analysis, including gravity sensitive Mg, Ca and TiO absorption features. We will also attempt to increase the mass range and include bins of higher galaxy mass. The restriction in mass of the present study mostly comes from the requirement of having a large galaxy sample for stacking in order to achieve the necessary high S/N ratio. We will benefit from the larger galaxy samples in future data releases of MaNGA to this end. Further ahead, closer to the final MaNGA sample size of $\sim 10,000$ galaxies, we expect to be able to extend the parameter space considerably and carry out an investigation of radial IMF gradients as a function of galaxy environment, galaxy mass and galaxy type.

\section{Conclusions}
We derive radial gradients within the half-light radius in the IMF slope, alongside stellar ages, metallicities, and [$\alpha$/Fe] and [Na/Fe] element abundance ratios, for a sample of 366 early-type galaxies in the mass range $9.9 - 10.8\;\log M_{\odot}$ using IFU observations from the SDSS-IV MaNGA survey. MaNGA provides spectra for a large wavelength range from $3,600\;$\AA\ to $10,000\;$\AA, so that key spectral features that are sensitive to stellar population parameters, element abundance ratios and dwarf-to-giant ratio can be measured simultaneously.

The spectra of individual spaxels are binned in two steps: radially within galaxies in bins with a width of $0.1\; R_{\rm e}$ and additionally in the three galaxy mass bins. The mass bins contain 122 galaxies each and are centred on $\log M/M_{\odot}$ of 10, 10.4, and 10.6, corresponding to a central velocity dispersion $\sigma$ of 130, 170, and $200\;$km/s, respectively. In this way we achieve a very high S/N of at least $\sim 1000$\; pixel$^{-1}$ out to the half-light radius around $8,500\;$\AA. Crucially, the spectral stacking mitigates the problem of contamination from sky line residuals, and therefore provides highly cleaned spectra at all wavelengths, except potentially at the very red end close to the region of the Wing-Ford band.

The stellar kinematics and the emission line component of the stacked spectrum are derived through spectral fitting. We then remove the emission line component of the spectrum and measure the key absorption indices H$\beta$, Mg$b$, <Fe>, NaD, NaI and FeH. Index measurements are corrected for the effect of velocity dispersion broadening and analysed with state-of-the-art stellar population models deriving gradients in IMF slope as well as age, metallicity, [$\alpha$/Fe], and [Na/Fe]. To assess possible systematic effects different sets of stellar population models are used \citep{ThomasD2011b,Maraston2011,Conroy2017b} with particular focus on differences in the modelling of the IMF-sensitive near-IR indices.

We find significant radial gradients in all absorption indices measured, except the Wing-Ford band FeH. The strengths of Mg$b$, Fe5270, Fe5335, NaD, and NaI increase, while the strengths of H$\beta$ and FeH decrease toward the galaxy centre. These profiles generally get stronger with increasing galaxy mass, a trend that is most pronounced in NaD. Most importantly, a clear gradient in NaI is detected, which also steepens with increasing galaxy mass.

The stellar population fittings of the optical absorption features yield gradients in age, metallicity and [$\alpha$/Fe] ratio. We find consistency with the literature for the age and metallicity gradients in sign, slope and zero-point, except for the lowest mass bin for which we measure a somewhat steeper negative age gradient. Our [alpha/Fe] values, which have no significant slope in all three mass bins, are also consistent with the literature.

The relatively steep gradient in NaD absorption interestingly leads to the derivation of significant gradients in Na abundance with $\nabla_{[{\rm Na}/{\rm Fe}]}\sim -0.25\;$dex per $R_{\rm e}$. The strength of the Na enhancement in galaxy centres increases with increasing galaxy mass with a maximal $[{\rm Na}/{\rm Fe}]\sim 0.5\;$dex. Such relatively high [Na/Fe] ratios are puzzling, as Milky Way stars have not exceeded a value of about $0.2\;$dex to date. The origin of this Na enhancement in the centres of early-type remains yet to be explained through chemical evolution modelling.

The observed index strengths of NaI are well reproduced with models of a mildly varying IMF slope. The IMF slope inferred in the centres is Milky Way-type (Kroupa) for the lowest mass bin ($\log M/M_{\odot}\sim 10$) and slightly more bottom-heavy close to Salpeter for the higher mass bins ($\log M/M_{\odot}\sim 10.4, 10.6$). These values are in good agreement with recent literature. VCJ models lead to steeper IMF slopes (super-Salpeter) in the centres. 

More specifically, our analysis based on the M11-MARCS models yields a galaxy mass-dependent radial gradient. While the lowest mass bin shows a shallow gradient in IMF slope around a Kroupa IMF, more massive galaxies are found, instead, to have a significant gradient in IMF slope with the values converging to a Kroupa IMF at the half-light radius for all masses. This implies that the emergence of a bottom-heavy IMF in more massive galaxies is a phenomenon restricted to galaxy centres, providing hints to the different formation and evolution processes of these regions. This trend further suggests decoupled formation scenarios for the centres and the outskirts, with galaxy centres probably growing and quenching rapidly and outskirts assembling over longer periods of time through minor mergers.

While the analysis based on the VCJ models leads to a super-Salpeter IMF in the centres of more massive galaxies, we also obtain a steeper IMF at large radii with an IMF half-way between Kroupa and Salpeter at the half-light radius. As a consequence, the derived gradient in the IMF slope agrees well with the gradient derived through M11-MARCS.

Finally, the Wing-Ford band turns out to be challenging for both models, and the strong positive gradient we measure in FeH implies radial trends of the IMF slope that are inconsistent with the results obtained through NaI. However, the FeH measurements might be affected by sky residuals not mitigated through stacking, and this may explain the contrived IMF slopes derived and why the models are unable to reproduce the strengths.

We are planning to extend the present analysis in several aspects in future work including further diagnostics available from MaNGA spectroscopy in the analysis and increasing the mass range, once larger MaNGA samples are available. Further ahead, closer to the final MaNGA sample size of $\sim 10,000$ galaxies, we expect to be able to extend the parameter space considerably and carry out an investigation of radial IMF gradients as a function of galaxy environment, galaxy mass and galaxy type.

\section*{Acknowledgements}
TP would like to thank P. Guarnieri and P. Carter for fruitful discussions. The authors are grateful to C. Conroy for providing latest versions of their stellar population models. We are also thankful to the referee for useful comments and suggestions which have improved the paper.

This research made use of the Python packages numpy \citep{VanDerWalt2011}, scipy \citep{Jones2001}, Matplotlib \citep{Hunter2007}, and Astropy \citep{Astropy2013}. This work also made use of Marvin, a core Python package and web framework for MaNGA data, developed by Brian Cherinka, Jos\'e S\'anchez-Gallego, and Brett Andrews. (MaNGA Collaboration, 2017).

TP is funded by a University of Portsmouth PhD bursary. The Science, Technology and Facilities Council is acknowledged for support through the Consolidated Grant Cosmology and Astrophysics at Portsmouth, ST/N000668/1. Numerical computations were performed on the Sciama High Performance Computer (HPC) cluster which is supported by the Institute of Cosmology of Gravitation, SEPnet and the University of Portsmouth.

Funding for the Sloan Digital Sky Survey IV has been provided by the Alfred P. Sloan Foundation, the U.S. Department of Energy Office of Science, and the Participating Institutions. SDSS-IV acknowledges
support and resources from the Center for High-Performance Computing at
the University of Utah. The SDSS web site is www.sdss.org.

SDSS-IV is managed by the Astrophysical Research Consortium for the 
Participating Institutions of the SDSS Collaboration including the 
Brazilian Participation Group, the Carnegie Institution for Science, 
Carnegie Mellon University, the Chilean Participation Group, the French Participation Group, Harvard-Smithsonian Center for Astrophysics, 
Instituto de Astrof\'isica de Canarias, The Johns Hopkins University, 
Kavli Institute for the Physics and Mathematics of the Universe (IPMU) / 
University of Tokyo, Lawrence Berkeley National Laboratory, 
Leibniz Institut f\"ur Astrophysik Potsdam (AIP), 
Max-Planck-Institut f\"ur Astronomie (MPIA Heidelberg), 
Max-Planck-Institut f\"ur Astrophysik (MPA Garching), 
Max-Planck-Institut f\"ur Extraterrestrische Physik (MPE), 
National Astronomical Observatories of China, New Mexico State University, 
New York University, University of Notre Dame, 
Observat\'ario Nacional / MCTI, The Ohio State University, 
Pennsylvania State University, Shanghai Astronomical Observatory, 
United Kingdom Participation Group,
Universidad Nacional Aut\'onoma de M\'exico, University of Arizona, 
University of Colorado Boulder, University of Oxford, University of Portsmouth, 
University of Utah, University of Virginia, University of Washington, University of Wisconsin, 
Vanderbilt University, and Yale University.


\bibliographystyle{mnras}
\bibliography{IMF}


\appendix
\section{Stacked spectra}
\label{sec:app_spectra}
\begin{figure*}
  \includegraphics[width=\linewidth]{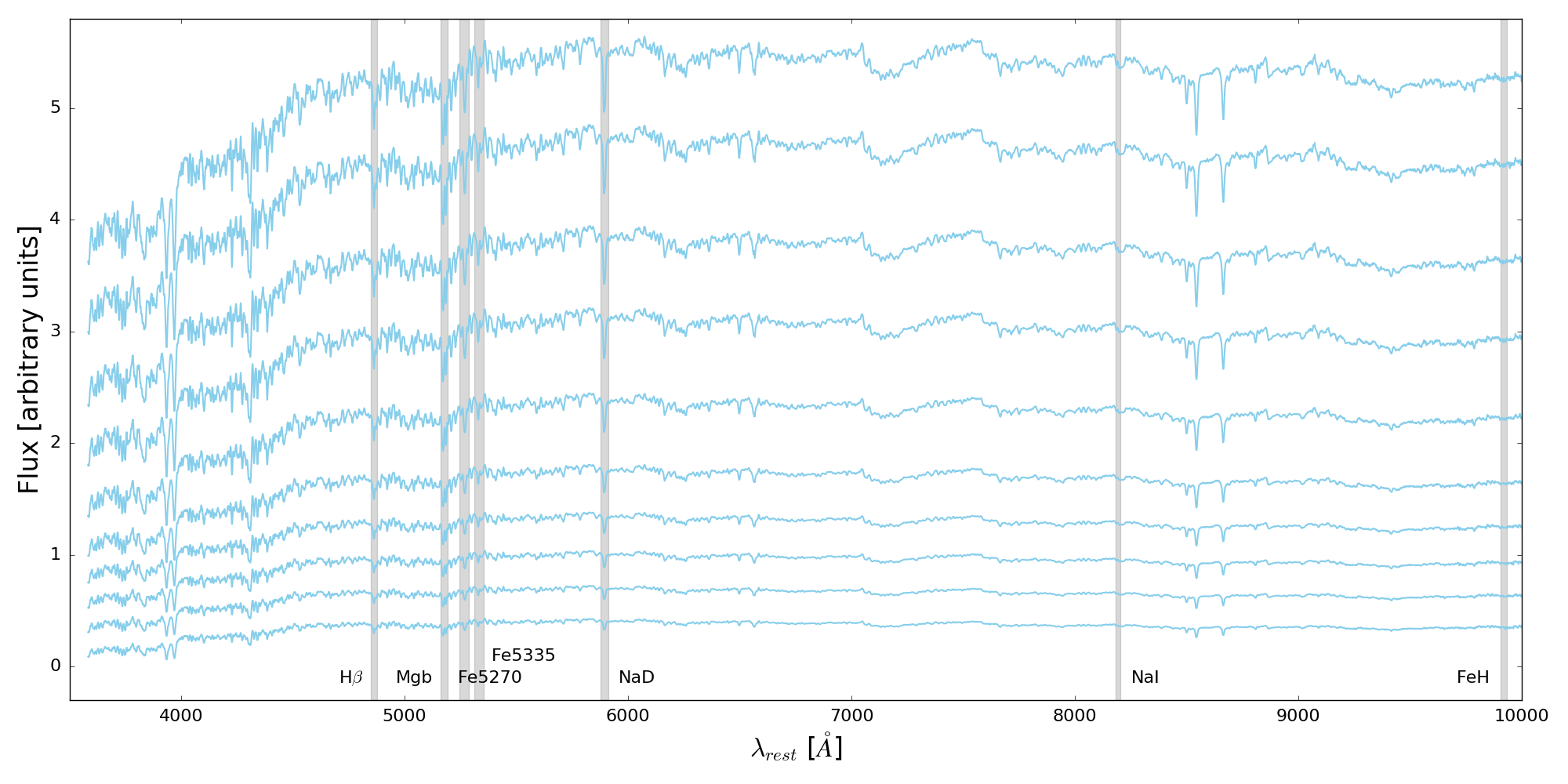}
  \caption{Stacked spectra for the low mass bin, $\log M/M_{\odot}=9.8 - 10.2$, top to bottom represents increasing radius out to 1 $R_\mathrm{e}$. The absorption features we use in our analysis are highlighted in grey.}
\label{fig:spectra2}
\end{figure*}

\begin{figure*}
  \includegraphics[width=\linewidth]{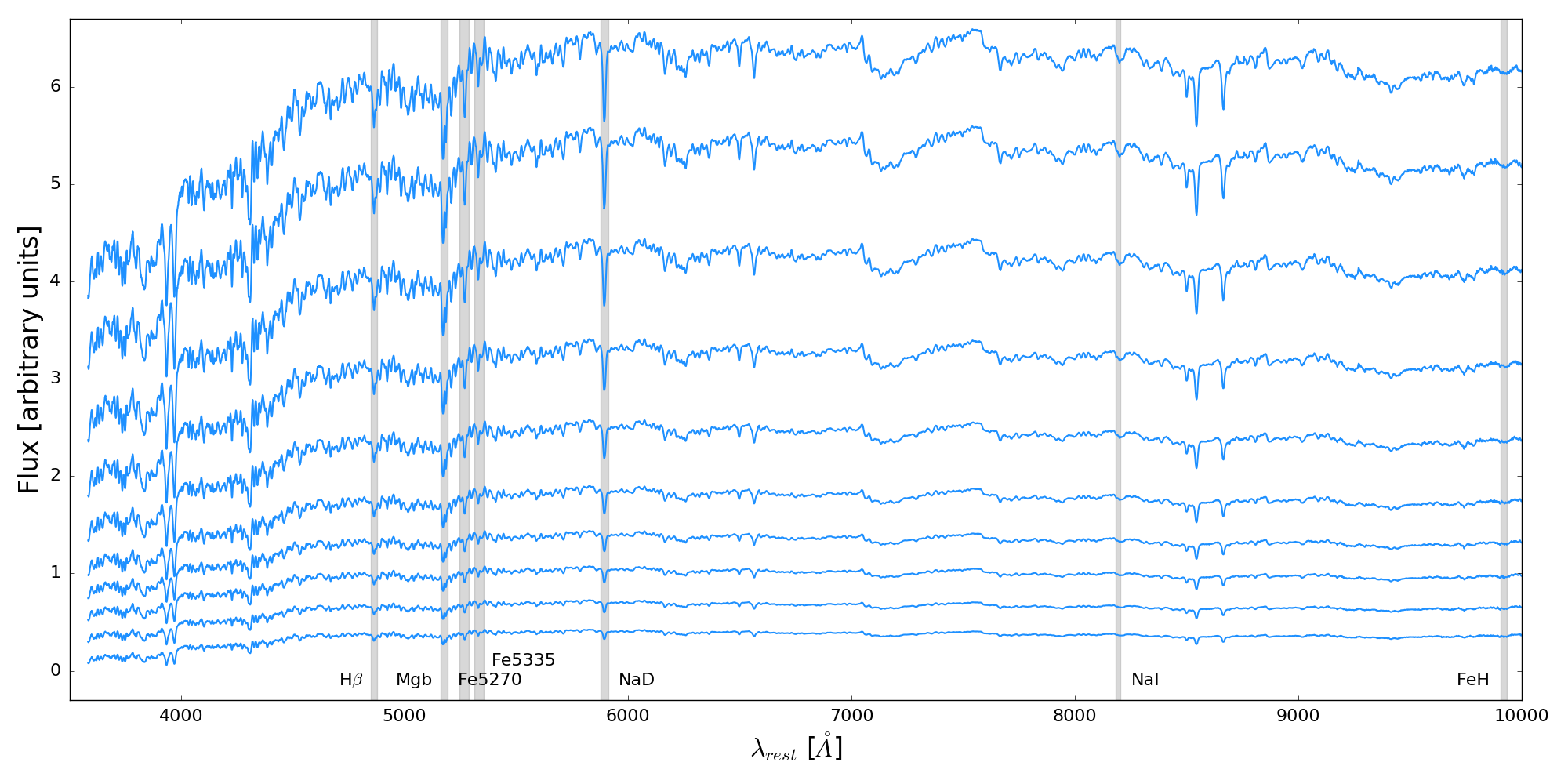}
  \caption{Same for the intermediate mass bin, $\log M/M_{\odot}=10.2 - 10.5$.}
\label{fig:spectra3}
\end{figure*}

\begin{figure*}
  \includegraphics[width=\linewidth]{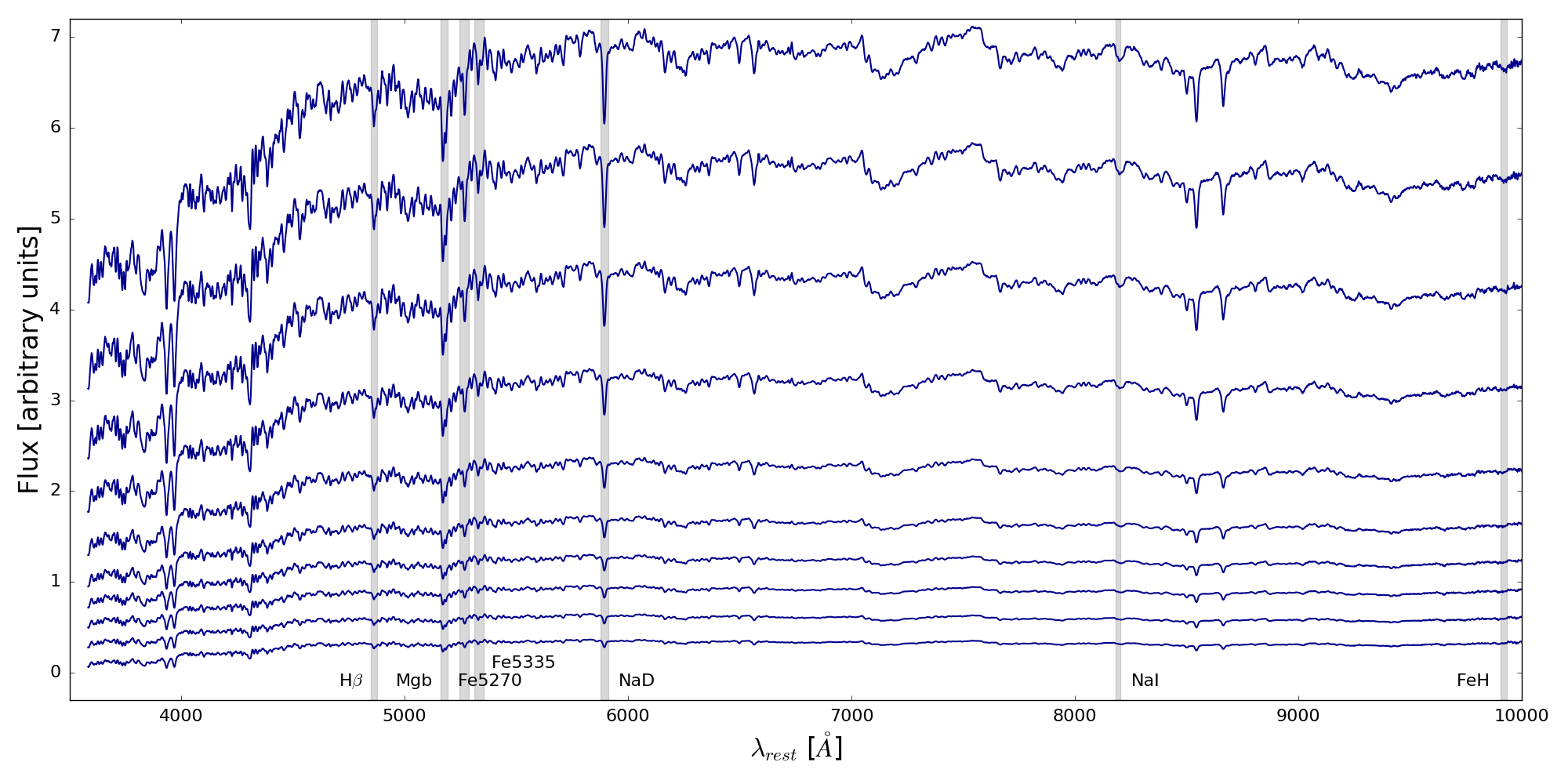}
  \caption{Same for the high mass bin, $\log M/M_{\odot}=10.5 - 10.8$.}
\label{fig:spectra4}
\end{figure*}
Figs.~\ref{fig:spectra2}, \ref{fig:spectra3} \& \ref{fig:spectra4} show our stacked spectra in order of increasing mass bin, from the central radial bin (top of plot) going out to 1 $R_\mathrm{e}$ (bottom of plot). The absorption indices used in this work are labelled and highlighted in grey shaded regions.

\section{Index measurements and stellar population parameters}
\label{sec:app_tables}
\begin{table*}
  \centering
  \caption{Measured equivalent widths at all radial bins from $0 - 1\;R_\mathrm{e}$ for the mass bins. Optical indices have been corrected to MILES resolution and NIR indices to $100\;$km/s.}
  \label{tab:ET_indices}
  \begin{tabular}{lcccccccccc}
  \hline
  $10^{9.9 - 10.2}\;M_{\odot}$ & 0.0-0.1 & 0.1-0.2 & 0.2-0.3 & 0.3-0.4 & 0.4-0.5 & 0.5-0.6 & 0.6-0.7 & 0.7-0.8 & 0.8-0.9 & 0.9-1.0 \\
  \hline
  H$\beta$ & 1.99$\pm$0.03    & 1.97$\pm$0.02    & 2.00$\pm$0.01    & 2.01$\pm$0.01    & 2.04$\pm$0.01    & 2.07$\pm$0.01    & 2.10$\pm$0.02    & 2.16$\pm$0.02    & 2.16$\pm$0.02    & 2.20$\pm$0.02    \\
  Mg$b$    & 3.76$\pm$0.02    & 3.77$\pm$0.01    & 3.75$\pm$0.01    & 3.74$\pm$0.01    & 3.65$\pm$0.01    & 3.61$\pm$0.01    & 3.5$\pm$0.01     & 3.41$\pm$0.01    & 3.42$\pm$0.01    & 3.35$\pm$0.02    \\
  Fe5270   & 3.13$\pm$0.02    & 3.10$\pm$0.01    & 3.09 $\pm$0.01   & 3.04$\pm$0.01    & 3.01$\pm$0.01   & 2.98$\pm$0.01    & 2.95$\pm$0.01    & 2.91$\pm$0.01    & 2.88$\pm$0.01    & 2.83$\pm$0.02    \\
  Fe5335   & 2.95$\pm$0.02    & 2.87$\pm$0.01    & 2.83$\pm$0.01    & 2.79$\pm$0.01    & 2.84$\pm$0.01    & 2.72$\pm$0.01    & 2.72$\pm$0.01    & 2.66$\pm$0.01    & 2.68$\pm$0.01    & 2.63$\pm$0.02    \\
  NaD      & 3.39$\pm$0.01    & 3.35$\pm$0.01    & 3.27$\pm$0.01    & 3.21$\pm$0.01    & 3.11$\pm$0.01    & 2.97$\pm$0.01    & 2.87$\pm$0.01    & 2.75$\pm$0.01    & 2.65$\pm$0.01    & 2.59$\pm$0.01    \\
  NaI      & 0.57$\pm$0.01    & 0.59$\pm$0.01    & 0.54$\pm$0.01    & 0.52$\pm$0.01    & 0.51$\pm$0.01    & 0.49$\pm$0.01    & 0.47$\pm$0.01    & 0.42$\pm$0.01    & 0.44$\pm$0.01    & 0.44$\pm$0.01    \\
  FeH      & 0.25$\pm$0.04    & 0.15$\pm$0.02    & 0.18$\pm$0.02    & 0.17$\pm$0.02    & 0.26$\pm$0.03    & -       & -       & -       & -       & -      \\
  \hline
  $10^{10.2 - 10.5}\;M_{\odot}$ & 0.0-0.1 & 0.1-0.2 & 0.2-0.3 & 0.3-0.4 & 0.4-0.5 & 0.5-0.6 & 0.6-0.7 & 0.7-0.8 & 0.8-0.9 & 0.9-1.0 \\
  \hline
  H$\beta$ & 1.82$\pm$0.02 & 1.87$\pm$0.01 & 1.90$\pm$0.01 & 1.94$\pm$0.01 & 1.91$\pm$0.01 & 1.87$\pm$0.01 & 1.91$\pm$0.01 & 1.94$\pm$0.01 & 2.02$\pm$0.01 & 1.99$\pm$0.01 \\
  Mg$b$    & 4.32$\pm$0.01 & 4.32$\pm$0.01 & 4.21$\pm$0.01 & 4.12$\pm$0.01 & 4.09$\pm$0.01 & 4.01$\pm$0.01 & 3.89$\pm$0.01 & 3.78$\pm$0.01 & 3.76$\pm$0.01 & 3.70$\pm$0.01 \\
  Fe5270   & 3.25$\pm$0.01 & 3.22$\pm$0.01 & 3.20$\pm$0.01 & 3.20$\pm$0.01 & 3.09$\pm$0.01 & 3.13$\pm$0.01 & 3.12$\pm$0.01 & 3.08$\pm$0.01 & 3.06$\pm$0.01 & 3.02$\pm$0.01 \\
  Fe5335   & 3.10$\pm$0.01 & 3.08$\pm$0.01 & 3.10$\pm$0.01 & 3.01$\pm$0.01 & 2.94$\pm$0.01 & 2.96$\pm$0.01 & 2.97$\pm$0.01 & 2.94$\pm$0.01 & 2.85$\pm$0.01 & 2.76$\pm$0.01 \\
  NaD      & 4.00$\pm$0.01 & 3.96$\pm$0.01 & 3.82$\pm$0.01 & 3.72$\pm$0.01 & 3.49$\pm$0.01 & 3.37$\pm$0.01 & 3.29$\pm$0.01 & 3.18$\pm$0.01 & 3.06$\pm$0.01 & 2.97$\pm$0.01 \\
  NaI      & 0.72$\pm$0.01 & 0.73$\pm$0.01 & 0.73$\pm$0.01 & 0.66$\pm$0.01 & 0.62$\pm$0.01 & 0.61$\pm$0.01 & 0.58$\pm$0.01 & 0.56$\pm$0.01 & 0.55$\pm$0.01 & 0.53$\pm$0.01 \\
  FeH      & 0.40$\pm$0.03 & 0.46$\pm$0.01 & 0.50$\pm$0.01 & 0.52$\pm$0.01 & 0.59$\pm$0.02 & -    & -    & -    & -    & -      \\
  \hline
  $10^{10.5 - 10.8}\;M_{\odot}$ & 0.0-0.1 & 0.1-0.2 & 0.2-0.3 & 0.3-0.4 & 0.4-0.5 & 0.5-0.6 & 0.6-0.7 & 0.7-0.8 & 0.8-0.9 & 0.9-1.0 \\
  \hline
  H$\beta$ & 1.82$\pm$0.01 & 1.80$\pm$0.01 & 1.84$\pm$0.01 & 1.89$\pm$0.01 & 1.90$\pm$0.01 & 1.91$\pm$0.01 & 1.91$\pm$0.01 & 1.92$\pm$0.01 & 1.95$\pm$0.01 & 1.95$\pm$0.01 \\
  Mg$b$    & 4.47$\pm$0.01 & 4.44$\pm$0.01 & 4.36$\pm$0.01 & 4.32$\pm$0.01 & 4.19$\pm$0.01 & 4.14$\pm$0.01 & 4.07$\pm$0.01 & 4.03$\pm$0.01 & 3.96$\pm$0.01 & 3.89$\pm$0.01 \\
  Fe5270   & 3.25$\pm$0.01 & 3.22$\pm$0.01 & 3.15$\pm$0.01 & 3.14$\pm$0.01 & 3.10$\pm$0.01 & 3.05$\pm$0.01 & 3.01$\pm$0.01 & 2.99$\pm$0.01 & 2.92$\pm$0.01 & 2.92$\pm$0.02 \\
  Fe5335   & 3.13$\pm$0.01 & 3.18$\pm$0.01 & 3.14$\pm$0.01 & 3.08$\pm$0.01 & 3.05$\pm$0.01 & 3.02$\pm$0.01 & 2.98$\pm$0.01 & 2.94$\pm$0.02 & 2.86$\pm$0.03 & 2.83$\pm$0.03 \\
  NaD      & 4.31$\pm$0.01 & 4.23$\pm$0.01 & 4.04$\pm$0.01 & 3.89$\pm$0.01 & 3.71$\pm$0.01 & 3.53$\pm$0.01 & 3.38$\pm$0.01 & 3.29$\pm$0.01 & 3.17$\pm$0.01 & 3.12$\pm$0.01 \\
  NaI      & 0.82$\pm$0.01 & 0.79$\pm$0.01 & 0.74$\pm$0.01 & 0.70$\pm$0.01 & 0.64$\pm$0.01 & 0.62$\pm$0.01 & 0.60$\pm$0.01 & 0.59$\pm$0.01 & 0.57$\pm$0.01 & 0.56$\pm$0.01 \\
  FeH      & 0.46$\pm$0.03 & 0.42$\pm$0.02 & 0.40$\pm$0.02 & 0.41$\pm$0.02 & 0.61$\pm$0.03 & -    & -    & -    & -    & -   \\
  \hline
  \end{tabular}
\end{table*}

\begin{table*}
  \centering
  \caption{Stellar population parameters and errors derived using the TMJ + M11-MARCS models and a combination of the optical indices + NaI.}
  \label{tab:parameters}
  \resizebox{\linewidth}{!}{%
  \begin{tabular}{lllllllllll}
  \hline
    Age (Gyr)        & 0-0.1                 & 0.1-0.2               & 0.2-0.3               & 0.3-0.4              & 0.4-0.5              & 0.5-0.6               & 0.6-0.7              & 0.7-0.8              & 0.8-0.9               & 0.9-1.0              \\
    \hline
    $9.9 - 10.2$  & 7.55$\pm$0.75 &  8.29$\pm$0.31 &  7.14$\pm$0.33 &  7.41$\pm$0.37 &  7.04$\pm$0.37 &  6.55$\pm$0.35 &  7.0$\pm$0.14 &  5.7$\pm$0.22 &  5.82$\pm$0.2 &  5.53$\pm$0.29 \\
    $10.2 - 10.5$ & 10.76$\pm$0.49 &  10.38$\pm$0.24 &  9.47$\pm$0.17 &  8.45$\pm$0.12 &  9.37$\pm$0.18 &  9.98$\pm$0.16 &  9.21$\pm$0.25 &  8.83$\pm$0.2 &  7.26$\pm$0.23 &  7.88$\pm$0.3 \\
    $10.5 - 10.8$ & 11.05$\pm$0.08 &  10.65$\pm$0.09 &  9.39$\pm$0.1 &  9.59$\pm$0.13 &  9.49$\pm$0.17 &  9.54$\pm$0.2 &  9.33$\pm$0.24 &  9.01$\pm$0.3 &  9.06$\pm$0.17 \\
      \hline
    [Z/H] (solar scaled)        & 0-0.1                 & 0.1-0.2               & 0.2-0.3               & 0.3-0.4              & 0.4-0.5              & 0.5-0.6               & 0.6-0.7              & 0.7-0.8              & 0.8-0.9               & 0.9-1.0              \\
      \hline
	$9.9-10.2$ & 0.022$\pm$0.019 &  -0.01$\pm$0.011 &  0.014$\pm$0.008 &  -0.007$\pm$0.009 &  -0.015$\pm$0.009 &  -0.025$\pm$0.01 &  -0.064$\pm$0.006 &  -0.051$\pm$0.01 &  -0.057$\pm$0.009 &  -0.07$\pm$0.013 \\
    $10.2 - 10.5$ & 0.062$\pm$0.015 &  0.07$\pm$0.01 &  0.077$\pm$0.004 &  0.082$\pm$0.005 &  0.021$\pm$0.005 &  -0.005$\pm$0.004 &  -0.006$\pm$0.007 &  -0.023$\pm$0.007 &  0.007$\pm$0.006 &  -0.034$\pm$0.005 \\
    $10.5 - 10.8$ & 0.098$\pm$0.002 &  0.077$\pm$0.003 &  0.063$\pm$0.003 &  0.086$\pm$0.003 &  0.045$\pm$0.004 &  0.027$\pm$0.005 &  0.001$\pm$0.006 &  -0.008$\pm$0.007 &  -0.031$\pm$0.008 &  -0.047$\pm$0.007 \\
	\hline
    [$\alpha$/Fe]        & 0-0.1                 & 0.1-0.2               & 0.2-0.3               & 0.3-0.4              & 0.4-0.5              & 0.5-0.6               & 0.6-0.7              & 0.7-0.8              & 0.8-0.9               & 0.9-1.0              \\
    \hline
    $9.9 - 10.2$ & 0.227$\pm$0.012 &  0.243$\pm$0.003 &  0.258$\pm$0.005 &  0.263$\pm$0.005 &  0.24$\pm$0.005 &  0.272$\pm$0.004 &  0.242$\pm$0.012 &  0.269$\pm$0.004 &  0.266$\pm$0.005 &  0.279$\pm$0.006 \\
    $10.2 - 10.5$ & 0.271$\pm$0.008 &  0.278$\pm$0.004 &  0.269$\pm$0.002 &  0.279$\pm$0.002 &  0.287$\pm$0.002 &  0.249$\pm$0.002 &  0.228$\pm$0.003 &  0.225$\pm$0.003 &  0.251$\pm$0.005 &  0.263$\pm$0.013 \\
    $10.5 - 10.8$ & 0.301$\pm$0.002 &  0.286$\pm$0.002 &  0.291$\pm$0.002 &  0.312$\pm$0.002 &  0.29$\pm$0.002 &  0.292$\pm$0.003 &  0.293$\pm$0.005 &  0.299$\pm$0.005 &  0.328$\pm$0.006 &  0.311$\pm$0.007 \\
	\hline
	[Na/Fe]         & 0-0.1                 & 0.1-0.2               & 0.2-0.3               & 0.3-0.4              & 0.4-0.5              & 0.5-0.6               & 0.6-0.7              & 0.7-0.8              & 0.8-0.9               & 0.9-1.0              \\
    \hline
    $9.9 - 10.2$ & 0.31$\pm$0.02 &  0.33$\pm$0.0 &  0.31$\pm$0.01 &  0.31$\pm$0.01 &  0.27$\pm$0.01 &  0.26$\pm$0.01 &  0.21$\pm$0.01 &  0.22$\pm$0.01 &  0.18$\pm$0.01 &  0.19$\pm$0.01 \\
    $10.2 - 10.5$ & 0.41$\pm$0.08 &  0.42$\pm$0.01 &  0.38$\pm$0.0 &  0.38$\pm$0.0 &  0.35$\pm$0.0 &  0.28$\pm$0.0 &  0.26$\pm$0.0 &  0.24$\pm$0.0 &  0.23$\pm$0.01 &  0.23$\pm$0.02 \\
    $10.5 - 10.8$ & 0.49$\pm$0.01 &  0.48$\pm$0.0 &  0.45$\pm$0.01 &  0.41$\pm$0.0 &  0.4$\pm$0.0 &  0.36$\pm$0.0 &  0.34$\pm$0.01 &  0.32$\pm$0.01 &  0.33$\pm$0.01 &  0.31$\pm$0.01 \\
	\hline
    IMF Slope        & 0-0.1                 & 0.1-0.2               & 0.2-0.3               & 0.3-0.4              & 0.4-0.5              & 0.5-0.6               & 0.6-0.7              & 0.7-0.8              & 0.8-0.9               & 0.9-1.0              \\
    \hline
    $9.9 - 10.2$ & 1.64$\pm$0.35 &  1.37$\pm$0.11 &  1.49$\pm$0.16 &  1.2$\pm$0.14 &  1.29$\pm$0.16 &  1.36$\pm$0.16 &  1.0$\pm$0.09 &  1.0$\pm$0.12 &  1.25$\pm$0.15 &  1.22$\pm$0.19 \\
    $10.2 - 10.5$ & 2.0$\pm$0.14 &  1.97$\pm$0.1 &  2.3$\pm$0.08 &  2.21$\pm$0.08 &  1.54$\pm$0.07 &  1.32$\pm$0.06 &  1.31$\pm$0.08 &  1.27$\pm$0.08 &  1.89$\pm$0.13 &  1.35$\pm$0.11 \\
    $10.5 - 10.8$ & 2.03$\pm$0.1 &  1.98$\pm$0.03 &  1.79$\pm$0.03 &  2.0$\pm$0.04 &  1.53$\pm$0.05 &  1.54$\pm$0.07 &  1.33$\pm$0.08 &  1.38$\pm$0.08 &  1.31$\pm$0.11 &  1.15$\pm$0.09 \\
    \hline
    \end{tabular}}
\end{table*}

\begin{table*}
  \centering
  \caption{Same as Table~\ref{tab:parameters} for the VCJ models.}
  \label{tab:parameters_vcj}
  \resizebox{\linewidth}{!}{%
  \begin{tabular}{lllllllllll}
  \hline
    Age (Gyr)        & 0-0.1                 & 0.1-0.2               & 0.2-0.3               & 0.3-0.4              & 0.4-0.5              & 0.5-0.6               & 0.6-0.7              & 0.7-0.8              & 0.8-0.9               & 0.9-1.0              \\
    \hline
    $9.9 - 10.2$  & 0.89$\pm$0.03 &  0.89$\pm$0.02 &  0.86$\pm$0.01 &  0.83$\pm$0.01 &  0.81$\pm$0.02 &  0.8$\pm$0.03 &  0.78$\pm$0.03 &  0.67$\pm$0.03 &  0.7$\pm$0.02 &  0.66$\pm$0.02 \\
    $10.2 - 10.5$ & 0.95$\pm$0.02 &  0.95$\pm$0.01 &  0.95$\pm$0.01 &  0.8$\pm$0.01 &  0.83$\pm$0.01 &  0.9$\pm$0.01 &  0.9$\pm$0.01 &  0.89$\pm$0.01 &  0.79$\pm$0.01 &  0.85$\pm$0.01 \\
    $10.5 - 10.8$ & 0.95$\pm$0.01 &  0.95$\pm$0.0 &  0.95$\pm$0.0 &  0.84$\pm$0.01 &  0.9$\pm$0.01 &  0.81$\pm$0.01 &  0.9$\pm$0.02 &  0.91$\pm$0.02 &  0.8$\pm$0.02 &  0.86$\pm$0.03 \\
      \hline
    [Z/H] (solar scaled)        & 0-0.1                 & 0.1-0.2               & 0.2-0.3               & 0.3-0.4              & 0.4-0.5              & 0.5-0.6               & 0.6-0.7              & 0.7-0.8              & 0.8-0.9               & 0.9-1.0              \\
      \hline
	$9.9-10.2$ & -0.0$\pm$0.029 &  0.0$\pm$0.013 &  0.016$\pm$0.01 &  0.03$\pm$0.005 &  0.019$\pm$0.01 &  0.01$\pm$0.016 &  -0.0$\pm$0.032 &  0.035$\pm$0.025 &  0.011$\pm$0.011 &  0.019$\pm$0.018 \\
    $10.2 - 10.5$ & 0.094$\pm$0.012 &  0.09$\pm$0.007 &  0.062$\pm$0.005 &  0.149$\pm$0.006 &  0.114$\pm$0.007 &  0.047$\pm$0.006 &  0.019$\pm$0.008 &  0.0$\pm$0.009 &  0.055$\pm$0.008 &  -0.0$\pm$0.013 \\
    $10.5 - 10.8$ & 0.12$\pm$0.005 &  0.115$\pm$0.003 &  0.093$\pm$0.003 &  0.166$\pm$0.005 &  0.083$\pm$0.011 &  0.137$\pm$0.006 &  0.051$\pm$0.013 &  0.035$\pm$0.015 &  0.089$\pm$0.01 &  0.033$\pm$0.019 \\
	\hline
    [$\alpha$/Fe]        & 0-0.1                 & 0.1-0.2               & 0.2-0.3               & 0.3-0.4              & 0.4-0.5              & 0.5-0.6               & 0.6-0.7              & 0.7-0.8              & 0.8-0.9               & 0.9-1.0              \\
    \hline
    $9.9 - 10.2$ & 0.178$\pm$0.019 &  0.207$\pm$0.013 &  0.223$\pm$0.011 &  0.25$\pm$0.01 &  0.218$\pm$0.01 &  0.244$\pm$0.012 &  0.229$\pm$0.014 &  0.263$\pm$0.013 &  0.241$\pm$0.012 &  0.267$\pm$0.018 \\
    $10.2 - 10.5$ & 0.269$\pm$0.007 &  0.273$\pm$0.004 &  0.234$\pm$0.005 &  0.305$\pm$0.005 &  0.314$\pm$0.005 &  0.243$\pm$0.005 &  0.205$\pm$0.009 &  0.185$\pm$0.01 &  0.238$\pm$0.009 &  0.235$\pm$0.01 \\
    $10.5 - 10.8$ & 0.287$\pm$0.004 &  0.279$\pm$0.002 &  0.278$\pm$0.002 &  0.339$\pm$0.004 &  0.275$\pm$0.009 &  0.33$\pm$0.006 &  0.276$\pm$0.011 &  0.269$\pm$0.007 &  0.347$\pm$0.013 &  0.291$\pm$0.017 \\
	\hline
	[Na/Fe]        & 0-0.1                 & 0.1-0.2               & 0.2-0.3               & 0.3-0.4              & 0.4-0.5              & 0.5-0.6               & 0.6-0.7              & 0.7-0.8              & 0.8-0.9               & 0.9-1.0              \\
    \hline
    $9.9 - 10.2$ & 0.27$\pm$0.03 &  0.29$\pm$0.02 &  0.26$\pm$0.02 &  0.25$\pm$0.01 &  0.22$\pm$0.02 &  0.22$\pm$0.02 &  0.21$\pm$0.01 &  0.2$\pm$0.05 &  0.16$\pm$0.01 &  0.16$\pm$0.02 \\
    $10.2 - 10.5$ & 0.33$\pm$0.01 &  0.33$\pm$0.0 &  0.29$\pm$0.01 &  0.32$\pm$0.01 &  0.29$\pm$0.01 &  0.23$\pm$0.01 &  0.21$\pm$0.02 &  0.2$\pm$0.02 &  0.18$\pm$0.01 &  0.2$\pm$0.01 \\
    $10.5 - 10.8$ & 0.39$\pm$0.01 &  0.37$\pm$0.0 &  0.35$\pm$0.0 &  0.34$\pm$0.01 &  0.32$\pm$0.02 &  0.29$\pm$0.01 &  0.26$\pm$0.02 &  0.25$\pm$0.02 &  0.26$\pm$0.01 &  0.24$\pm$0.03 \\
	\hline
	IMF slope        & 0-0.1                 & 0.1-0.2               & 0.2-0.3               & 0.3-0.4              & 0.4-0.5              & 0.5-0.6               & 0.6-0.7              & 0.7-0.8              & 0.8-0.9               & 0.9-1.0              \\
    \hline
    $9.9 - 10.2$ & 1.17$\pm$0.5 &  1.61$\pm$0.19 &  1.06$\pm$0.4 &  1.05$\pm$0.15 &  0.94$\pm$0.53 &  1.04$\pm$0.68 &  0.72$\pm$0.6 &  0.64$\pm$0.41 &  1.03$\pm$0.68 &  1.32$\pm$0.58 \\
    $10.2 - 10.5$ & 2.71$\pm$0.07 &  2.79$\pm$0.04 &  2.73$\pm$0.04 &  2.65$\pm$0.02 &  2.56$\pm$0.03 &  2.4$\pm$0.03 &  2.03$\pm$0.14 &  1.78$\pm$0.17 &  2.47$\pm$0.08 &  1.46$\pm$0.18 \\
    $10.5 - 10.8$ & 2.98$\pm$0.04 &  2.9$\pm$0.02 &  2.68$\pm$0.02 &  2.87$\pm$0.05 &  2.3$\pm$0.37 &  2.61$\pm$0.03 &  2.3$\pm$0.33 &  2.3$\pm$0.0 &  2.55$\pm$0.06 &  2.3$\pm$0.49 \\
    \hline
    \end{tabular}}
\end{table*}
We provide our index-strength measurements for the optical indices H$\beta$, Mg$b$, Fe5270, Fe5335, NaD and the NIR indices NaI and FeH for all radial bins in the mass ranges $9.9 - 10.2\;\log M/M_{\odot}$, $10.2 - 10.5\;\log M/M_{\odot}$ \& $10.5 - 10.8\;\log M/M_{\odot}$ in Table~\ref{tab:ET_indices}. The optical indices have been corrected to MILES resolution and the NIR indices to $100\;$km/s. Due to the extremely high S/N of our stacked spectra, the Monte-Carlo based errors on the index measurements are negligible for a lot of measurements and have been rounded to the nearest significant digit.

Also provided are the following parameters at each radial bin: age, metallicity, [$\alpha$/Fe], and [Na/Fe], and the IMF slope, derived using TMJ + M11-MARCS models in Table~\ref{tab:parameters} and the VCJ models in Table~\ref{tab:parameters_vcj}.

\section{Extinction from full spectrum fitting}
\label{sec:app_dust}
\begin{figure}
 \includegraphics[width=.9\linewidth]{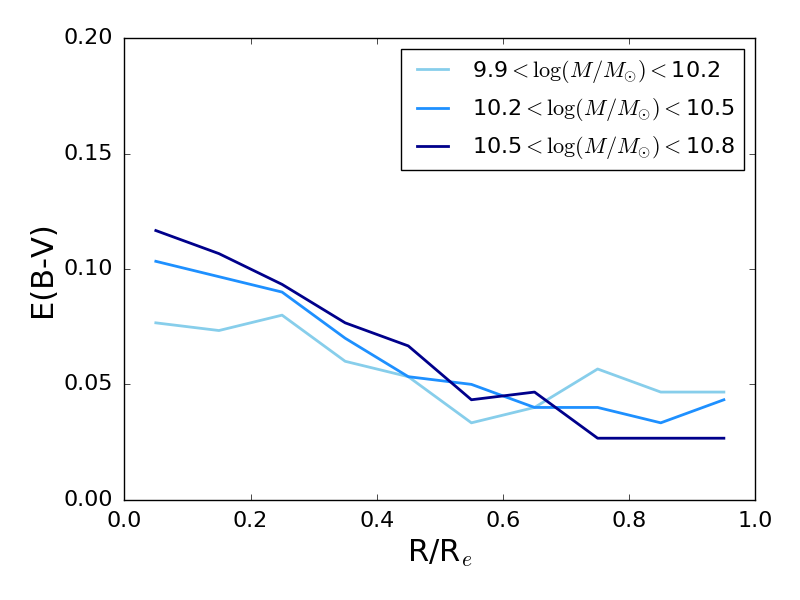}
\caption{E(B-V) derived using the full spectral fitting code FIREFLY.}
\label{fig:extinction}
\end{figure}
As mentioned in Section~\ref{sec:na}, we investigate the reddening effect on our stacked spectra by carrying out full spectrum fitting via the FIREFLY code \citep{Wilkinson2015,Wilkinson2017}. Using M11-MILES models and a Salpeter IMF, the derived E(B-V) values are plotted for the mass bins as a function of radius in Fig.~\ref{fig:extinction}.

Generally, the dust values are low, with a maximum $\sim$ 0.1. The values do increase towards galaxy centres, and more for the high mass bins. Therefore, although the effect on NaD may be negligible, our [Na/Fe] values could be affected. Assuming contamination due to interstellar absorption, we would derive even steeper IMF slopes and steeper radial gradients in the IMF for the high mass bins.

\begin{figure*}
  \includegraphics[width=.32\linewidth]{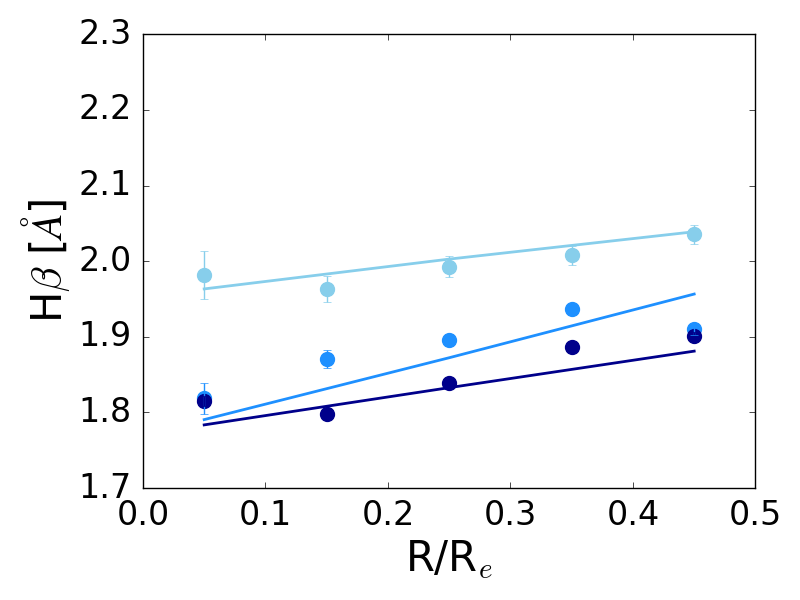}
  \includegraphics[width=.32\linewidth]{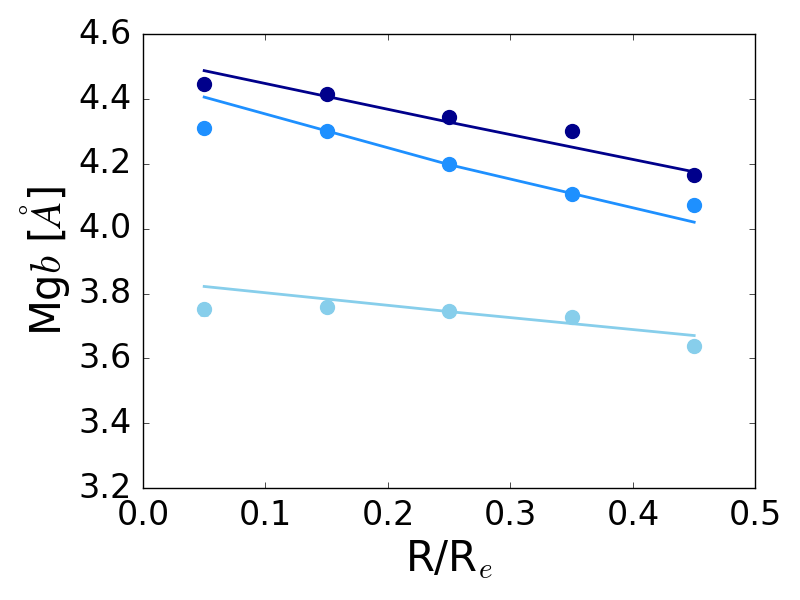}
  \includegraphics[width=.32\linewidth]{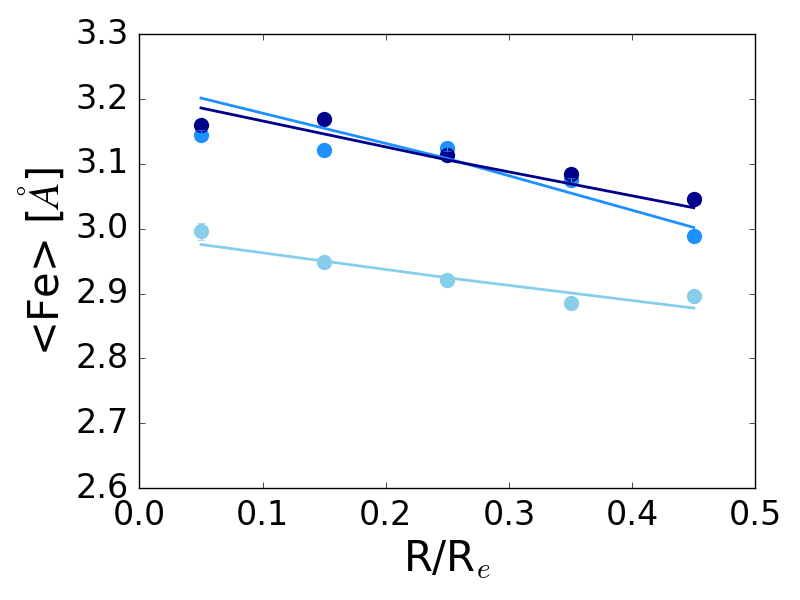}
  \includegraphics[width=.32\linewidth]{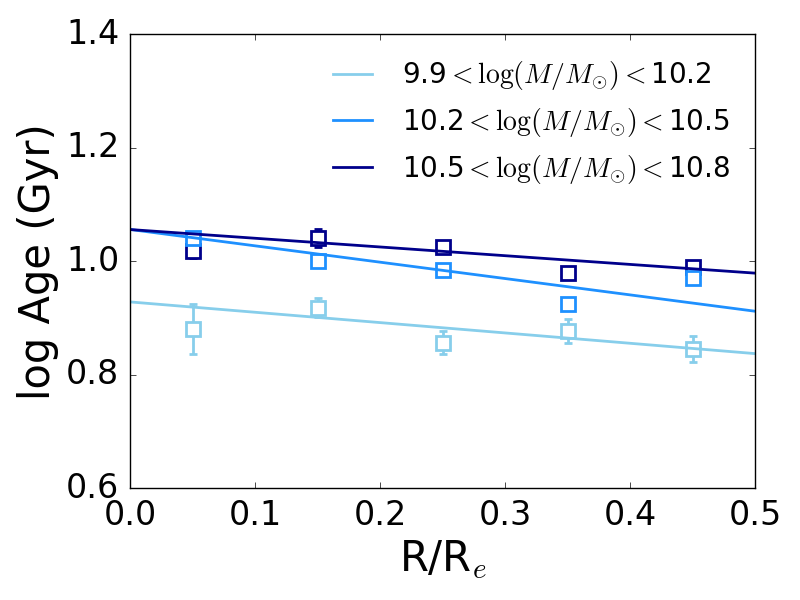}
  \includegraphics[width=.32\linewidth]{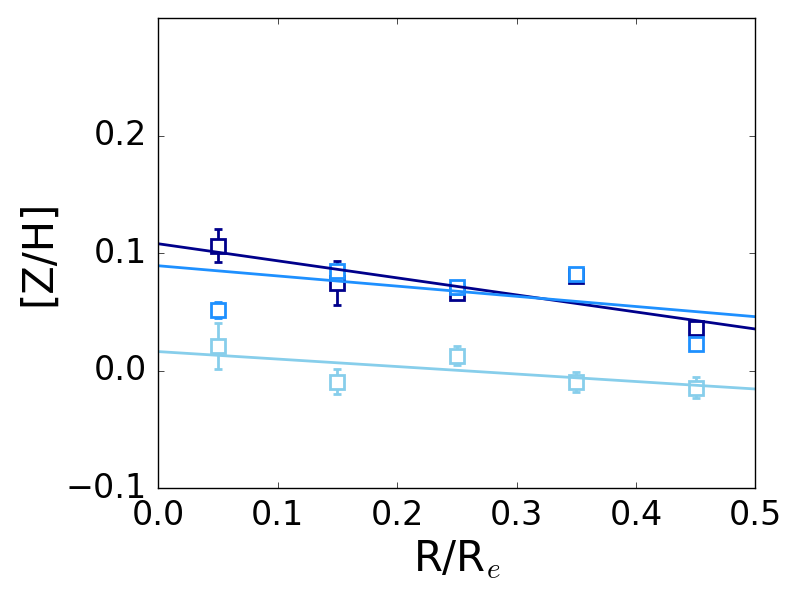}
  \includegraphics[width=.32\linewidth]{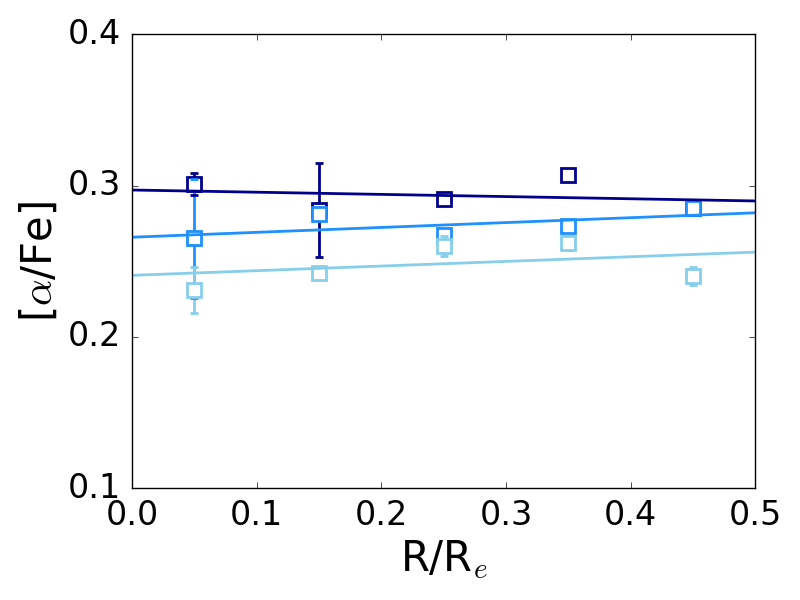}
\caption{{\it Top row}: H$\beta$, Mg$b$ and <Fe> for each mass bin as a function of radius.
Line-strengths are measured on the stacked spectra with 1-$\sigma$ errors calculated using a Monte Carlo-based analysis. The lines are TMJ model fittings derived through chi-squared minimisation (see Section~\ref{sec:results}). {\it Bottom row}: TMJ-derived age, metallicity and [$\alpha$/Fe] as a function of radius for different mass bins. The lines are calculated by fitting a straight line to the derived parameter at each radius in order to match the data.}
\label{fig:indices_FeH}
\end{figure*}

\begin{figure*}
  \includegraphics[width=.32\linewidth]{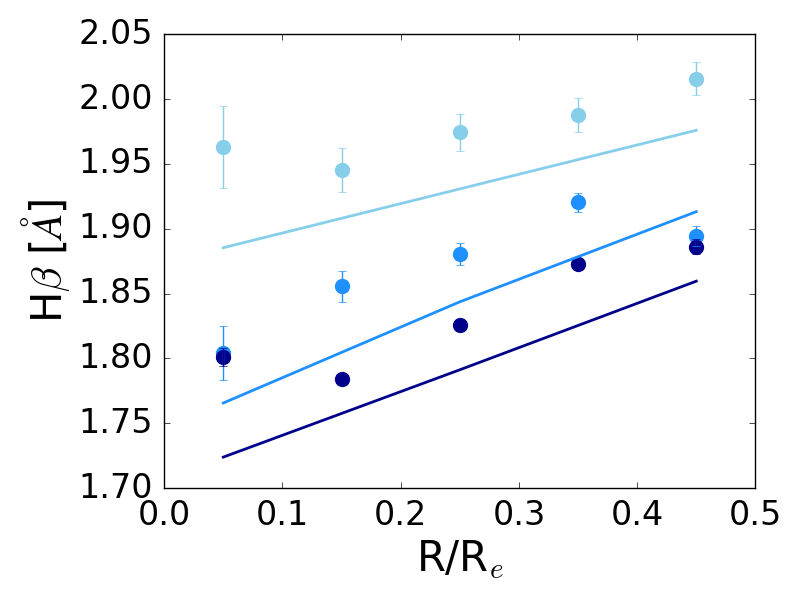}
  \includegraphics[width=.32\linewidth]{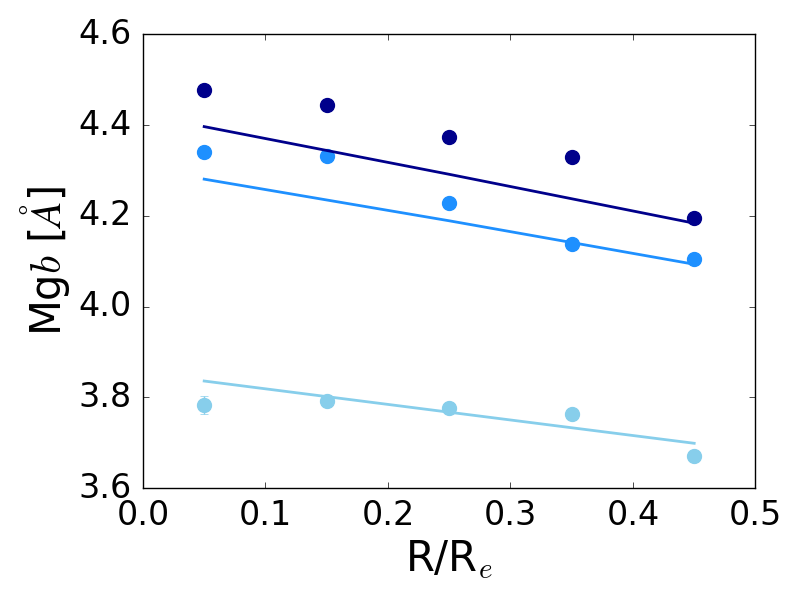}
  \includegraphics[width=.32\linewidth]{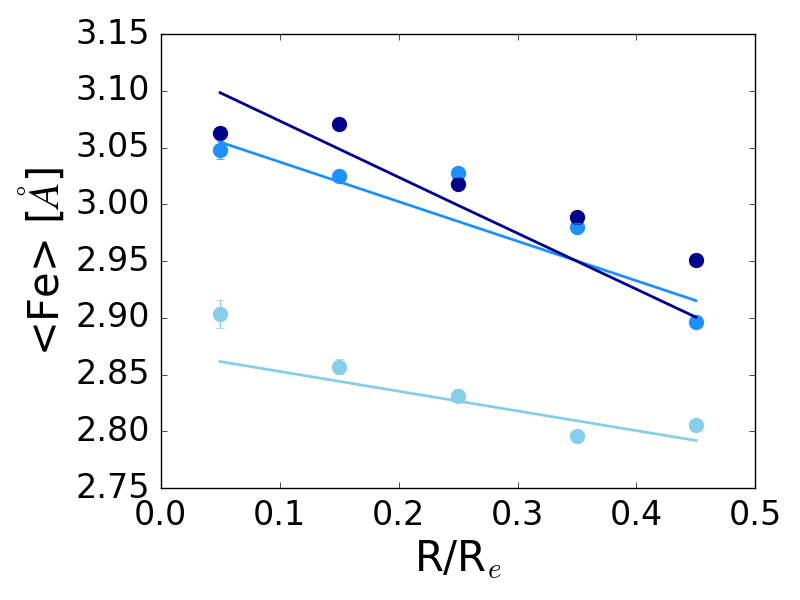}
  \includegraphics[width=.32\linewidth]{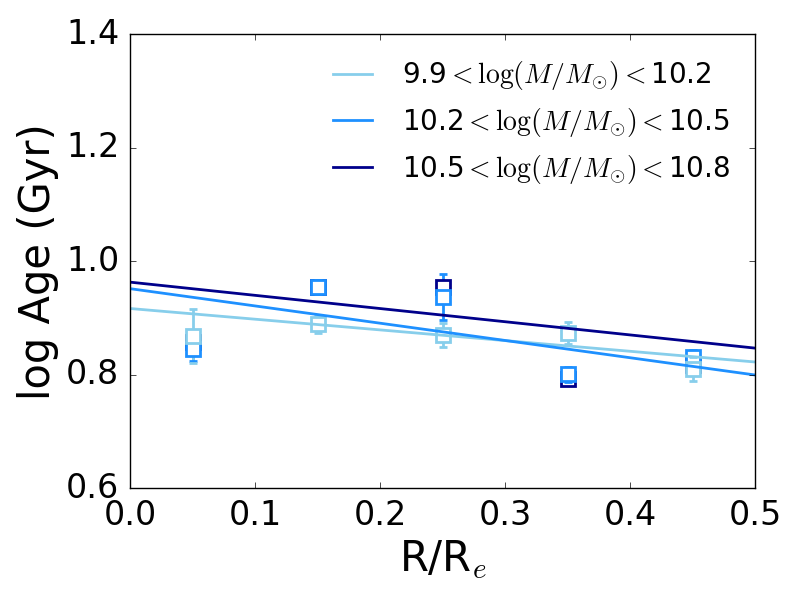}
  \includegraphics[width=.32\linewidth]{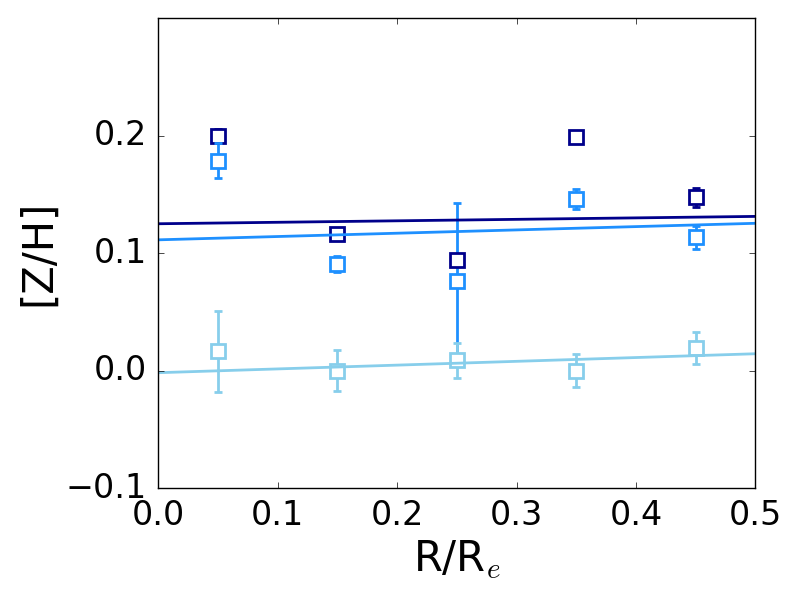}
  \includegraphics[width=.32\linewidth]{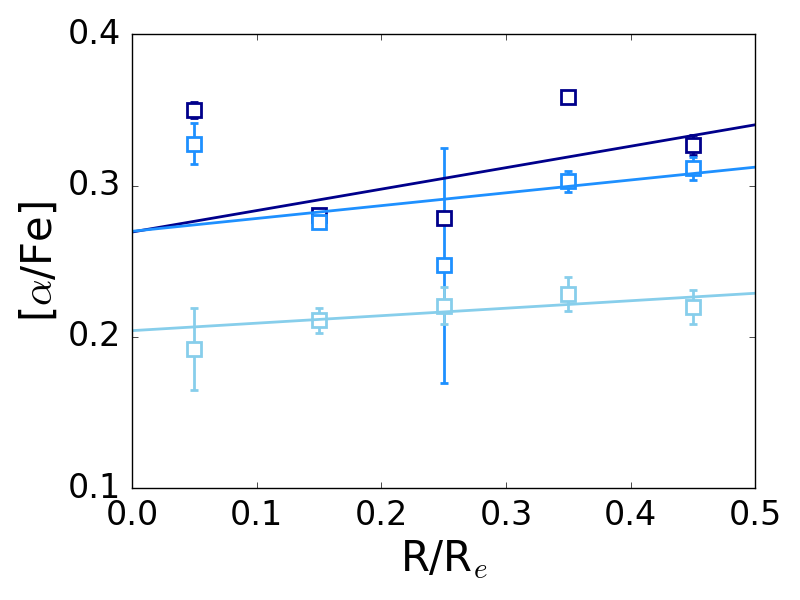}
\caption{Same as Fig.~\ref{fig:indices_FeH} for VCJ models.}
\label{fig:indices_Conroy_FeH}
\end{figure*}

\begin{figure*}
  \includegraphics[width=.35\linewidth]{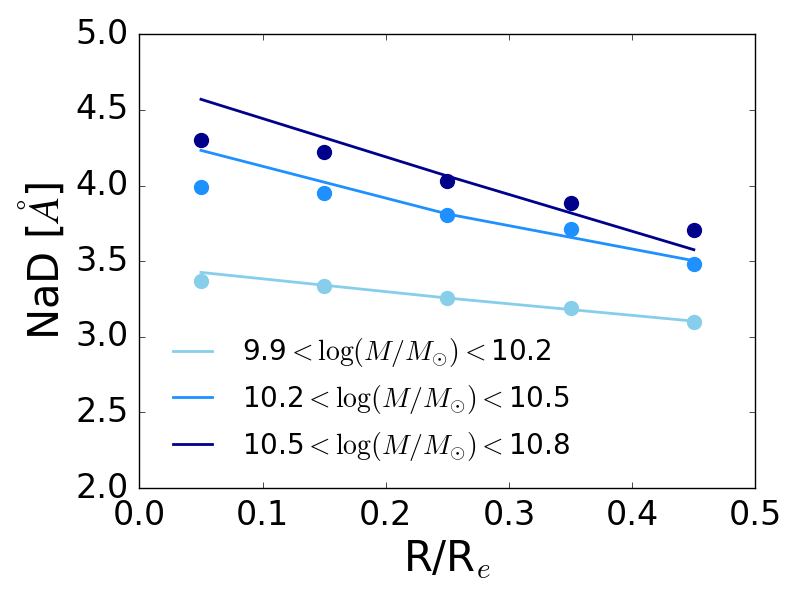}
  \includegraphics[width=.35\linewidth]{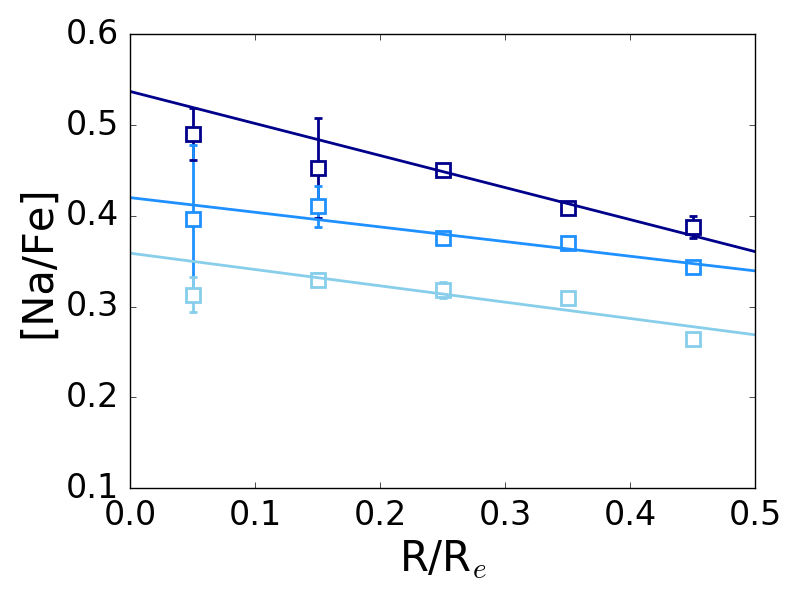}
  \includegraphics[width=.35\linewidth]{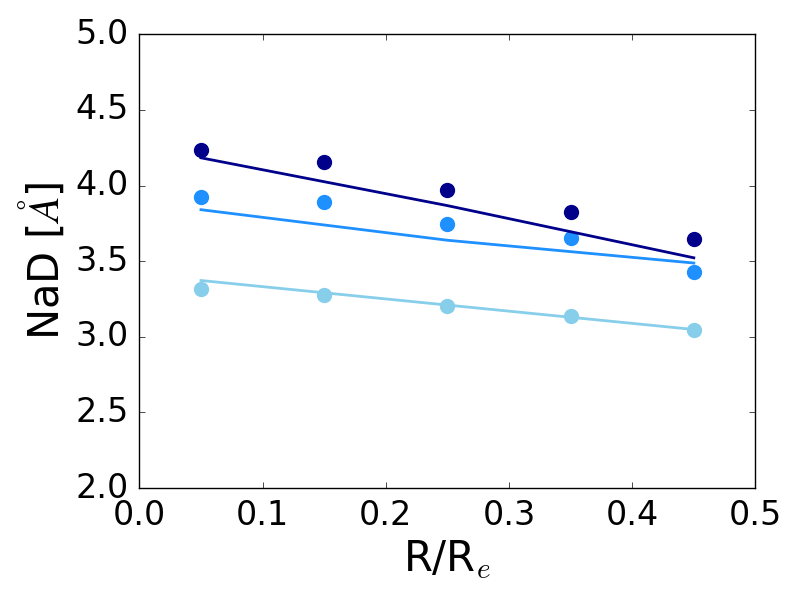}
  \includegraphics[width=.35\linewidth]{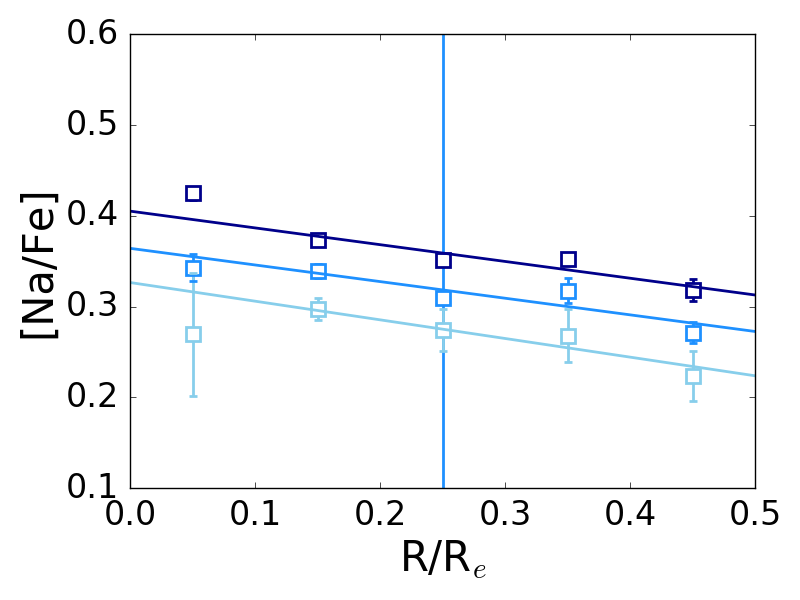}
\caption{NaD equivalent width as a function of radius for the mass bins, with 1-$\sigma$ errors calculated using a Monte Carlo-based analysis, and stellar population fittings using TMJ (top panels) and VCJ (bottom panels). Left: Lines are models with the simultaneously derived parameters of age, metallicity, [$\alpha$/Fe], [Na/Fe] and IMF slope. Right: [Na/Fe] abundance (including [$\alpha$/Fe] given before) for the best-fit models.}
\label{fig:indices_nad_FeH}
\end{figure*}

\section{Optical indices and stellar population parameters using FeH analysis}
\label{sec:app_feh}
Shown here are the indices H$\beta$, Mg$b$, <Fe>, and NaD with the model fits using TMJ and VCJ. The derived parameters, different to Section~\ref{sec:age_elements} \& \ref{sec:na} since FeH is included in the simultaneous fitting procedure rather than NaI, are shown in Fig.~\ref{fig:indices_FeH}, Fig.~\ref{fig:indices_Conroy_FeH} \& Fig.~\ref{fig:indices_nad_FeH}.


\bsp	
\label{lastpage}
\end{document}